\documentclass[aps,prx,twocolumn,showpacs,superscriptaddress]{revtex4-2}
\usepackage{latexsym}
\usepackage{amssymb}
\usepackage{graphicx}
\usepackage{amsmath}
\usepackage{tikz}
\usepackage{bm}
\usepackage[colorlinks,linkcolor=red,citecolor=blue,urlcolor=blue]{hyperref}
\usepackage{verbatim}
\usepackage{bbding}
\usepackage{mathrsfs}
\usepackage{extarrows}
\usepackage{comment}
\usepackage{mathtools,slashed}
\usepackage{soul}
\usepackage[toc,page]{appendix}
\usepackage[vcentermath]{youngtab}
\usepackage{multirow}
\usepackage[width=.75\textwidth]{caption}
\usepackage{atbegshi,picture}
\usepackage{lipsum}

\newcommand{\mr}{moir\'{e} }

\makeatletter
\renewcommand\@makecaption[2]{
	\par
	\vskip\abovecaptionskip
	\begingroup
	\small\rmfamily
	\begingroup
	\samepage
	\flushing
	\let\footnote\@footnotemark@gobble
	\@make@capt@title{#1}{#2}\par
	\endgroup
	\endgroup
	\vskip\belowcaptionskip
}
\makeatother

\AtBeginShipoutNext{\AtBeginShipoutUpperLeft{
		\put(\dimexpr\paperwidth-1cm\relax,-1.5cm){\makebox[0pt][r]}
}}

\makeatletter
\begin{document}

	\title{Dual topological nonlinear sigma models of $\text{QED}$ theory by dimensional reduction and monopole operators }

\author{Zhengzhi Wu}
\email{wuzz18@mails.tsinghua.edu.cn}
\thanks{Both authors contributed equally to this work.}
\affiliation{Institute for Advanced Study, Tsinghua University, Beijing 100084, China}
\author{Linhao Li}     
\email{linhaoli601@issp.u-tokyo.ac.jp}
\thanks{Both authors contributed equally to this work.}
\affiliation{Institute for Solid State Physics, The University of Tokyo. Kashiwa, Chiba 277-8581, Japan}

\date{\today}

\begin{abstract}
Nonlinear $\sigma$ models (NLSM) with topological terms, i.e., Wess-Zumino-Witten (WZW) terms, or topological NLSM, are potent descriptions of many critical points and phases beyond the Landau paradigm. These critical systems include the deconfined quantum critical points (DQCP) between the Neel order and valance bond solid, and the Dirac spin liquid, in which the topological NLSMs are dual descriptions of the corresponding fermionic models or $\text{QED}$ theory.  In this paper, we propose a dimensional reduction scheme to derive the $\text{U}(1)$ gauged topological NLSM in $n$-dimensional spacetime on a general target space represented by a Hermitian matrix from the dual QED theory. Compared with the famous Abanov-Wiegmann (AW) mechanism, which generally requires the fermions to be Dirac fermions in the infrared (IR), our method is also applicable to non-relativistic fermions in IR, which can have quadratic dispersion or even a Fermi surface.  As concrete examples, we construct several two dimensional lattice models, whose IR theories are all the $N_f=4$ $\text{QED}_3$ with fermions of quadratic dispersion and show that its topological NLSM dual description has level-2 WZW terms on the Grassmannian manifold $\frac{\text{U(4)}}{\text{U(2)}\times \text{U(2)}}$ coupled with a dynamical $\text{U(1)}$ gauge field. 
We also study 't Hooft anomaly matching and the same effect of defects in both theories, such as interface, gauge monopoles and vortexes, which further support our duality. Finally, we discuss how the macroscopic symmetries act on the $\text{U(1)}$ monopole operators and the corresponding quantum number. 

\end{abstract}
	
	\maketitle
	\section{Introduction}
	
One central issue of quantum many-body physics is understanding quantum phases and phase transitions.
	In the past several decade of years, gapped phases, such as spontaneously symmetry breaking phases, phases with topological order (TO) \cite{KITAEV20032,KITAEV20062,doi:10.1080/00018739500101566,doi:10.1142/S0217979290000139,PhysRevB.65.165113,PhysRevB.39.11413,PhysRevLett.96.110405,Wen2013,RevModPhys.89.041004,Xiao:803748,vijay2016fracton,Haah2011Local,pretko2020fracton,PhysRevResearch.4.L032008,https://doi.org/10.48550/arxiv.2204.03045,https://doi.org/10.48550/arxiv.2205.09545} and symmetry protected topological (SPT) phases \cite{PhysRevB.81.064439,chen2011complete,hasan2010colloquium,RevModPhys.83.1057, doi:10.1126/science.1227224} have been relatively well understood through exactly solvable models, topological field theory and systematic mathematical classification, etc. In comparison, the phase transitions between them are far from being well understood. Many exotic quantum critical points have been discovered or proposed in the vicinity of various intertwined orders. 
For example, DQCP \cite{doi:10.1126/science.1091806,PhysRevB.70.144407,PhysRevLett.89.247201,PhysRevB.80.180414,PhysRevB.82.174428,PhysRevLett.107.110601,PhysRevLett.111.087203,PhysRevX.5.041048,doi:10.1126/science.aad5007,PhysRevX.7.031051,huang2022competing,huang2022non} and the Dirac spin liquid \cite{RevModPhys.78.17,PhysRevB.63.014413,PhysRevB.72.104404,PhysRevB.77.224413,PhysRevLett.98.117205,PhysRevB.93.144411,doi:10.1126/sciadv.aat5535,PhysRevX.10.011033,song2019unifying,https://doi.org/10.48550/arxiv.2103.13405,PhysRevX.7.031020} have been proposed as critical points and phases between the antiferromagnetic order (AFM) and the valance bond solid (VBS) order. Interestingly, these critical points or phases are argued to be second-order phase transitions without fine tuning \cite{doi:10.1126/science.1091806,PhysRevB.70.144407} (or near a fixed point of the renormalization group flow \cite{PhysRevX.5.041048,PhysRevX.7.031051,PhysRevB.102.020407}); or even flow to conformal field theories \cite{PhysRevD.94.065026,PhysRevD.93.045020,PhysRevD.100.054514}, although the broken symmetry group of one ordered phase is not the subgroup of the other and thus they are beyond the traditional Landau-Ginzburg paradigm. Further, the DQCP of the spin-$\frac{1}{2}$ models on the square lattice has nontrivial properties with respect to real space singularities of ordered phases, which are also known as the topological defects \cite{PhysRevB.74.064405}. Concretely, the defects of one ordered phase carry the nontrivial quantum number of the other ordered phase, which is theoretically captured by WZW terms in the nonlinear $\sigma$ model description. For instance, the vortex of the VBS order parameter carries a
spinor representation of the onsite spin rotation symmetry and the skyrmion
of the AFM has a nonzero lattice momentum.  Theoretically, one field theory description in IR of these critical points and phases is the 2+1d $\text{QED}$ theory with four flavours of Dirac fermions, or the $N_f=4$ $\text{QED}_3$ theory \cite{PhysRevX.10.011033,PhysRevB.72.104404,PhysRevB.77.224413}, where the emergent Dirac fermions and dynamic gauge fields come from the parton treatment of spin models, which fractionalizes the spin degrees of freedom(DOF) into emergent fermions and gauge field \cite{Xiao:803748}.  This field theory has been proposed to have many dual field theory descriptions.  The topological NLSM with WZW terms, which is also the main interest of this work, is one important dual description\cite{PhysRevX.11.031043}. This duality can be formally derived through the Abanov Weigman mechanism \cite{ABANOV2000685} if the fermions in the QED theory are Dirac fermions. The bosonic field of the
nonlinear sigma model lives in the target space of the ground states manifold of the nearby ordered phases (AFM and VBS). For example, the target space of the above DQCP is $S^4$ \cite{PhysRevX.7.031051}, and that of the above Dirac spin liquid is the Grassmannian manifold $\frac{\text{U(4)}}{\text{U(2)}\times\text{U(2)}}$ \cite{PhysRevB.97.195115}.  The 'intertwinement' between order parameters is precisely captured by
the topological WZW terms.

Meanwhile, the duality between different field theories in both the ultraviolet (UV) and IR has attracted longstanding interest and played an important role in the study of strongly correlated systems \cite{PhysRevB.39.2756,PhysRevLett.47.1556,PESKIN1978122,PhysRevX.6.031043,SEIBERG2016395,PhysRevX.5.031027,PhysRevB.93.245151,PhysRevX.5.041031,polyakov1988fermi,shaji1990bose,PhysRevLett.120.016602,KITAEV20062}. The most well-known example is the 
particle-vortex duality in 2+1 dimensions \cite{PhysRevB.39.2756,PhysRevLett.47.1556,PESKIN1978122}, which maps interacting boson theory to the dual vortex theory on the lattice. The particle-vortex duality in IR states that the fixed point of scalar QED theory is the same as the Wilson-Fisher fixed point.  Besides, the boson-fermion dualities have also been proposed in gapless systems, which state that scalar QED theory with Chern-Simons terms is dual to free Dirac fermions \cite{polyakov1988fermi,shaji1990bose,PhysRevX.6.031043,SEIBERG2016395,PhysRevLett.120.016602}, which later leads to a duality web \cite{PhysRevX.5.031027,PhysRevB.93.245151,PhysRevX.5.041031}. The correspondence mentioned above between the $\text{QED}$ theory and the topological nonlinear sigma model is also a field theory duality. For example, the topological NLSM on the Grassmannian manifold and the $N_f=4$ $\text{QED}$ theory are believed to flow to the same strong coupling fixed point in IR \cite{PhysRevX.11.031043}. The current theoretical understanding of this duality in DQCP and Dirac spin liquid generally starts from the IR $\text{QED}$ theory, such as the $N_f=4$ $\text{QED}_3$ theory of Dirac spin liquid \cite{PhysRevX.10.011033,PhysRevB.72.104404,PhysRevB.77.224413}. And we couple this effective theory with dynamical bosonic fields, which represent the smooth fluctuations of the order parameters of nearby phases, in the demanded target space. Topological terms, such as the WZW terms, can be arrived at by integrating out the fermions, which is known as the AW mechanism \cite{ABANOV2000685,nagaosa1996chiral} . However, the AW mechanism only applies for Dirac fermions, and the direct application of the AW mechanism is generally expected to fail for fermions with generic infrared dispersion \cite{PhysRevB.82.245117}, which is commonly encountered in condensed matter systems. This means that the integration of fermions is totally different if the infrared fermions are not relativistic, and a general theoretical framework is needed for these systems, especially when a dynamical gauge field is not confined in the low energy.

 In this work, we propose a general dimensional reduction method for the kinematic duality between the topological NLSM models with WZW terms and QED theory with generic dispersive fermions \footnote{Strictly speaking, the duality here means the two theories have the same kinematic properties, i.e. the same local operators, same global symmetries and same 't Hooft anomalies, which is also known as the 'weak duality' in the literature \cite{PhysRevX.7.031051,SEIBERG2016395,SENTHIL20191}. If the topological NLSM model and the \text{QED} theory flow to the same IR fixed point, these two theories have the so-called 'strong duality'\cite{PhysRevX.7.031051,SEIBERG2016395,SENTHIL20191}. The determining answer of the IR dynamical behaviour of the topological NLSM is beyond the scope of this work. But we comment that the topological NLSM is theoretically possible to have the same IR behaviour with the $\text{QED}$ theory since they share the same kinematic properties, if all the symmetry-allowed couplings in the topological NLSM are allowed to tune.}. The application of the dimensional reduction method to condensed matter systems in the literature has been mainly focused on gapped systems \cite{PhysRevB.78.195424,ryu2010topological,PhysRevX.7.011020,PhysRevB.96.195101,PhysRevB.101.085137,PhysRevB.104.075111,PhysRevB.101.165129}. In the seminal paper \cite{PhysRevB.78.195424}, it was proposed that lower dimensional topological insulators and their topological response terms can be derived from even dimensional Chern insulators by compacting the extra dimensions and viewing the momenta in these dimensions as parameters. The topological response terms of lower dimensional topological insulators can also be derived from the higher dimensional Chern-Simons terms. The dimensional reduction method in gapless or critical systems, especially those described by the field theories with topological terms in IR mentioned above, is much less explored. The dimensional reduction method has been applied to the fermionic systems coupled with dynamical bosonic fields on $S^n$, in the previous work \cite{PhysRevB.82.245117}. This method is qualitatively different from that \cite{PhysRevB.78.195424} in topological insulators. It can be applied to DQCP where the gauge field is confined, and only fermion survives in the low energy. But how to arrive at the topological terms from QED theory with generic dispersive fermions or even a Fermi surface in IR, with more general target spaces beyond $S^n$ is a long-standing open question. In this work, we propose that the dual WZW terms of $\text{QED}$ theory with fermions of general dispersion in $n$ dimensional spacetime can be derived from a series of Chern-Simons terms in $2k+1$ dimensional spacetime,  where $k=[\frac{n}{2}],[\frac{n}{2}]+1,...n$ and there are $2k+1-n$ extended dimensions besides the physical $n$ spacetime dimensions. This series of Chern-Simons terms are the low energy response theory of the fermions gapped by the order parameter in the demanded target space. The $2k+1-n$ extended dimensions serve as parameters in the Chern-Simons terms and as a result we dub this scheme as the dimensional reduction scheme. The dimensional reduction scheme proposed in this work is applicable to a general target manifold represented by a Hermitian matrix, including the $n$-sphere $S^n$ and Grassmannian manifold.

The organization of the rest of the paper is as follows.
	In section \ref{sec 2}, we give a brief review of the reason why the AW mechanism
may break down for fermions beyond relativistic dispersion using a toy model with quadratic dispersive fermions in $1+1$d. Then we give a summary of our dimensional reduction method to derive the dual topological NLSM model of $\text{QED}$ theory, which is the central result of this paper, in section \ref{sec 3}. In section \ref{sec 4}, as a warm-up, we first apply our method to the 2+1 dimensional $N_f=4$ QED theory with Dirac fermions, to arrive at the level-1 WZW terms on the Grassmannian manifold  $\frac{\text{U(4)}}{\text{U(2)}\times \text{U(2)}}$, which is consistent with previous results. After that, we study three fermionic lattice models with four quadratic band touching (QBT) points coupled with a U(1) gauge field and viewed them as parton mean field theories of spin models. Their low energy theories are all the $N_f=4$ $\text{QED}$ theory with quadratic dispersive fermions.  Then we show that this $\text{QED}$ theory is dual to a topological NLSM with level-2 WZW terms on the Grassmannian manifold  $\frac{\text{U(4)}}{\text{U(2)}\times \text{U(2)}}$. To support this result, we perform several theoretical consistency checks in section \ref{sec 5}. First, both theories share the same global symmetry and 't Hooft anomalies. Then we consider an interface with codimension-1, and show that both theories give $1+1$ dimensional $\text{SU(2)}_2$ WZW theory on the interface. Moreover, under the insertion of gauge monopoles and vortexes with codimension-2,  the quantum numbers of zero modes are consistent in both theories.     Finally, we calculate the quantum numbers of gauge-invariant monopole operators and their behavior under macroscopic symmetries in section \ref{sec 6}. These results can guide future investigations on nearby ordered phases.
 
\section{A brief review of the background}\label{sec 2}
We first give a brief review of the Goldstone-Wilcezk model \cite{PhysRevLett.47.986}, which is the simplest application of the AW mechanism \cite{ABANOV2000685}. Then we discuss a simple one dimensional lattice model with target space $S^1$ first proposed in \cite{PhysRevB.82.245117} to show why the AW mechanism may break down for nonrelativistic  fermions generally.

The Goldstone-Wilcezk model describes the 1+1d Dirac fermions coupled with a chiral vector field $(\Delta_1,\Delta_2)$ :
\begin{equation}
    \mathscr{L}_{1}=\bar{\psi}\left[i \hat{D}+i m\left(\Delta_{1}+i \gamma_{5} \Delta_{2}\right)\right] \psi, \quad \Delta_{1}+i \Delta_{2}=e^{i \phi},
\end{equation}
where $\hat{D}=\gamma_0(\partial_{t}-iA_0)+\gamma_1(\partial_{x}-iA_1)$ and $A$ is a background U(1) gauge field.

 If we integrate out the Dirac fermions and expand the resultant action, we arrive at the dual topological nonlinear $\sigma$ model:
\begin{equation}
\begin{aligned}
    \mathscr{W}_{1}&=-\ln \operatorname{det}\left(i \hat{D}+i  e^{i \gamma_{5} \phi}\right)\\
    &=\frac{i}{2\pi} \int A \wedge d\phi+i \pi  Q \oint d t \frac{\partial_{t} \phi}{2 \pi}+\int d^{2} x \frac{1}{8 \pi}\left(\partial_{\mu} \phi\right)^{2}.
    \end{aligned}
\end{equation}
Physically, the first term has the consequence of spectral flow in the spectrum of fermion, which means that there is one energy level crossing $E=0$ given a spatial soliton configuration of the mass term with $2\pi$ winding as $x$ cycles the base manifold $S^1$ .

 Now let's move to the lattice model with quadratic dispersion fermions in IR. We consider the following Bloch Hamiltonian:
\begin{equation}
    H(k)=2\left(\begin{array}{cc}0 & \left(\cos \frac{k}{2}+i \delta t \sin \frac{k}{2}\right)^{2} \\ \left(\cos \frac{k}{2}-i \delta t \sin \frac{k}{2}\right)^{2} & 0\end{array}\right),
    \label{1+1}
\end{equation}
where the parameter $\delta t $ is the hopping amplitude. This lattice model can be viewed as the effective model of a two-leg ladder tight-binding model as shown in \cite{PhysRevB.82.245117}. If we couple the system with a vector bosonic field $\left(m_{1}, m_{2}\right) \rightarrow m(\cos \theta(x), \sin \theta(x))$, with the base manifold $S^1$:
\begin{equation}
    H(k)=2\left(\begin{array}{cc}m_{2} & \left(\cos \frac{k}{2}+i m_{1} \sin \frac{k}{2}\right)^{2} \\ \left(\cos \frac{k}{2}-i m_{1} \sin \frac{k}{2}\right)^{2} & -m_{2}\end{array}\right)
    \label{lattice},
\end{equation}
then there are two energy levels cross the $E=0$ for each $2\pi$ winding of $\theta$ when $x$ cycles the base manifold $S^1$. Thus if we introduce a background U(1) gauge field $A$, it is natural to conjecture the effective action of $\theta$ to describe this generalized Goldstone-Wilczek effect to be $S_{B}=-\frac{i}{2 \pi} 2 \int d x A_{0} \partial_{x} \theta \label{wil}$. However, we cannot directly use the AW mechanism to prove this duality between the fermion lattice model (\ref{lattice}) and the topological term $S_B$. The continuous Hamiltonian of (\ref{lattice}) is:
\begin{equation}
    H=\left(\begin{array}{cc}2 m_{2}(x) & {\left[i \partial_{x}+i 2 m_{1}(x)\right]^{2}} \\ {\left[i \partial_{x}-i 2 m_{1}(x)\right]^{2}} & -2 m_{2}(x)\end{array}\right).
\end{equation}
The mass $m_1$ is intrinsically coupled with the spatial derivative due to the cross term in $\left[i \partial_{x}+i 2 m_{1}(x)\right]^{2}$, but $m_2$ is not, so we are not expected to arrive at $S_B$ with derivatives solely to $\theta$ directly using the AW mechanism. Moreover, we cannot find a chiral rotation \cite{PhysRevLett.42.1195,nagaosa1996chiral} to rotate the vector $(m_1,m_2)$ into a phase, such as $\Delta e^{i\gamma_5\theta }$ in the Goldstone-Wilcezk model, as the two masses $m_1,m_2$ are not on the equal footing.

This is a typical example of the correspondence between topological NLSM and  nonrelativistic fermionic sigma model \footnote{The fermionic sigma models here are defined as fermions coupled with order parameters, as in \cite{PhysRevB.82.245117}. }, where the AW mechanism generally fails. As a result, a new general method is required to derive this correspondence.  The correspondence between
 topological NLSM whose order parameter manifold is $S^n$ and fermionic sigma model (without the constraint of Dirac fermions) is proposed in \cite{PhysRevB.82.245117}. Motivated by this correspondence, we propose a dimensional reduction scheme which can be used to derive the dual topological NLSM models with a general order parameter manifold of $\text{QED}$ theories with fermions of general dispersion. The rigorous proof for the above correspondence is
presented in appendix \ref{General Theory}.
\section{Summary of the dimensional reduction method}\label{sec 3}
The dimensional reduction method proposed in this
paper to derive the dual topological nonlinear $\sigma$ model description of QED theory is summarized as follows.
We start from a general  $n-1$ dimensional lattice QED theory:
\begin{equation}\label{QED}
\begin{aligned}
S_0&=S_e+S_a,\\
S_e&=\int dt  \sum_{\vec{j}}\sum_{l=1}^{D} \psi^{\dagger}_{l}(\vec{j})(\partial_{t}-ia_0) \psi_{l}(\vec{j})\\
&+\sum_{\vec{i},\vec{j}}\sum_{l,l^{\prime}}\psi^{\dagger}_l(\vec{i})e^{-i\int_{\vec{i}}^{\vec{j}} \vec{a}\cdot d\mathbf{l} }h_{l,l^{\prime}}(\vec{i},\vec{j})\psi_{l^{\prime}}(\vec{j}),
\end{aligned}
\end{equation}
where $l$ and $l^{\prime}$ label the onsite DOF of the fermions, such as sublattice or spin DOF, etc, and $\vec{i},\vec{j}$ are the lattice sites. The fermion kinetic energy $\hat{H}_0=\sum_{\vec{i},\vec{j}}\sum_{l,l^{\prime}} \psi^{\dagger}_l(\vec{i})h_{l,l^{\prime}}(\vec{i},\vec{j})\psi_{l^{\prime}}(\vec{j})$ preserves the lattice translation symmetries and its Bloch Hamiltonian can have nodal Fermi points or Fermi surface in IR. $S_a$ is the kinetic energy action of the $\text{U}(1)$ gauge field $a$, whose concrete expression does not affect our following discussions and conclusions.

We assume the $\text{QED}$ theory in Eq.\eqref{QED} describes certain critical points or phases with nearby xintertwining orders, e.g., the lattice regularization of the IR $N_f=4$ $\text{QED}$ theory of the $2+1d$ Dirac spin liquid. To describe the quantum fluctuation of the nearby ordered phases, we introduce a dynamical bosonic field, physically the quantum fluctuation of the order parameters, on the manifold $\mathbf{M}$, i.e. the target space \footnote{Here the broken symmetries of order parameters can be determined in the IR theory.}. We use a matrix field $M_{i,k}(\vec{j},t)$ as the representation of $\mathbf{M}$. Following the basic logic of the AW mechanism, we couple the bosonic field to the fermions as mass terms \footnote{Here for the convenience of notation, we denote the mass term as onsite coupling, but our following formalism can be directly applied to non-onsite mass terms as long as it preserves the lattice translation symmetries.  } : 
\begin{equation}\label{order parameter}
  S_c= \int dt \sum_{\vec{j},i,k}\psi^{\dagger}_{i}(\vec{j},t) m_{i,k}(\vec{j},t)\psi_k(\vec{j},t).
\end{equation}
 Here we remark that for a general $\hat{H}_0$, it is challenging to write down a general expression between the mass $m(\vec{j},t)$ and the bosonic field $M(\vec{j},t)$. In this paper, we assume $h_{l,l^{\prime}}(\vec{i},\vec{j})$ can be spanned by the $\gamma$ matrices $\gamma_\mu$ ($\mu=0,\cdots,n-1$), which is a representative class of systems to illustrate our method. In this case, when the spacetime dimension $n$ is odd, we can simply take the mass term as $m=M\gamma_0$ where $M$ only acts on the flavor index. When $n$ is even, we take it as $m=M\gamma_0=M_1\gamma_0+iM_2\gamma_1...\gamma_{n-1}$, where $M_1$ and $M_2$ only act on the flavor index \footnote{For the Dirac fermions and quadratic dispersive fermions in IR, this kind of mass term is the only choice, if we impose the continuous spatial rotation symmetry. }. For a more general Bloch Hamiltonian, our following method is still applicable as long as the order parameter $M$ and its coupling to the fermions are specified in that special system.  
 
Due to the existence of the gauge field $a$, it is possible to have a WZW term corresponding to each integer $k$ between $[\frac{n}{2}]$ and $n$, where $M(\vec{j},t)$ is coupled with the gauge field, after we integrate out the fermions. To determine whether the WZW term exists for each $k$, we create the following $2k$  dimensional Bloch Hamiltonian: 
 \begin{eqnarray}
   \label{Bloch hamiltonian2}
 H_{2k}(k_1,..k_{n-1},p_1,...p_{2k-n+1})=h({k_1,...,k_{n-1}})&&\nonumber\\+\lambda m(p_1,p_2,...,p_{2k-n+1}),
 \end{eqnarray}
 where $m(\vec{p})$ is a mass term and we assume the Hamiltonian is fully gapped in the  Brillouin zone at least in a finite range of $\lambda$. Moreover, we demand the Pontryagin index $P_{2k-n+1}$ of
the map $T^{2k-n+1} \to \mathbf{M}$ is nonzero, where the $2k-n+1$ dimensional torus $T^{2k-n+1}$ is the base manifold of the extended momentum $\vec{p}$. We note that for each fixed $\vec{p}$ in Eq.\eqref{Bloch hamiltonian2}, the mass term of the Hamiltonian \eqref{Bloch hamiltonian2} gives a fixed configuration to the order parameter $M$. We also introduce a gauge field $A$ in the Hamiltonian \eqref{Bloch hamiltonian2} with extended dimensions and take it as:
 \begin{equation}\label{background}
    \begin{aligned}
    &A_{\mu}\left(t, j_{1}, \ldots, j_{2 k}\right)=a_{\mu}\left(t, j_{1}, \ldots, j_{n-1}\right), \quad \\
    &\text { for } \mu=0, \ldots, n-1, \\ 
    &A_{i+n-1}\left(t, j_{1}, \ldots, j_{2 k}\right)=\theta_{i}\left(t, j_{1}, \ldots, j_{n-1}\right), \quad \\
    &\text { for } i=1, \ldots, 2k-n+1,
    \end{aligned}
\end{equation}  
where $\vec{\theta}$ is the background gauge field.
  The order parameter is now reparametrized as $M(\vec{p}+\vec{\theta}(\vec{j},t))$ to reproduce the configuration $M_{i,k}(\vec{j},t)$ in the action $S=S_0+S_c$. As a result, the integration of fermions in the Hamiltonian \eqref{Bloch hamiltonian2} for any fixed $\vec{p}$ is the same as the integration of fermions in the original action $S=S_0+S_c$ with the order parameter configuration $M(\vec{p}+\vec{\theta}(\vec{j},t))$ \footnote{Actually, the mass in Eq.\eqref{Bloch hamiltonian2} can also depend on $\vec{k}$, as long as the Pontryagin indexes are the same for each $\vec{k}$. We give the details about this point in the Appendix \ref{General Theory}. }. Moreover, since the Eq.\eqref{Bloch hamiltonian2} is fully gapped and the  Hamiltonians with different $\vec{p}$ can be adiabatically deformed into one another without
closing the energy gap, the topological terms obtained for each $\vec{p}$ are the same after integrating out the fermions. As a result, the original problem of integrating out fermions in the action $S=S_0+S_c$, which is a strongly coupled lattice gauge theory, is turned into integrating out the fermions in the Hamiltonian\eqref{Bloch hamiltonian2}, and the cost is the increase of the dimensions.
  
 Eq.\eqref{Bloch hamiltonian2} describes a $2k$-dimensional Chern insulator, whose low energy response theory is the $2k+1$ dimensional Chern-Simons theory with level $C_k$ after integrating out the fermions. The level $C_k$ is the $k$-th Chern number:
  \begin{eqnarray}
   \label{chern number}
C_k=\frac{1}{k!(4\pi)^k}\int d^{2k}p \epsilon^{i_1 i_2 \cdots i_{2k}}\mathbf{Tr}[F_{i_1 i_2 }F_{i_3 i_4 }\cdots F_{i_{2k-1} i_{2k}}].\nonumber\\
~
 \end{eqnarray}
 Here $F_{ij}$ is the Berry curvature of the occupied bands in the momentum space. 
 
  Moreover, with the gauge field configuration as in Eq.\eqref{background}, the $2k+1$ dimensional Chern-Simons term is the summation of the same topological terms in $n$-dimensional subsystems labeled by different momenta $\vec{p}$. In the Appendix \ref{General Theory}, we rigorously prove that the topological term of each fixed $\vec{p}$ is the level-$\frac{C_k}{P_{2k-n+1}}$ WZW term on the target manifold M in the physical system with spatial dimension $n-1$.  And in the next section, we also give the detailed derivation of the above statement in concrete two dimensional lattice models and we take the target manifold M to be the Grassmannian manifold $\frac{\text{U(4)}}{\text{U(2)}\times \text{U(2)}}$ there. Written explicitly, the above statement is:
 \begin{equation}
\begin{aligned}
S_{\mathsf{CS}}=\sum_{\vec{p}} S^{k}_{(n-1)\text{d}}(\vec{p}=0)&=\sum_{\vec{p}} S^{k}_{\text{WZW}}\left(M(\vec{\theta}(\vec{j},t))\right),
\end{aligned}
\end{equation} 
 which means the Chern-Simons term in the dimension-extended system Eq.\eqref{Bloch hamiltonian2} can be reduced to the WZW term in the original physical system by the reduction of the extended dimensions $\vec{p}$. The expressions of the WZW terms are different in odd and even spacetime dimensions and we list their results below for later convenience.

 First of all, we can assume that the mass term $m$
 satisfies  $m^{\dagger}m=1$ and thus $M^{\dagger}M=1$. This is due to that the mass of fermion $m$ should be non-degenerate; thus all the eigenvalues are non-zero and we can deform
the absolute values of them all to $1$ without crossing zero. 

When the spacetime dimension $n$ is odd,  the matrix $M$  only acts on the flavour indices. Thus the unitarity of the theory requires $M$ to be Hermitian and it further satisfies $M^2=1$. 
We can arrive at the following WZW terms after integrating out the
fermions under our dimensional reduction method:
\begin{eqnarray}\label{odd}
   S_{\text{WZW}}=&&2\pi i\sum^{n}_{k=\frac{n+1}{2}}\frac{ C_{k}\theta_{2k-n+1}}{P_{2k-n+1}(n-k)!}\int dt d^{n-1}x\int_0^1 du \nonumber\\
   &&\mathbf{Tr}[M^{-1}(dM)^{2k-n+1} ](\frac{f}{2\pi})^{n-k},
\end{eqnarray}
where $u$ is the auxiliary dimension to define the WZW term and the bosonic field $M(\vec{x},t,u)$ smoothly interpolates between $M(\vec{x},t,u=1)=M_0$ and $M(\vec{x},t,u=0)=\mathbb{I}$, where $M_0$ is a constant matrix belonging to the representation of the target manifold $\mathbf{M}$. The Pontryagin index $P_{2k-n+1}$ of the map $T^{2k-n+1}\rightarrow M(p_1,p_2,...p_{2k-n+1})$ is now given by \cite{PhysRevB.102.245113}:
\begin{equation}
   P_{2k-n+1}=\theta_{2k-n+1}\int_{T^{2k-n+1}} d^{2k-n+1}p \mathbf{Tr}[M^{-1}(dM)^{2k-n+1}],
\end{equation}
where $2\pi\theta_{2k-n+1}=\frac{2^{\frac{2k-n-1}{2}}}{4^{2k-n}\pi^{\frac{2k-n-1}{2}}\Gamma(\frac{2k-n+1}{2})(2k-n+1)}$ and the trace is only over flavor indices.

When the spacetime dimension $n$ is even,  M satisfies  $M^{+}M=1$. After integrating out the
fermion field in the action $S=S_0+S_c$, we obtain the following WZW terms:
\begin{equation}\label{even}
\begin{aligned}
S_{\text{WZW}}=2\pi i\sum^{n}_{k=\frac{n}{2}}&\frac{ C_{k}\theta_{2k-n+1}}{2^{\frac{n-1}{2}} P_{2k-n+1}(n-k)!}\int dt d^{n-1}x\int du \nonumber\\
   &\mathbf{Tr}[M^{-1}dM(d\!\!\!/M)^{2k-n}](\frac{f}{2\pi})^{n-k},
\end{aligned}
\end{equation}
where the Pontryagin index $P_{2k-n+1}$ of the map $T^{2k-n+1}\rightarrow M(p_1,p_2,...p_{2k-n+1})$ \cite{PhysRevB.102.245113} is:
\begin{eqnarray}
   P_{2k-n+1}&&=\frac{\theta_{2k-n+1}}{2^{\frac{n}{2}}}\int_{T^{2k-n+1}} d^{2k-n+1}p \mathbf{Tr}[M^{-1}dM(d\!\!\!/M)^{2k-n}].\nonumber\\
   ~
 \end{eqnarray}  
We remark that in the Eq. \eqref{even}, the trace is over flavour indices and gamma matrix indices.

\section{Lattice models}\label{sec 4}
In this section, we first prove the relativistic $N_f=4$ $\text{QED}_3$, which is the IR theory of the Dirac spin liquid, is kinematically dual to the topological NLSM with level-1 $\frac{\text{U}(4)}{\text{U}(2)\times \text{U}(2)}$ WZW terms coupled with a dynamical U(1) gauge field using our dimensional reduction method, which is consistent with the results from AW mechanism. The bosonic matrix field on the Grassmannian manifold $\frac{\text{U}(4)}{\text{U}(2)\times \text{U}(2)}$ captures the quantum fluctuation of the nearby intertwining orders, e.g. the $120^{\circ}$ coplanar magnetic orders and the $\sqrt{12}\times \sqrt{12}$ valance-bond-solid orders on the triangular lattice\cite{PhysRevX.11.031043,song2019unifying,PhysRevX.10.011033,PhysRevB.97.195115,PhysRevB.72.104404,PhysRevB.77.224413}.  After that, we construct three lattice models whose IR theories all have four QBT cones. We view these three fermionic lattice models as the parton construction of lattice spin models, so a dynamical $\text{U(1)}$ gauge field must be complemented to conserve the fermion number on each site.  And the low energy theories of these three lattice models are all $N_f=4$ $\text{QED}_3$ with quadratic dispersive fermions. The Lagrangian density of this $\text{QED}$ theory is:
\begin{equation}\label{qbt_qed3}
    \mathcal{L}=\sum_{f=1}^4{\psi}^{\dagger}_fD_t\psi+{\psi}^{\dagger}[(D_x^2-D_y^2)\mu_1+2D_xD_y\mu_2]\psi_f,
\end{equation}
where $D_{\mu}=\partial_{\mu}+ia_{\mu}$ and $a_{\mu}$ is the dynamical U(1) gauge field. Inspired by the intertwining orders of the Dirac spin liquid,  we view this nonrelativistic $N_f=4$ $\text{QED}_3$ as a critical theory and also take the Grassmannian manifold $\frac{\text{U}(4)}{\text{U}(2)\times \text{U}(2)}$ as the target space of the nearby fluctuating order parameters. We show this critical theory is dual to the topological NLSM with level-2 WZW terms on the Grassmannian manifold using our dimensional reduction method. These level-1 and level-2 WZW terms are also the same topological terms of the $N=6$ Stiefel liquid with $k=1$ and $k=2$ respectively  \cite{PhysRevX.11.031043}.
\subsection{Target space and topological properties}
Before diving into the concrete lattice models, we first explain the topological properties of the target space we investigate in the following sections. The target space we investigate in all the lattice models here is the Grassmannian manifold \cite{PhysRevB.97.195115,huang2021non,huang2022competing,huang2022non}: 
\begin{equation}
    \begin{aligned}
   \mathbf{M}=\frac{\text{U}(4)}{\text{U}(2)\times \text{U}(2)}.
    \end{aligned}
\end{equation}
It has the matrix representation \cite{PhysRevB.97.195115}:
\begin{equation}
    \begin{aligned}
   M=\sum^3_{a,b=1}  N^a_m N^b_e \tau_a\otimes \sigma_b+\sum^3_{a=1} M^a_m \tau_a\otimes \sigma_0+\sum^3_{b=1} M^b_e \tau_0\otimes \sigma_b,
    \end{aligned}
\end{equation}
where
\begin{equation}
    \begin{aligned}
  M^2=\mathbb{I}_{4\times 4},\quad \Vec{N}_e\cdot \Vec{M}_e=\Vec{N}_m\cdot \Vec{M}_m=0.
    \end{aligned}
\end{equation}
Physically,  $\Vec{N}_e$ and $\Vec{M}_e$ can describe a general AFM order parameter. Meanwhile, $\Vec{N}_m$ and  $\Vec{M}_m$ correspond to a general VBS order parameter \footnote{They can describe the usual Neel order and VBS order around the DQCP point. Besides, they can also describe the noncollinear order around Dirac spin liquids on the triangular lattice and kagome lattice. }.

In 2+1d systems, there are at most two WZW terms on the Grassmannian manifold $\mathbf{M}$ in the action, corresponding to the two nontrivial homotopy groups $\pi_4$ and $\pi_2$. Under the formalism of our dimensional reduction method, the levels of these two WZW terms are $\frac{C_3}{P_4}$ and $\frac{C_2}{P_2}$ respectively, where the numerators are the Chern number of the Chern insulators in the extended spacetime and the denominators are the Pontryagin indexes of the mass terms in the target manifold. Technically, these topological number can be arrived by investigating the submanifolds of the target space. If we take the mapping from the base manifold of the extended momenta to the submanifold of the target space with Pontryagin indexes $P_2$ or $P_4$ equal to one, then the Chern number $C_2,C_3$ with the mass term in the submanifold are all we need to get the quantized levels of the topological terms. As a result, we only need to consider two submanifolds $S^4$ and $S^2$ to arrive at the levels of the topological terms, since there exists a map from $T^4(T^2)$ to $S^4(S^2)$ whose Pontryagin index $P_4$($P_2$) is 1. 

We first discuss the $S^4$ submanifold, where we set $\Vec{N}_m=(0,0,1)$ and $M^3_m=0$. Then the matrix representation becomes:
\begin{equation}\label{so(5)}
    \begin{aligned}
   M=\sum^3_{b=1} N^b_e \tau_3\otimes \sigma_b+\sum^2_{a=1} M^a_m \tau_a\otimes \sigma_0.
    \end{aligned}
\end{equation}
Moreover, $\mathbf{M}$ has another submanifold $S^2$ where we take $\Vec{N}_e=\Vec{M}_m=(0,0,0)$. Then the
configuration of  $\mathbf{M}$ is given by:
\begin{equation}
    \begin{aligned}
   M= \sum^3_{a=1} M^a_m \tau_a\otimes \sigma_0.
    \end{aligned}
\end{equation}
In the concrete two dimensional lattice models we investigate in the following sections, we extend the physical system to six dimensional and four dimensional Chern insulators with the mass terms(or order parameters) in the above submanifolds $S^4$ and $S^2$. We denote the Chern number of the two Chern insulators as $C_3$ and $C_2$ respectively, and the WZW terms after integrating out
the fermions can be written as:
\begin{eqnarray}\label{odd}
   S_{\text{WZW}}=&&2\pi i\sum^{3}_{k=2}\frac{ C_{k}\theta_{2k-2}}{P_{2k-2}(3-k)!}\int dt d^2x\int du \nonumber\\
   &&\mathbf{Tr}[M^{-1}(dM)^{2k-2} ](\frac{F}{2\pi})^{3-k},
   \end{eqnarray}
   where the two Pontryagin indexes $P_4$ and $P_2$ are taken to be one in the following constructions.
\subsection{Dirac spin liquid on the honeycomb lattice}
\label{sec:dirac}
In this section, we discuss the topological NLSM description of the Dirac spin liquid on the honeycomb lattice. The IR theory of the Dirac spin liquid is the relativistic $N_f=4$ $\text{QED}_3$ theory \cite{PhysRevX.11.031043,PhysRevX.10.011033,PhysRevB.97.195115,PhysRevB.72.104404,PhysRevB.77.224413}, and we show that it duals to the nonlinear $\sigma$ model with level-1 WZW terms on the Grassmannian manifold using our dimensional reduction method. 

We take the parton mean field ansatz as uniform nearest-neighbour hopping(coupled with dynamical U(1) gauge field) for the Dirac spin liquid. The quadratic part of the parton mean field Hamiltonian can be written as $\hat{H}_{2d}=\sum_kf^{\dagger}(\mathbf{k})h_{2d}(\mathbf{k})f(\mathbf{k})$ in the four-component spinon operator basis, including the sublattice and valley DOF:
\begin{equation}
    f(\mathbf{k})=(c_A(\mathbf{k}),c_B(\mathbf{k}),c_B(-\mathbf{k}),c_A(-\mathbf{k}))^T.
    \label{ope_basis}
\end{equation}
The two dimensional Bloch Hamiltonian takes the form:
\begin{equation}
\begin{aligned}
 h_{2d}(k)&=[1+\cos(k_1-k_2)+\cos(k_1)]\Gamma_1\\
 &-[\sin(k_1-k_2)+\sin(k_1)]\Gamma_2\\
 & k_1=\frac{\sqrt{3}}{2}k_x+\frac{1}{2}k_y,\quad k_2=\frac{\sqrt{3}}{2}k_x-\frac{1}{2}k_y,\\
 &\Gamma_1=\mu_1\otimes \tau_0\otimes \sigma_0, \quad \Gamma_2=\mu_2\otimes \tau_0\otimes \sigma_0,
\end{aligned}
\end{equation}
where the Pauli matrices $\mu$, $\tau$ and $\sigma$ act on the sublattice, valley and spin DOF respectively. We take the lattice constant of the A sublattice to be one, and the momenta $k_1,k_2$ here take values in $[0,2\pi]$ in the  Brillouin zone. 


 Following the general dimensional reduction scheme summarized above, we first extend the Hamiltonian to six spatial dimensions and only retain the order parameter configuration on $S^4$ with a nonzero Pontryagin index to arrive at the level of the first WZW term in which the gauge field does not appear. The gapped 6d Bloch Hamiltonian is given by\footnote{The mass term of the Hamiltonian \eqref{dirac_6} can also be written in the spinon basis $ f(\mathbf{k})=(c_A(\mathbf{k}+\mathbf{K}),c_B(\mathbf{k}+\mathbf{K}),c_A(\mathbf{k}-\mathbf{K}),c_B(\mathbf{k}-\mathbf{K}))^T$, where $\pm\mathbf{K}$ are the momentum of the two Dirac cones. The mass term under this new basis gives a clear physical meaning of the order parameter as density wave on the lattice. The Chern number is the same as the Hamiltonian \eqref{dirac_6}, since both mass terms gap out the Dirac cones and the gap is not closed when we deform one Chern insulator into the other. As a result, these two Chern insulators in turn give the same quantized levels of the WZW terms as $\frac{C_3}{P_4}$. This also indicates the level of the WZW term is intrinsically determined by the low energy fermions in the QED theory.}:
    \begin{equation}\label{dirac_6}
\begin{aligned}
 h_{6d}(k)&=[1+\cos(k_1-k_2)+\cos(k_1)]\Gamma_1\\
 &-[\sin(k_1-k_2)+\sin(k_1)]\Gamma_2\\
 &+m_3\Gamma_3+m_4\Gamma_4+m_5\Gamma_5+m_6\Gamma_6\\
 &+(m+\cos k_3+\cos k_4\\
 &+\cos k_5+\cos k_6)\Gamma_7,\\
 &m_3=\sin k_3,\quad m_4=\sin k_4,\quad m_5=\sin k_5,\\
 &m_6=\sin k_6,\\
 &
   \Gamma_3=\mu_3\otimes \tau_1\otimes \sigma_0\quad\Gamma_4=\mu_3\otimes \tau_2\otimes \sigma_0\quad
   \\
   &\Gamma_5=\mu_3\otimes \tau_3\otimes \sigma_1\quad
   \Gamma_6=\mu_3\otimes \tau_3\otimes \sigma_2\quad\\
   &\Gamma_7=\mu_3\otimes \tau_3\otimes \sigma_3.
\end{aligned}
\end{equation}
If we set $m=-3.5$, both the Chern number $C_3$ and the Pontryagin index $P_4=\frac{1}{256\pi^2}\int \text{Tr}[M^{\dagger}(dM)^4]$ equal to 1. The Chern-Simons response action of this Chern insulator is: 
\begin{equation}
    S_{\text{CS}}=i\frac{C_3}{(2\pi)^3}\int A\wedge(dA)^3=\frac{i}{(2\pi)^3}\int A\wedge(dA)^3.
\end{equation}
We take the configurations of $A$ as: $A_{0,1,2}=a_{0,1,2}(x,y,t), A_{3,4,5,6}=\theta_{1,2,3,4}(x,y,t)$, where $x,y,t$ are the physical spacetime coordinates of the 2+1d lattice system. Since the Chern-Simons action $S_{\text{CS}}$ does not depend on the extended coordinates, we can integrate them out. We further introduce the auxiliary dimension $u\in [0,1]$ and let $\theta(x,y,t,u)$ smoothly interpolate between $\theta_i(x,y,t,u=1)=\theta_i(x,y,t)$ and $\theta_i(x,y,t,u=0)=0 $. As a result, the Chern-Simons action becomes:
\begin{eqnarray}
     S_{\text{CS}}&&=\frac{i}{(2\pi)^3} \frac{L^4}{4!}\int dxdydt \int_0^1 du\nonumber\\
     &&\epsilon^{ijk}\epsilon^{b_1b_2b_3b_4}\partial_u [\theta_{b_1} \partial_i \theta_{b_2}\partial_j \theta_{b_3}\partial_k \theta_{b_4}],\label{second}\\
     &&=\frac{i}{(2\pi)^3} \frac{L^4}{4!}\int dxdydtdu \nonumber\\
     &&\epsilon^{a_1a_2a_3a_4}\epsilon^{b_1b_2b_3b_4}\partial_{a_1}[\theta_{b_1}\partial_{a_2}\theta_{b_2}\partial_{a_3}\theta_{b_3}\partial_{a_4}\theta_{b_4}]\label{third}\\
     &&=\frac{i}{(2\pi)^3} \frac{L^4}{4!}\int dxdydtdu \nonumber\\
     &&\epsilon^{a_1a_2a_3a_4}\epsilon^{b_1b_2b_3b_4}\partial_{a_1}\theta_{b_1}\partial_{a_2}\theta_{b_2}\partial_{a_3}\theta_{b_3}\partial_{a_4}\theta_{b_4},
     \end{eqnarray}
where $L$ is the length scale of the extended dimensions. The indexes $b_i=1,\cdots 4$ represents the components of $\theta$ field, and $a_i=0,\cdots,3$ represents the spacetime coordinate $u,t,x,y$. 

Now we insert the Pontryagin index $P_4$ in both the numerator and the denominator:
    \begin{eqnarray}
     S_{\text{CS}}&&=\frac{i}{(2\pi)^3} \frac{L^4}{4!}\frac{1}{256\pi^2P_4}\int dxdydt  du\epsilon^{a_1a_2a_3a_4}\epsilon^{b_1b_2b_3b_4}\nonumber\\
     &&\partial_{a_1}\theta_{b_1}\partial_{a_2}\theta_{b_2}\partial_{a_3}\theta_{b_3}\partial_{a_4}\theta_{b_4}\int d^4p \text{Tr}[M^{\dagger}(d_p M)^4]\\
     &&=\frac{i}{(2\pi)^3} \frac{L^4}{4!}\frac{1}{256\pi^2}\int d^4p \int dxdydtdu  \nonumber\\
     &&\epsilon^{a_1a_2a_3a_4}\epsilon^{b_1b_2b_3b_4}\partial_{a_1}\theta_{b_1}\partial_{a_2}\theta_{b_2}\partial_{a_3}\theta_{b_3}\partial_{a_4}\theta_{b_4}\nonumber\\
     &&\text{Tr}[M^{\dagger}(\vec{p}+\vec{\theta})[d_{\theta} M(\vec{p}+\vec{\theta})]^4].
     \label{diffp}
     \end{eqnarray}
It is important to note that the Pontryagin index of $M$ does not depend on $\theta$, so we do the variable replacement in Eq.\eqref{diffp}. 

Further, since the gauge field doesn't depend on the coordinate $(x_{n},\cdots,x_{2n+1})$,  the $2n$ dimensional system (coupled with gauge field) still preserves translation symmetry on these directions and the $2n$ dimensional system decouples into ($n$-1) dimensional subsystems labeled by different $\vec{p}$ with the Hamiltonian of each subsystem  $h_{6d}(k_1, k_2; \vec{p} + \vec{\theta}(t, \vec{x}))$.  As the Hamiltonian with different $p$ can be adiabatically deformed into each other without closing the gap, the WZW terms contributed by different $p$ slices should be the same. Thus we can just take the momentum $\vec{p}$ of $M(\vec{p} + \vec{\theta})$ in Eq.\eqref{diffp} to be zero:
\begin{equation}
\begin{aligned}
S_{\mathsf{CS}}=\sum_{\vec{p}} S_{2d}(\vec{p}=0)&=\sum_{\vec{p}} S_{\text{WZW}}\left(M(u,t,x,y)\right)
\end{aligned}
\end{equation}
where $M(u,t,x,y)= M(\vec{\theta}(u,t,x,y))$.

With the chain rule of derivatives, we finally arrive at the level-1 WZW term on the Grassmannian manifold:
\begin{equation}
    \begin{aligned}
    S^{k=3}_{\text{WZW}}&=\frac{2\pi i}{4!}\frac{1}{256\pi^2}\int \epsilon^{a_1a_2a_3a_4}\epsilon^{b_1b_2b_3b_4}\partial_{a_1}\theta_{b_1}\partial_{a_2}\theta_{b_2}\partial_{a_3}\theta_{b_3}\partial_{a_4}\theta_{b_4}\\
     &\epsilon^{j_1j_2j_3j_4} \text{Tr}[M^{\dagger}\partial_{\theta_{j_1}} M\partial_{\theta_{j_2}} M\partial_{\theta_{j_3}} M\partial_{\theta_{j_4}} M]\\
     &=\frac{2\pi i}{256\pi^2}\int \epsilon^{a_1a_2a_3a_4}\partial_{a_1}\theta_{j_1}\partial_{a_2}\theta_{j_2}\partial_{a_3}\theta_{j_3}\partial_{a_4}\theta_{j_4}\\
     & \text{Tr}[M^{\dagger}\partial_{\theta_{j_1}} M\partial_{\theta_{j_2}} M\partial_{\theta_{j_3}} M\partial_{\theta_{j_4}} M]\\
     & =\frac{2\pi i}{256\pi^2} \int  \text{Tr}[M^{\dagger}(d M)^4].
    \end{aligned}
\end{equation}

What's more, there is another topological term corresponding to the skyrmion current: $j_{\text{skyrmion}}\wedge da$, due to the nontrivial lower dimensional homotopy group of the Grassmannian manifold $\pi_2(M)=\mathbb{Z}$. We extend the parton Hamiltonian to a 4d topological insulator as:
 \begin{equation}
\begin{aligned}
 h_{4d}(k)&=[1+\cos(k_1-k_2)+\cos(k_1)]\Gamma_1\\
 &-[\sin(k_1-k_2)+\sin(k_1)]\Gamma_2\\
 &+m_3\Gamma_3+m_4\Gamma_4\\
 &+(m+\cos k_3+\cos k_4)\Gamma_5,\\
 &m_3=\sin k_3,\quad m_4=\sin k_4,\\
 &m_5=m+\cos k_3+\cos k_4.
\end{aligned}
\end{equation}
The Chern number $C_2$ and Pontryagin index $P_2=\frac{1}{16\pi} \int \text{Tr}[M^{\dagger} (dM)^2]$ are both 1 for $m=-1.5$. As a result, we can obtain  the five-dimensional Chern-Simons response term after integrating out the fermions:
\begin{equation}
    S_{\text{CS}}=\frac{iC_2}{(2\pi)^2}\int A\wedge (dA)^2=\frac{i}{(2\pi)^2}\int A\wedge (dA)^2,
\end{equation}
where the configurations of $A$ are taken as: $A_{0,1,2}=a_{0,1,2}(x,y,t), A_{3,4}=\theta_{1,2}(x,y,t)$.

We also integrate out the extended dimensions and introduce the auxiliary dimension $u$. Then the Chern-Simons term becomes:
\begin{equation}
\begin{aligned}
    S_{\text{CS}}&=\frac{iL^2}{(2\pi)^2}\frac{1}{2!}\int dxdydtdu \epsilon^{b_1b_2b_3b_4}\epsilon^{c_1c_2}\partial_{b_{1}} a_{b_2} \partial_{b_{3}} \theta_{c_1} \partial_{b_{4}} \theta_{c_2},
    \label{4dcs}
    \end{aligned}
\end{equation}
where the index $c_{1,2}\in \{1,2\}$ denote the components of $\theta$ and $b_{1,2,3,4}\in\{0,1,2,3\}$ denote the $u,t,x,y$ directions. 

Then we put the Pontryagin index $P_2$ in both the numerator and denominator in Eq.\eqref{4dcs}:
\begin{equation}
    \begin{aligned}
    S_{\text{CS}} &=\frac{iL^2}{2!(2\pi)^2P_2}\int dxdydtdu \epsilon^{b_1b_2b_3b_4}\epsilon^{c_1c_2}\\
    &\partial_{b_{1}} a_{b_2} \partial_{b_{3}} \theta_{c_1} \partial_{b_{4}} \theta_{c_2} \frac{1}{16\pi} \int d^2p \text{Tr}[M^{\dagger}(\vec{p}) (dM(\vec{p}))^2]\\
    &=\frac{iL^2}{2!(2\pi)^2}\int d^2p \int dxdydtdu \epsilon^{b_1b_2b_3b_4}\epsilon^{c_1c_2}\\
    &\partial_{b_{1}} a_{b_2} \partial_{b_{3}} \theta_{c_1} \partial_{b_{4}} \theta_{c_2}\frac{1}{16\pi}  \text{Tr}[M^{\dagger}(\vec{\theta}) (d_{\theta}M(\vec{\theta}))^2]\\
    &=\frac{iL^2}{2!(2\pi)^2}\int d^2 p \int dxdydtdu \epsilon^{b_1b_2b_3b_4}\epsilon^{c_1c_2}\\
    &\partial_{b_{1}} a_{b_2} \partial_{b_{3}} \theta_{c_1} \partial_{b_{4}} \theta_{c_2}\frac{1}{16\pi} \text{Tr}[M^{\dagger}(\vec{\theta}) (d_{\theta}M(\vec{\theta}))^2]\\
   & =\frac{iL^2}{16\pi (2\pi)^2} \int d^2p \int dxdydtdu \epsilon^{b_1b_2b_3b_4} \\& f_{b_1b_2}\text{Tr}[M^{\dagger} \partial_{b_3}M\partial_{b_4}M]\\
     &=\frac{iL^2}{16\pi (2\pi)^2}\int d^2 p\int  f\wedge \text{Tr}[M^{\dagger} (dM)^2],
    \end{aligned}
\end{equation}
where $f_{ij}=\partial_i a_j-\partial_j a_i$ is the curvature of the $U(1)$ gauge field $a$. Therefore, the topological WZW term on each subsystem labeled by different $\vec{p}$ is given by 
\begin{equation}
    \begin{aligned}
    S^{k=2}_{\text{WZW}} =\frac{i}{16\pi}\int  f\wedge \text{Tr}[M^{\dagger} (dM)^2].
    \end{aligned}
\end{equation}

As a result, the dual theory has the WZW term on the Grassmannian manifold with level-1, whose action is given by:
\begin{eqnarray}
   S^{\text{level-1}}_{\text{ DSL}}=&&2\pi i\int dt d^2x\int du \lbrace\theta_{3}\mathbf{Tr}[M^{-1}(dM)^{3} ]\nonumber\\
   && +\theta_{2}\mathbf{Tr}[M^{-1}(dM)^{2} ](\frac{f}{2\pi})\rbrace,
\end{eqnarray}
where $\theta_3=\frac{1}{256\pi^2}, \theta_2=\frac{1}{16\pi}$. This result is consistent with the IR duality between the $N_f=4$ $\text{QED}_3$ and topological NLSM with level-1 WZW terms on the Grassmannian manifold (equivalently $N=6,k=1$ Stiefel liquid), which derives from the AW mechanism \cite{PhysRevX.11.031043,PhysRevX.10.011033,PhysRevB.97.195115}. In the following sections, we go beyond the AW mechanism and construct three lattice models on the AB stacked bilayer honeycomb lattice, AA stacked bilayer checkerboard lattice and bilayer triangular lattice whose low energy theories are all $N_f=4$ nonrelativistic $\text{QED}_3$ theory with quadratic dispersive fermions. We also apply the dimensional reduction method to derive their dual WZW descriptions.

\subsection{AB stacked bilayer honeycomb lattice}
\begin{figure}[htpb] 
\centering 
\includegraphics[width=0.45\textwidth]{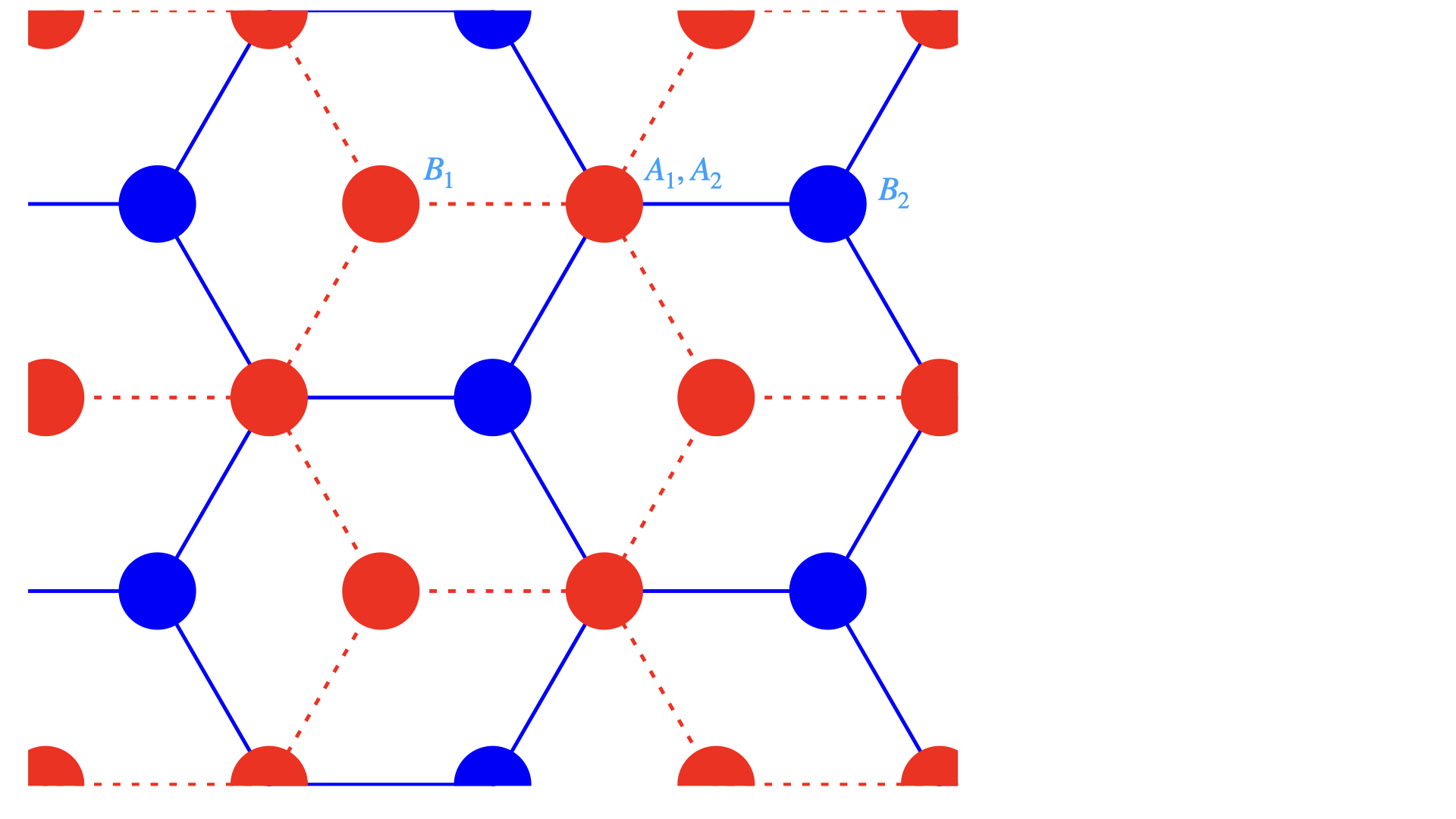} 
\caption{ Bilayer honeycomb lattice with Bernal stacking. The red lines and dots are the lattice bonds and sites of the first layer, and the blue lines and dots are the lattice bonds and sites of the second layer. A/B sublattice index is labeled on the lattice site. } 
\label{blg} 
\end{figure}

In this section, we investigate a lattice model whose IR theory is $N_f=4$ $\text{QED}_3$ theory with quadratic dispersive fermions on the AB stacked (Bernal stacked) bilayer honeycomb lattice, and the lattice structure is shown in Fig \ref{blg}. The parton Hamiltonian is:
\begin{equation}
    \begin{aligned}
    &H_{0}=H_{\|}+H_{\perp}\\ 
    &H_{\|}=-t \sum_{\langle i, j\rangle, m, \sigma}\left(a_{m, i, \sigma}^{\dagger} b_{m, j, \sigma}+h . c .\right), \\ 
    &H_{\perp}=-t_{\perp} \sum_{i, \sigma}\left(a_{1, i, \sigma}^{\dagger} a_{2, i, \sigma}+h . c .\right), \\
     & H_{0}=\sum_{\mathbf{k}} \chi_{\mathbf{k}}^{\dagger}\left(\begin{array}{cccc}0 & -t f_{\mathbf{k}}^{*} & 0 & 0\\ -t f_{\mathbf{k}} & 0 & -t_{\perp} & 0 \\ 0 & -t_{\perp} & 0 & -t f_{\mathbf{k}}^{*} \\ 0 & 0 & -t f_{\mathbf{k}} & 0\end{array}\right) \chi_{\mathbf{k}},\\
    &\chi_{\mathbf{k}}=\left(b_{1}(\mathbf{k}), a_{1}(\mathbf{k}), a_{2}(\mathbf{k}), b_{2}(\mathbf{k})\right)^{T},
    \end{aligned}
\end{equation}
where $f(k)=1+e^{-ik_1}+e^{-ik_1+ik_2}$.

When $t_{\perp}$ is much larger than $t$, the $a_1$ and $a_2$ bands are gapped. We can project out the two gapped bands by second order perturbation in $\frac{t}{t_{\perp}}$, and are left with an effective two band model: 
\begin{equation}
    \begin{aligned}
    &H_{eff}=-\frac{t^2}{t_{\perp}}\sum_{\mathbf{k}} \chi_{\mathbf{k}}^{\dagger}
    \left(\begin{array}{cc}0 &  (f_{\mathbf{k}}^{*})^{2} \\  f_{\mathbf{k}}^2 & 0  \end{array}\right)\chi_{\mathbf{k}},\\
    &\chi_{\mathbf{k}}=\left(b_1(\mathbf{k}),  b_2(\mathbf{k})\right)^{T}.
    \end{aligned}
\end{equation}

This Hamiltonian has two QBT points at $\pm \mathbf{K}= \frac{2\pi}{3}( \sqrt{3},\pm 1)$, and the low energy Hamiltonian around these two valleys is: 
\begin{equation}\label{low}
\begin{aligned}
H_{\text{QBT}}=(k^2_x-k^2_y)\mu_1\otimes \tau_0\otimes \sigma_0+2k_x k_y \mu_2 \otimes \tau_0\otimes \sigma_0,
\end{aligned}
\end{equation}
where $\tau$ acts on the two valleys $\pm\bf{K}$. 

Now we can extend the Hamiltonian to 6d by only retaining the $S^4$ configuration with nonzero coverings to arrive at the level of the WZW term. As a result, the gapped 6d Bloch Hamiltonian of a topological insulator in the basis $f(\vec{k})=(b_1(\mathbf{k})),b_2(\mathbf{k})),b_2(-\mathbf{k})),b_1(-\mathbf{k})))^T$ is:
\begin{equation}
\begin{aligned}
 h_{6d}(k)&=[1+\cos(2k_1)+\cos(2k_2-2k_1)+2\cos(k_1)\\
 &+2\cos(k_2-k_1)+2\cos(k_2-2k_1)]\Gamma_1\\
 &+[\sin(2k_2-2k_1)-2\sin(k_1)-\sin(2k_1)\\
 &+2\sin(k_2-k_1)+2\sin(k_2-2k_1)]\Gamma_2\\
 &+m_3\Gamma_3+m_4\Gamma_4+m_5\Gamma_5+m_6\Gamma_6\\
 &+(m+\cos k_3+\cos k_4\\
 &+\cos k_5+\cos k_6)\Gamma_7,\\
 &m_3=\sin k_3,\quad m_4=\sin k_4,\quad m_5=\sin k_5,\\
 &m_6=\sin k_6.
\end{aligned}
\end{equation}

When $-4<m<-2$, the Pontryagin index $P_4$ of the map
$(p_3,p_4,p_5,p_6) \to (m_3,m_4,m_5,m_6,m_7)/\sqrt{\sum_{i=3}^7m_i^2}$ is 1 where $m_7=(m+\cos k_3+\cos k_4+\cos k_5+\cos k_6)$ \cite{RevModPhys.83.1057}. The Chern number of the above 6d Hamiltonian can be calculated numerically. For example, if we take $m=-3.5$, straightforward
computation shows the third Chern number is 2. 

Now we move to the level of the second WZW term, and we extend the effective two band Hamiltonian to 4d:
\begin{equation}\label{4dham}
    \begin{aligned}
 h_{4d}(k)&=[1+\cos(2k_1)+\cos(2k_2-2k_1)+2\cos(k_1)\\
 &+2\cos(k_2-k_1)+2\cos(k_2-2k_1)]\Gamma_1\\
 &+[\sin(2k_2-2k_1)-2\sin(k_1)-\sin(2k_1)\\
 &+2\sin(k_2-k_1)+2\sin(k_2-2k_1)]\Gamma_2\\
 &+m_3\Gamma_3+m_4\Gamma_4\\
 &+(m+\cos k_3+\cos k_4)\Gamma_5,\\
 &m_3=\sin k_3,\quad m_4=\sin k_4.
\end{aligned}
\end{equation}
Similarly, the Pontryagin index $P_2$ of the map
$(p_3,p_4) \to (m_3,m_4)$ is 1 when $-2<m<0$ \cite{RevModPhys.83.1057}. And if we take $m=-1.5$, the second Chern number of Eq.\eqref{4dham} is also 2. As a result, the critical theory is dual to the level-2 WZW model on the Grassmannian manifold using our dimensional reduction method. The derivation is similar to that in the section \ref{sec:dirac} except that the two Chern numbers are replaced with 2. The WZW term is given by:
\begin{eqnarray}\label{DSL}
   S^{\text{level-2}}_{\text{WZW}}=&&4\pi i\int dt d^2x\int du \lbrace\theta_{3}\mathbf{Tr}[M^{-1}(dM)^{3} ]\nonumber\\
   && +\theta_{2}\mathbf{Tr}[M^{-1}(dM)^{2} ](\frac{f}{2\pi})\rbrace.
\end{eqnarray}.

\subsection{AA stacked bilayer checkerboard lattice}
\begin{figure}[htpb] 
\centering 
\includegraphics[width=0.7\textwidth]{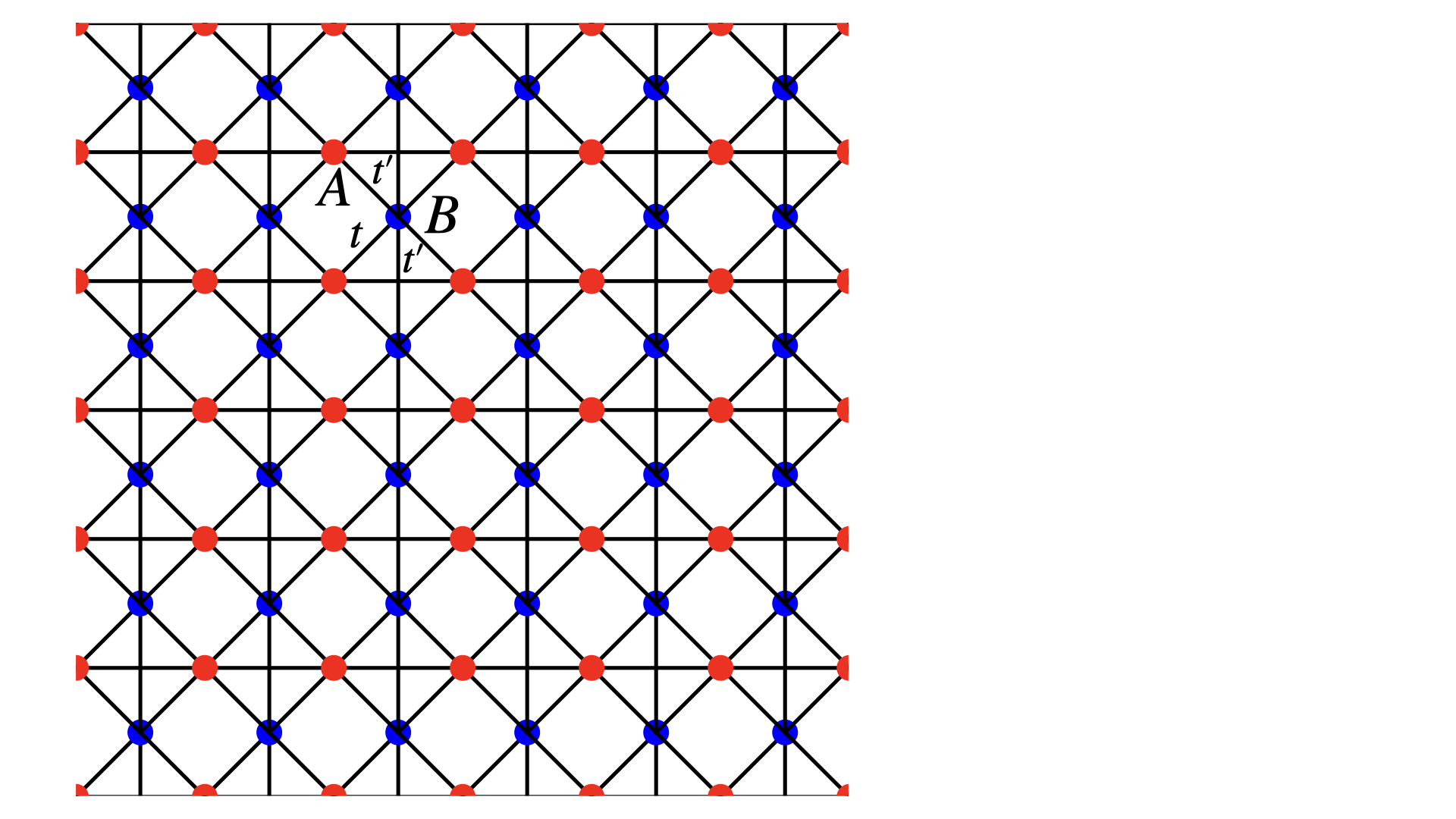} 
\caption{ Single layer checkerboard lattice. The two sublattices are labeled as A and B. The parton construction is taken as uniform nearest neighbour hopping with hopping amplitude $t$, and uniform next-nearest neighbour hopping with amplitude $\pm t^{\prime}$ if the two next-nearest neighbour sites are connected(or not connected) by a bond. } 
\label{ckb} 
\end{figure}

In this section, we investigate a lattice model whose IR theory is Eq.\eqref{qbt_qed3} on the AA stacked bilayer checkerboard lattice whose Bravais lattice is a square lattice, and there are two sites in each unit cell, as shown in Fig \ref{ckb}. This is also the 2d lattice formed by the Cu atoms in a $\text{CuO}_2$ plane of the
cuprates. The band structure has one QBT at half-filling. This QBT with $2\pi$ Berry flux is protected by the $C_4$ lattice rotation and time reversal symmetry \cite{PhysRevLett.103.046811}, thus robust to the long range hoppings preserving these symmetries. In bilayer checkerboard lattice, there are two QBTs per spin in the low energy limit if the system has $C_4$, time reversal and particle-hole symmetries \cite{https://doi.org/10.48550/arxiv.2111.12107}. 

As a result,  the $N_f=4$ QED theory with quadratic dispersion spinons is a natural candidate low energy theory to describe the 
critical point or phase of spin models on the bilayer checkerboard lattice. We take the basis as follows:
\begin{equation}
    f_k^{\dagger}=(f_{A_1}^{\dagger}(\mathbf{k}),f_{B_1}^{\dagger}(\mathbf{k}),f_{A_2}^{\dagger}(\mathbf{k}),f_{B_2}^{\dagger}(\mathbf{k})).
\end{equation}
The particle-hole symmetric parton mean field under this basis is:
\begin{equation}
\begin{aligned}
    &\hat{H}=\sum_k f_k^{\dagger}h_{2d}(k)f_k,\\
   & h_{2d}(k)=d_x(k)\mu_1\otimes\tau_0\otimes \sigma_0+d_z(k)\mu_3\otimes\tau_0\otimes\sigma_0,\\
    &d_x(k)=4t\cos(\frac{k_x}{2})\cos(\frac{k_y}{2}),d_z(k)=2t^{\prime}(\cos k_x-\cos k_y),
    \end{aligned}
\end{equation}
where the $\mu,\tau,\sigma$ act on the sublattice, layer(valley) and spin DOF respectively. It can be directly verified that the low energy Hamiltonian is \eqref{low} after the basis transformation $U=e^{-i\frac{\pi}{4}\mu_1}e^{-i\frac{\pi}{4}\mu_2}$ and we set $t=-2t^{\prime}=2$. 

Following the general scheme, we extend the above Hamiltonian to the 6d and 4d Chern insulators as:
\begin{equation}
\begin{aligned}
 h_{6d}(k)&=2\cos(\frac{k_x}{2})\cos(\frac{k_y}{2})\Gamma_1-(\cos k_x-\cos k_y)\Gamma_2\\
 &+m_3\Gamma_3+m_4\Gamma_4+m_5\Gamma_5+m_6\Gamma_6\\
 &+(m_{6d}+\cos k_3+\cos k_4\\
 &+\cos k_5+\cos k_6)\Gamma_7,\\
 h_{4d}(k)&=2\cos(\frac{k_x}{2})\cos(\frac{k_y}{2})\Gamma_1-(\cos k_x-\cos k_y)\Gamma_2\\
 &+m_3\Gamma_3+m_4\Gamma_4+(m_{4d}+\cos k_3+\cos k_4)\Gamma_5,
\end{aligned}
\end{equation}
with $m_{6d}=-3.5,m_{4d}=-1.5 $ to guarantee the Pontryagin indices equal one. Both Chern numbers are equal to 2, which means that the critical theory is the same as the Eq.\eqref{DSL}.

\subsection{AA stacked bilayer triangular lattice}
\begin{figure}[htpb] 
\centering 
\includegraphics[width=0.5\textwidth]{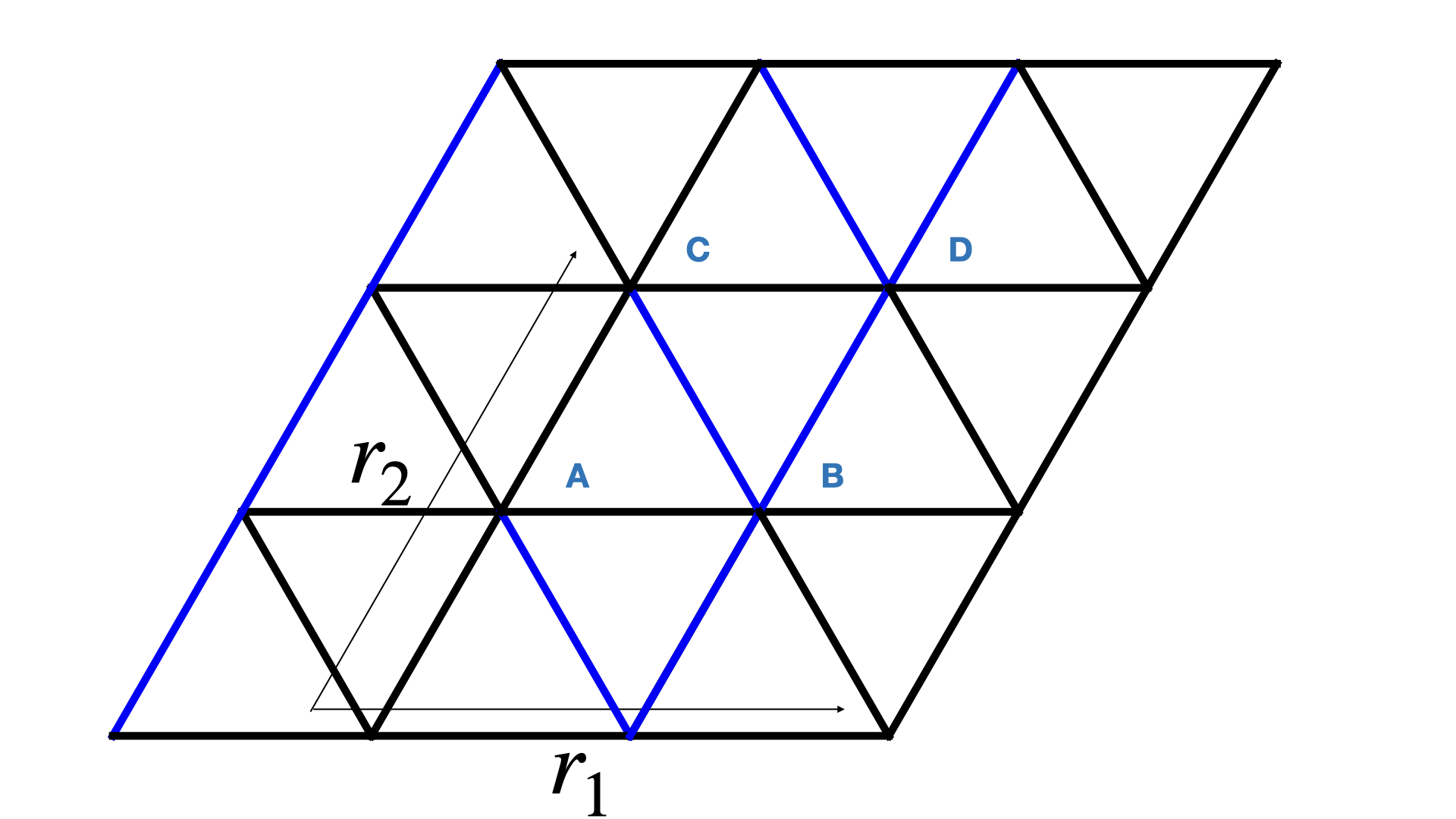} 
\caption{ Staggered $\pi$ flux Hamiltonian on the triangular lattice. The hopping parameter differs by a phase $\pi$ on the black and blue bonds with the same amplitude $t$.} 
\label{triangular}
\end{figure}

Besides the above lattice models on the bilayer bipartite lattice, we can also create a lattice model on the bilayer triangular lattice whose IR theory is Eq.\eqref{qbt_qed3}. We take the bilayer triangular lattice to be AA stacking and the parton Hamiltonian for each layer $H_{\|}$ has the staggered $\pi$ flux pattern, and the hopping pattern is shown in Fig.\ref{triangular}. Motivated by the gapping potential in the AB stacked bilayer graphene, we also include the interlayer hopping $H_{\perp}$ as the gapping term :
\begin{eqnarray}
    &&H_{0}=H_{\|}+H_{\perp},\nonumber\\ 
    &&H_{\|}= \sum_{a=1,2,\langle i, j\rangle, \sigma}-t_{ij}\left(f_{ a,i, \sigma}^{\dagger} f_{ a,j, \sigma}+h . c .\right), \nonumber\\ 
    && H_{\perp}=t_{\perp}\sum_{i}(-f^+_{1,A,i}f_{2,A,i}+f^+_{1,B,i}f_{2,B,i}+h.c.).
    \label{bitri}
\end{eqnarray}
After a layer rotation $U=\exp(-i\pi\tau_y/4)\otimes \mathcal{I}_{4\times 4}$, where $\tau_y$ acts on the layer degrees of freedom, one can obtain the following Bloch Hamiltonian:
\begin{widetext}
\begin{equation}
    \begin{aligned}
      &H_{0}=\sum_{\mathbf{k}} \chi_{\mathbf{k}}^{\dagger}\left(\begin{array}{cccccccc}tg_k & 0 & 0 & -t f_{\mathbf{k}}^{*}&0&t_{\perp}&0&0\\ 0&-tg_k&-t f_{\mathbf{k}} & 0 &t_{\perp}&0 &0 & 0 \\ 0  & -t f_{\mathbf{k}}^{*}&tg_k&0&0&0&0&0\\  -t f_{\mathbf{k}} & 0&0&-tg_k&0&0&0&0\\0&t_{\perp}&0&0&tg_k & 0 & 0 & -t f_{\mathbf{k}}^{*}\\t_{\perp}&0&0&0&0&-tg_k&-t f_{\mathbf{k}} & 0\\0&0&0&0&0  & -t f_{\mathbf{k}}^{*}&tg_k&0\\0&0&0&0&-t f_{\mathbf{k}} & 0&0&-tg_k
      \end{array}\right) \chi_{\mathbf{k}},\nonumber\\
      &\chi_{\mathbf{k}}=U\left(f_{A,1}(\mathbf{k}), f_{B,1}(\mathbf{k}),f_{C,1}(\mathbf{k}), f_{D,1}(\mathbf{k}). f_{A,2}(\mathbf{k}), f_{B,2}(\mathbf{k}),f_{C,2}(\mathbf{k}), f_{D,2}(\mathbf{k})\right)^{T},
    \end{aligned}
\end{equation}
\end{widetext}
where $f_{\mathbf{k}}=-\cos(\frac{k_x}{2}+\frac{\sqrt{3}k_y}{2})+i\sin(\frac{k_x}{2}-\frac{\sqrt{3}k_y}{2})$ and $g_k=\cos k_x$.

The $A_i,B_i$ DOF are gapped out and the effective Hamiltonian after second order perturbation is given by:
\begin{equation}
    \begin{aligned}
    &H_{eff}=-\frac{t^2}{t_{\perp}}\sum_{\mathbf{k}} (\chi'_{\mathbf{k}})^{\dagger}
    \left(\begin{array}{cccc}0 &  (f_{\mathbf{k}}^{*})^{2} & 0 & 0\\  f_{\mathbf{k}}^2 & 0 & 0 & 0 \\ 0 & 0 & 0 &  (f_{\mathbf{k}}^{*} )^2\\ 0 & 0 &  f_{\mathbf{k}}^2 & 0\end{array}\right)\chi'_{\mathbf{k}},\\
    &\chi'_{\mathbf{k}}=\left(\chi_{3}(\mathbf{k}), \chi_{8}(\mathbf{k}), \chi_{7}(\mathbf{k}), \chi_{4}(\mathbf{k})\right)^{T}.
    \end{aligned}
    \label{qbt}
\end{equation}

The spectrum is gapless at $(k_x,\sqrt{3}k_y)=(\frac{\pi}{2},\frac{\pi}{2})$, and the low energy action also takes the form as Eq.\eqref{low} after we rescale $k_y$ to $\frac{k_y}{\sqrt{3}}$.

Now we extend the Hamiltonian to 6d and 4d by retaining the $S^4$ and $S^2$ configuration with nonzero coverings. The gapped 6d and 4d Bloch Hamiltonians are:
\begin{equation}
\begin{aligned}
 h_{6d}(k)&=(f_1^2-f_2^2)\Gamma_1+2f_1f_2\Gamma_2\\
 &+m_3\Gamma_3+m_4\Gamma_4+m_5\Gamma_5+m_6\Gamma_6\\
 &+(m+\cos k_3+\cos k_4\\
 &+\cos k_5+\cos k_6)\Gamma_7,\\
  h_{4d}(k)&=(f_1^2-f_2^2)\Gamma_1+2f_1f_2\Gamma_2\\
 &+m_3\Gamma_3+m_4\Gamma_4\\
 &+(m+\cos k_3+\cos k_4)\Gamma_5,\\
 &f_1=\text{Re}f_{\mathbf{k}},\quad f_2=\text{Im}f_{\mathbf{k}},\quad m_3=\sin k_3,\\
 &m_4=\sin k_4,\quad m_5=\sin k_5,\quad m_6=\sin k_6.
\end{aligned}
\end{equation}

The Pontryagin indexes $P_4$ and $P_2$ both equal one when the mass $m$ equals -3.5 and -1.5 respectively. The two Chern numbers $C_3$ and $C_2$ of the 6d and 4d Hamiltonian are both two, so the dual topological action is the same as Eq.\eqref{DSL}.
\section{Interface theory and anomalies}\label{sec 5}
In this section, we perform serval theoretical nontrivial consistency checks of our correspondence between the topological NLSM and the \text{QED} theory. The first consistency check is the interface theory, which is a generalized Jackiw-Rebbi problem. A similar discussion for Dirac fermions and their corresponding topological NLSMs is investigated in \cite{PhysRevB.102.245113}. We take one component of the order parameter as the domain wall configuration, and investigate the interface theory from both the $\text{QED}$ theory \eqref{qbt_qed3} and the topological NLSM \eqref{DSL}. We show that they both give the 1+1d $\text{SU}(2)_2$ WZW model localized on the interface. The second consistency check is the 't Hooft anomaly matching. We show that there is a $\mathbb{Z}_2$ anomaly with respect to the global symmetry of the $\text{QED}$ theory \eqref{qbt_qed3} and the topological NLSM \eqref{DSL}.
\subsection{Interface theory}
We start from a single QBT valley. We take the domain wall potential as $\delta(y)=\tanh(y/u)$ and the interface is at $y=0$:
\begin{equation}\label{interface_qbt}
   \begin{array}{l}H_{\mathsf{interface}}=\left(\begin{array}{cc}2k_x k_y &  k_x^2-(k_y-i\delta t(y))^2 \\  k_x^2-(k_y+i\delta t(y))^2& -2k_x k_y  \end{array}\right).
   \end{array}
\end{equation}
  A physical way to understand this domain wall construction is that when $k_x =0$, this Hamiltonian is reduced to IR theory of the 1+1d system Eq.\eqref{1+1} with a domain wall.

There are two zero modes for the Hamiltonian \eqref{interface_qbt} with zero $k_x$ without normalization:
\begin{equation}
   \begin{array}{l}
   f_i=\left(\begin{array}{cc}  0 \\  \nu_i(y) \end{array}\right),
   \end{array}
\end{equation}
where the two  linearly independent solutions of $\nu_i$ are:
\begin{equation}
\nu_1=\exp(-\int_0^y \delta t(z)dz),\quad\quad \nu_2=y\exp(-\int_0^y \delta t(z)dz).
\end{equation}
The effective Hamiltonian projected on these two solutions is:
\begin{equation}
   \begin{array}{l}H_{\mathsf{eff}}=A\left(\begin{array}{cc}0 &  -ik_x\\ ik_x& 0 \end{array}\right),
   \end{array}
\end{equation}
where $A=-2\int_{-\infty}^{+\infty} dy \nu_1 \partial_y\nu_2$.

After the inclusion of the spin and valley DOF to the interface Hamiltonian \eqref{interface_qbt}, there are 8 zero modes considering the valley and spin DOF with the same domain wall as in \eqref{interface_qbt}. Now we move to the discussion of the kinematic consistency between the WZW terms in \eqref{DSL} and the $\text{QED}$ theory \eqref{qbt_qed3}. We start from the kinematic consistency discussion between the first WZW term and the $\text{QED}$ theory \eqref{qbt_qed3}, where the $\text{U}(1)$ gauge field does not play a role. We break the global symmetry of the $
\text{QED}$ theory \eqref{qbt_qed3}  into $\text{SO}(4)$ by introducing the five-component mass term: $(\tilde{n}_1,\tilde{n}_2,\tilde{n}_3,\tilde{n}_4,\delta t)$, where the first four components $\tilde{n}_i$ only depend on the spacetime coordinates $t,x$. Now the $\text{U}(1)$ gauge field is confined in IR, and the interface Hamiltonian becomes:
\begin{equation}\label{interface}
   \begin{aligned}
   H^{\mathsf{mass}}_{\mathsf{interface}}&=2k_x k_y \mu_3\otimes\tau_0\otimes \sigma_0+(k^2_x-k^2_y+\delta t^2(y))\mu_1\otimes\tau_0\otimes \sigma_0\\&+(k_y\delta t+\delta tk_y)\mu_2\otimes \tau_3\otimes \sigma_0+\sum^3_{i=1}\tilde{n}_i \mu_2\otimes \tau_1\otimes \sigma_i\\&+2\tilde{n}_4 k_y \mu_2\otimes \tau_2\otimes \sigma_0.
   \end{aligned}
\end{equation}
 There are 8 zero modes considering the valley and spin DOF. The effective Hamiltonian can be arrived at by projecting $ H^{\mathsf{mass}}_{\mathsf{interface}}$ on the zero modes:
 \begin{equation}
   \begin{aligned}
   H_{\mathsf{eff}}&=k_x \mu_2\otimes \tau_3\otimes \sigma_0+\tilde{n}_4\mu_2\otimes \tau_1\otimes \sigma_0\\
   &+\sum^3_{i=1}\tilde{n}_i \mu_0\otimes \tau_2\otimes \sigma_i,
   \end{aligned}
\end{equation}
where we have set $A=1$. This is nothing but the 1+1d Dirac fermion Hamiltonian with $S^3$ target space. We can arrive at the standard representation by a unitary transformation:
 \begin{equation}\label{1+1 Dirac}
 \begin{aligned}
  U_1&=(1\oplus \mu_1)\otimes \sigma_0,\quad  U_2=\mu_0\otimes \exp(\frac{\pi i}{4}\tau_1)\otimes \sigma_0,\\
  U_3&=(1\oplus \mu_2)\otimes \sigma_0,\\
H_{\mathsf{Dirac}}&=(U_3 U_2 U_1)^{\dagger}H_{\mathsf{eff}}(U_3 U_2 U_1)\\
&=k_x \mu_2\otimes \tau_0\otimes \sigma_0+\tilde{n}_4\mu_3\otimes \tau_0\otimes \sigma_0\\
&+\sum^3_{i=1}\tilde{n}_i \mu_1\otimes \tau_0\otimes \sigma_i.
\end{aligned}
\end{equation}
This 1+1d theory is equivalent to $\text{SU}(2)_2$ WZW model after integrating out the fermions using the AW mechanism. 


Now we consider the interface theory of the WZW model. Since the Hamiltonian Eq.\eqref{interface} has quadratic dispersive fermions, we first put it on the lattice and use our dimensional reduction method to derive the WZW terms.  The honeycomb lattice \footnote{Here take unit vectors as $a_1= (-\frac{\sqrt{3}}{2},\frac{1}{2})a$ and $a_2=(0,1)a$. And $k_1=-\frac{\sqrt{3}}{2}k_x+\frac{1}{2}k_y,k_2=k_y$ here.}
regularization of the Hamiltonian Eq.\eqref{interface} is:
\begin{equation}
\begin{aligned}
 h(k)&=[1+\cos(2k_1)+\cos(2k_2-2k_1)+2\cos(k_1)\\
 &+2\cos(k_2-k_1)+2\cos(k_2-2k_1)+\tilde{n}_5^2]\Gamma_1\\
 &+[\sin(2k_2-2k_1)-2\sin(k_1)-\sin(2k_1)\\
 &+2\sin(k_2-k_1)+2\sin(k_2-2k_1)]\Gamma_2\\
 &+\frac{2}{3}\sin (3k_2)\tilde{ n}_5\Gamma_3+\frac{2}{3}\sin (3k_2) \tilde{n}_4\Gamma_4+\tilde{n}_1\Gamma_5+\tilde{n}_2\Gamma_6\\
 &+\tilde{n}_3 \Gamma_7,
\end{aligned}
\end{equation}
where $\tilde{n}_5=\delta t(y)$. After integrating out the fermions through our dimensional reduction method, we can arrive at the level-2 SO(5) WZW term in 2+1 dimension:
\begin{equation}
   S_{\text{SO(5)}}=\frac{12\pi i}{8\pi^2}\int^1_0 du\int d^3 x \epsilon^{ijklm} n^i \partial_u n^j\partial_x n^k\partial_y n^l\partial_t n^m,
\end{equation}
where $n_a=\frac{\tilde{n}_a}{\sqrt{\sum^5_{a=1}\tilde{n}^2_a}}$.

To describe the interface in the WZW model, we take $n_5=\cos\theta(y)$, $n_i=\sin\theta(y)m_i(x,t,u),i=1,2,3,4$. When $y \to \pm \infty$, $\theta\to 0/\pi$ and $n_5\to\pm 1$, which is consistent with the behavior of $\delta t(y)$. Thus the action becomes 1+1d $\text{SU}(2)_2$ WZW model:
\begin{eqnarray}
   S_{\text{SO(5)}}&&=\frac{12\pi i}{8\pi^2}\int d\theta\sin^3  \theta \int^1_0 du\int dxdt\epsilon^{ijkl} m^i \partial_u m^j\partial_x m^k\partial_t m^l\nonumber\\
   &&=\frac{4\pi i}{2\pi^2} \int^1_0 du\int dxdt \epsilon^{ijkl} m^i \partial_u m^j\partial_x m^k\partial_t m^l,
\end{eqnarray}
which is consistent with the interface theory obtained from the QBT model.

Now we move to the discussion of the kinematic consistency between the second WZW term in \eqref{DSL} and the $\text{QED}$ theory \eqref{qbt_qed3}. We breaks the global symmetry of the $\text{QED}$ theory into $\text{SO}(2)$ by introducing the mass term $(\tilde{n}_1,\tilde{n}_2,\delta t)$ :
\begin{eqnarray}\label{interface2}
H^{\text{U(1)}}_{\mathsf{interface}}&&=2(k_x-a_x) k_y \mu_3\otimes \tau_0\otimes \sigma_0 \nonumber\\&&+[(k_x-a_x)^2-k^2_y+\delta t^2(y)]\mu_1\otimes \tau_0\otimes \sigma_0\nonumber\\&&+(k_y\delta t+\delta t k_y)\mu_2\otimes \tau_3\otimes \sigma_0+\tilde{n}_1 \mu_2\otimes \tau_2\otimes \sigma_0\nonumber\\&&+2\tilde{n}_2 k_y \mu_2\otimes \tau_2\otimes \sigma_0,\nonumber
\end{eqnarray}
where interface is also at $y=0$ and $\delta(y)=\tanh(y/u)$. Similarly,  the 1+1d effective theory can be arrived at by projecting Eq.\eqref{interface2} to the eight zero modes :
\begin{equation}
\begin{aligned}
H^{\text{U(1)}}_{\mathsf{eff}}&=(k_x-a_x) \mu_2\otimes \tau_3\otimes s_0+\tilde{n}_1\mu_0\otimes \tau_2\otimes s_0\\
&+\tilde{n}_2 \mu_2\otimes \tau_1\otimes s_0.
\end{aligned}
\end{equation}
After integrating out the Dirac fermions and expanding the resultant action through the AW mechanism, we will obtain the topological term:
\begin{eqnarray}\label{1+1 Dirac1}
S=\frac{i}{2\pi}\times 4 \int_{M_3} da \wedge d\phi
\end{eqnarray}
where $\tilde{n}_1 +i\tilde{n}_2 = |\tilde{n}|e^{i\phi}$.

To see if this result is consistent with the second WZW term, we consider the Bloch Hamiltonian which is the regularization of the Hamiltonian \eqref{interface2}  on the honeycomb lattice :
\begin{equation}
\begin{aligned}
 h'(k)&=[1+\cos(2k_1)+\cos(2k_2-2k_1)+2\cos(k_1)\\
 &+2\cos(k_2-k_1)+2\cos(k_2-2k_1)+\tilde{n}_3^2]\Gamma_1\\
 &+[\sin(2k_2-2k_1)-2\sin(k_1)-\sin(2k_1)\\
 &+2\sin(k_2-k_1)+2\sin(k_2-2k_1)]\Gamma_2\\
 &+\frac{2}{3}\sin (3k_2) \tilde{n}_3\Gamma_3+\frac{2}{3}\sin (3k_2) \tilde{n}_1\Gamma_4+\tilde{n}_2\Gamma_5,
\end{aligned}
\end{equation}then the second WZW term is given by
\begin{equation}
   S_{\text{SO(3)}}=\frac{4i}{8\pi}\int \epsilon^{\mu\nu\rho\lambda}\partial_{\nu}a_{\mu}\epsilon^{ijk}n_{i}\partial_{\rho}n_j\partial_{\lambda}n_k 
\end{equation}
where $n_a=\frac{\tilde{n}_a}{\sqrt{\sum^3_{a=1}\tilde{n}^2_a}}$.

Moreover, the interface configuration corresponding to Eq.\eqref{interface2} should satisfy: $n_3=\cos\theta(y)$, $n_i=\sin\theta(y)m_i(x,t,u)$, where as $y \to \pm \infty$, $\theta\to 0/\pi$ and $n_3\to\pm 1$. Under this configuration, the second WZW term is exactly Eq.\eqref{1+1 Dirac1}:
\begin{eqnarray}
   S_{\text{SO(3)}}&&=\frac{4i}{4\pi}\int d\theta\sin  \theta\int^1_0 du\int dxdt \epsilon^{\mu\nu\rho}\partial_{\nu}a_{\mu}\epsilon^{ij}m_{i}\partial_{\rho}m_j \nonumber\\
   &&=\frac{4i}{2\pi} \int du\int dxdt \epsilon^{\mu\nu\rho}\partial_{\nu}a_{\mu}\partial_{\rho}\phi .
\end{eqnarray}
\subsection{Anomaly matching}
The global symmetry group $G$ of the topological NLSM Eq.\eqref{DSL}, or more generally the level-$k$ $N=6$ Stiefel liquid is \cite{PhysRevX.11.031043}:
\begin{eqnarray}\label{totalsym}
   G=\frac{\text{SO}(6)\times \text{U}(1)_{\text{top}}}{\mathbb{Z}_2},
\end{eqnarray}
   where $\text{SO(6)}$ is the projective representation of $\text{SU(4)}$ rotation on Grassmannian manifold. In the QED theory, the $\text{SU(4)}$ rotation acts on the valley and spin indices of the fermion. 
   The $\text{U}(1)_{\text{top}}$ is the $\text{U}(1)$ topological symmetry of the gauge field, whose conserved current is \begin{eqnarray}
  j^\mu =\frac{k}{2\pi}\epsilon^{\mu\nu\lambda}\partial_{\nu}a_{\lambda}.
\end{eqnarray}
The quotient $\mathbb{Z}_2$ of the global symmetry $G$ is the common center of the $\text{SO}(6)$ and $\text{U}(1)_{\text{top}}$. In the QED theory with Dirac fermion, this quotient can be understood by how $G$ acts on the monopole operators, which are the gauge invariant physical operators \cite{PhysRevX.10.011033}: 
\begin{equation}
\vec{\Phi}= (f_{\alpha, s}^{\dagger}\left(\epsilon \vec{\tau}\right)^{\alpha \beta} \epsilon^{s s^{\prime}} f_{\beta, s^{\prime}}^{\dagger}M^{\dagger}_{\text{bare}},i f_{\alpha, s}^{\dagger}(\epsilon)^{\alpha \beta}\left(\epsilon \vec{s}\right)^{s s^{\prime}} f_{\beta, s^{\prime}}^{\dagger} M^{\dagger}_{\text{bare}}) ,
\end{equation}
where the Pauli matrices $\tau$ and $\sigma$ act on the valley and spin indexes respectively, and $\epsilon$ is the second rank antisymmetric tensor. $f^{\dagger}_{\alpha,s}$ creates a fermion in the zero mode with the valley and spin indexes $\alpha$ and $s$. $M^{\dagger}_{\text{bare}}$ creates 'bare' flux quanta without
filling any zero mode. This six components monopole operator $\vec{\Phi}$ forms the vector representation of $\text{SO}(6)$ and is odd under both $\mathbb{Z}_2$  center of $\text{SO}(6)$ and $U(1)_{\text{top}}$: $\Phi_{ij}\rightarrow-\Phi_{ij}$.  Moreover, one
can show that any local operator is simultaneously odd or even under the two $\mathbb{Z}_2$  center.

From the level-$k$ WZW model perspective, the self anomalies of $\text{SO(6)}$ and $\text{U}(1)_{\text{top}}$ are both $k \ \text{mod} \ 2$ and the mixed anomaly between them is $k \ \text{mod}\ 4$ \cite{PhysRevX.11.031043}. This implies that only the mixed anomaly is nontrivial for the even level $k=2$ in the Eq. \eqref{DSL}. 

 Let's start from the simpler case of the QED theory with Dirac fermions as a warm up. The global symmetry can be rewritten as: 
\begin{equation}
    G=\frac{\text{SO}(6)\times \text{U}(1)_{\text{top}}}{\mathbb{Z}_2}=\text{PSU}(4)\ltimes \text{U}(1)_{\text{top}}.
\end{equation}
Now if we turn on the background gauge field of $G$, the Dirac quantization of the $\text{U(1)}$ bundle is modified to  which is similar to the Abelian-Higgs model with four flavours \cite{10.21468/SciPostPhys.6.1.003}:
\begin{eqnarray}
\frac{da}{2\pi}=\frac{u_2 (BG)}{4}+\frac{w^{\text{TM}}_2 }{2} \ \text{mod}\ 1,
\end{eqnarray}
where $u_2 (BG)\in H^2(BG,\mathbb{Z}_4)$ and $BG$ is the classifying space of the $G$ group. Besides, $w^{\text{TM}}_2$ belongs to the second Stiefel-Whitney (SW) class $ H^2(\text{Spin}(3),\mathbb{Z}_2)$ and classify the spin structure of the Dirac fermion. Therefore, after turning on the background gauge field of $G$, the inflow action of self anomaly of $\text{U}(1)_{\text{top}}$ and the mixed anomaly of $\text{SO(6)}$ and $\text{U}(1)_{\text{top}}$ is given by
\begin{eqnarray}
S_{ \text{DSL-inflow}}&&=\int_{M_4} 2\pi i\frac{dA_{\text{top}}}{2\pi}\frac{da}{2\pi}\nonumber\\&&=\int_{M_4} \frac{\pi i}{2}\frac{dA_{\text{top}}}{2\pi}(w_2 (G)+2w^{\text{TM}}_2 )\nonumber\\&&=\int_{M_4}\frac{\pi i}{2}\frac{dA_{\text{top}}}{2\pi}(w_2 (G)+2\frac{dA_{\text{top}}}{2\pi} ).
\end{eqnarray}
And the self anomaly of SO(6) comes from Pauli-Villars regulator of fermions  \cite{song2019unifying}.

Now let's move to the QED theory with fermions of quadratic dispersion. We expect that there is a $\mathbb{Z}_2$ mixed anomaly between $\text{SO(6)}$ and $\text{U}(1)_{\text{top}}$, which matches the anomaly of the level-2 WZW model Eq. \eqref{DSL}.  We first show the $\text{SO}(6)$ symmetry is anomaly-free. The anomaly of the $\text{SO}(6)$ symmetry is generally determined by the inflow action \cite{PhysRevX.11.031043}:
\begin{equation}
\begin{aligned}
S_{ \text{SO(6)-inflow}}&=i\pi a \int_{M_4} w^{\text{SO(6)}}_4,
\end{aligned}
\end{equation}
where $w^{\text{SO(6)}}_4$ is the fourth SW classes and $a=0,1$ represents this symmetry is anomaly-free or anomalous respectively. To determine the coefficient $a$, we use the trick of breaking the global symmetry $\text{SO}(6)$ into $\text{SO(3)}_s \times\text{SO(3)}_v$ \cite{PhysRevX.11.031043}, which is the spin and valley rotation symmetry.  Then we have $w^{\text{SO(6)}}_4=w^{\text{SO(3)}_s}_2 \cup w^{\text{SO(3)}_v}_2$ where $w^{\text{SO(3)}_{s/v}}_2$ is the second SW classes of the corresponding bundles. If we insert a spin skyrmion in the $\text{QED}$ theory, which is a unit monopole of the $\text{SO}(3)_s$ gauge field and induces a nontrivial $w^{\text{SO(3)}_{s}}_2$, there are six zero modes belonging to the spin 1 and spin 0 representations of valley $\text{SO}(3)_v$ rotation (which is investigated in detail in the next section). Since these two representations are not projective representations, the coefficient $a$ must be even. As a result, the $\text{SO}(6)$ symmetry of the $\text{QED}$ theory is anomaly-free, which matches the anomaly (no anomaly in this case) of the topological NLSM in Eq.\eqref{DSL}.

Moreover, the $\text{U}(1)_{\text{top}}$ conserved current of the QBT model is:
\begin{eqnarray}
  j^\mu =2\frac{1}{2\pi}\epsilon^{\mu\nu\lambda}\partial_{\nu}a_{\lambda},
\end{eqnarray}
where coefficient 2 comes from the two layers.

Since the fermion of quadratic dispersion has an angular momentum $l =1$ under the spatial rotation SO(2), it carries the faithful representation of spatial rotation. As a result, the Dirac quantization of the U(1) bundle is modified to:
\begin{eqnarray}
\frac{da}{2\pi}=\frac{u_2 (BG)}{4} \ \text{mod}\ 1.
\end{eqnarray}
 Then the inflow action of self anomaly of $\text{U}(1)_{\text{top}}$ and the mixed anomaly of $\text{SO(6)}$ and $\text{U}(1)_{\text{top}}$ is given by
\begin{equation}
\begin{aligned}
S_{ \text{QBT-inflow}}&=\int_{M_4} 4\pi i\frac{dA_{\text{top}}}{2\pi}\frac{da}{2\pi}\\
&=\int_{M_4} i\pi \frac{dA_{\text{top}}}{2\pi} u_2 (BG)
\end{aligned}
\end{equation}
which implies  only the mixed anomaly is nontrivial and a $\mathbb{Z}_2$ phase.

\section{Zero modes of monopoles and vortexes}\label{sec 6}
In this section, in order to figure out the physical consequences of the topological terms, we discuss the quantum number of zero modes under the insertion of $\text{U(1)}$ monopoles and vortexes.
\subsection{Zero modes of $U(1)$ monopole}
We discuss the quantum number of zero modes in the ground state with the inclusion of a U(1) gauge monopole in QBT models. This can explain the physical meaning of the second topological term in \eqref{DSL} corresponding to the lower dimensional winding number. 

We first give a short review on the zero mode solutions of Dirac fermion on $S^2$, with a monopole at the center of $S^2$. The sphere is the compactification of the plane $R^2$, and the latter can be obtained from $S^2$ by stereographic projection. The Hamiltonian is \cite{imura2012spherical}:
\begin{equation}\label{dirac_monopole}
    \begin{array}{l}
    H^{\mathsf{Dirac}}_{\mathsf{mon}}=\left(\begin{array}{cc}0 &  -D_\theta+iD_{\phi}-\frac{\cot\theta}{2}\\ D_\theta+iD_{\phi}+\frac{\cot\theta}{2}& 0  \end{array}\right),
    \end{array}
\end{equation}
where 
\begin{equation}
\begin{aligned}
&D_{\theta}=\partial_\theta-ia_{\theta},\quad D_{\phi}=\frac{\partial_{\phi}}{\sin\theta}-ia_{\phi}.
\end{aligned}
\end{equation}
We take the monopole configuration as follows:
\begin{displaymath}
a_{\phi} = \left\{ \begin{array}{ll}
\frac{1}{2}\tan\frac{\theta}{2}, & 0<\theta<\frac{\pi}{2}\\
-\frac{1}{2}\cot\frac{\theta}{2}, & \frac{\pi}{2}<\theta<\pi.
\end{array} \right.
\end{displaymath}
Since the Hamiltonian \eqref{dirac_monopole} has rotation symmetry in the z-direction, we assume the zero mode solution in the northern hemisphere is:
\begin{equation}
    \begin{array}{l}
    \psi_{N}=\left(\begin{array}{cc}f_{N}(\theta)\\ g_{N}(\theta) \end{array}\right)e^{im\phi},
    \end{array}
\end{equation}

where $m\in \mathbb{Z}$ and the wave function satisfies:
\begin{equation}
\begin{aligned}
    &( d_\theta-\frac{2m-1}{2\sin\theta})f_{N}(\theta)=0,\\
   & -(d_\theta+\frac{2m-1}{2\sin\theta}+\cot\theta)g_{N}(\theta)=0.
    \end{aligned}
\end{equation}
  The solution is:
\begin{equation}
\begin{aligned}
    &f_{N}(\theta)=(\tan\frac{\theta}{2})^{m-\frac{1}{2}},\\
    &g_N(\theta)=\frac{1}{\sin\theta}(\tan\frac{\theta}{2})^{-m+\frac{1}{2}}.
    \end{aligned}
\end{equation}
Since the gauge field configurations in the northern and southern hemisphere differ with a gauge transformation on the equator: $A^{N}_{\phi}d\phi-A^{S}_{\phi}d\phi=d\phi$, the wavefunctions also differs by a gauge transformation: $\psi_{N}|_{\text{equator}}=\psi_{S}\exp(i\phi)|_{\text{equator}}$.
Thus the zero mode in the southern hemisphere is:
\begin{equation}
    \begin{array}{l}
    \psi_{S}=\left(\begin{array}{cc}f_{S}(\theta)\\ g_{S}(\theta) \end{array}\right)e^{i(m-1)\phi},
    \end{array}
\end{equation}
and satisfy the following equations:
\begin{equation}
\begin{aligned}
    &( d_\theta-\frac{2m-3}{2\sin\theta}-\frac{1}{\sin\theta})f_{S}(\theta)=0,\\
    &-(d_\theta+\frac{2m-3}{2\sin\theta}+\cot\frac{\theta}{2})g_S(\theta)=0.
    \end{aligned}
\end{equation}
The solution is:
\begin{equation}
\begin{aligned}
    &f_S(\theta)=(\tan\frac{\theta}{2})^{m-\frac{1}{2}},\\
    &g_S(\theta)=\frac{1}{\sin^2\frac{\theta}{2}}(\tan\frac{\theta}{2})^{-m+\frac{3}{2}}.
    \end{aligned}
\end{equation}
It is direct to verify that there is only one normalizable solution with $m=0$. In fact, this result is guaranteed by the Atiysh-Singer index theorem.

Now we move to the QBT model. The Hamiltonian with the same monopole configuration is: 
\begin{equation}
    \begin{array}{l}H^{\mathsf{QBT}}_{\mathsf{mon}}=\left(\begin{array}{cc}0 &  (-D_\theta+iD_{\phi}-\frac{\cot\theta}{2})^2 \\ (D_\theta+iD_{\phi}+\frac{\cot\theta}{2})^2& 0  \end{array}\right). 
    \end{array}
\end{equation}
Following the procedure similar to that of Dirac fermions, we find that there are two zero modes. The first solution is the same as that of Dirac fermion:
\begin{equation}
\begin{aligned}
    &\begin{array}{l}
    \psi^1_{N}=\left(\begin{array}{cc}1\\\frac{1}{\cos\theta+1} \end{array}\right)(\tan\frac{\theta}{2})^{-\frac{1}{2}},
    \end{array}\\
     &\begin{array}{l}
    \psi^1_{S}=\left(\begin{array}{cc}1\\\frac{2}{\cos\theta+1} \end{array}\right)(\tan\frac{\theta}{2})^{-\frac{1}{2}}e^{-i\phi}.
    \end{array}
    \end{aligned}
\end{equation}
And the second one is given by:
\begin{equation}
\begin{aligned}
    &\begin{array}{l}
    \psi^2_{N}=\left(\begin{array}{cc}1\\\frac{1}{\cos\theta+1} \end{array}\right)(\tan\frac{\theta}{2})^{-\frac{1}{2}}\theta,
    \end{array}\\
    & \begin{array}{l}
    \psi^2_{S}=\left(\begin{array}{cc}1\\\frac{2}{\cos\theta+1} \end{array}\right)(\tan\frac{\theta}{2})^{-\frac{1}{2}}\theta e^{-i\phi}.
    \end{array}
        \end{aligned}
\end{equation}

Now if we include the spin and valley DOF, there will be eight zero mode solutions in total and four zero modes solutions per valley. In principle, there are total $2^8$
possible ways to occupy them in the ground state. However, physical ground states must be gauge invariant. In other words, the physical ground states must be charge neutral. As a result, the ground states can only occupy four zero modes. We focus on the valley polarized occupation, which means that the ground state occupies two zero modes per valley, and there are $C_4^2=6$ possible ways per valley. These zero mode doublets are charge neutral solitons:  three are spin 1 and the other three are spin 0. Actually, the second WZW term \eqref{DSL} can describe the spin-1 quantum number (per valley) carried by the $\text{U}(1)$ monopole. If we consider the AFM order parameter: $\vec{N}_e=\vec{N}_m=\vec{M}_m=0, \vec{M}_e=(n_1,n_2,n_3)$, then the second WZW term becomes:
\begin{eqnarray}\label{quan_num_gauge}
   S&&=4\times\frac{2i}{16\pi}\int^1_{0} du\int dxdydt \epsilon^{\mu\nu\rho\lambda}\partial_{\nu}a_{\mu}\epsilon^{ijk}n_{i}\partial_{\rho}n_j\partial_{\lambda}n_k\nonumber\\
   &&=i\int^1_{0} du\int dt \epsilon^{\rho\lambda}\epsilon^{ijk}n_{i}\partial_{\rho}n_j\partial_{\lambda}n_k.
\end{eqnarray}
Here the extra 4 in the first line comes from the trace over the spin and valley DOF. This exactly describes the two copies of spin-1 quantum number contributed by the two valleys on the core of the $\text{U(1)}$ monopole.
\subsection{Zero mode of valley vortexes }
The physical consequence of the first WZW term in \eqref{DSL} can be reflected from the quantum number of the zero modes under the inclusion of  a valley vortex. The  vortex Hamiltonian is:
\begin{equation}
    \begin{aligned}
   H_{\text{vortex}}&=H_{\text{QBT}}+ \Delta(r)\cos\phi \mu_3\otimes\tau_1\otimes \sigma_0\\&+\Delta(r)\sin\phi \mu_3\otimes\tau_2\otimes \sigma_0.
    \end{aligned}\label{gauge field}
\end{equation}
where $(r,\phi)$ are the polar coordinates and the amplitude $\Delta(r)\rightarrow 0$ when $r\rightarrow 0$.

There are two zero modes \cite{2012Zero} per spin. Similar to the argument in the last section, the physical ground states must be charge neutral and only occupy half of the total four
zero modes. As a result, there are three charge neutral solitons with spin 1 and another three charge neutral solitons with spin 0 of the SO(3) spin rotation symmetry. The spin 1 quantum number is consistent with the first topological term in the WZW term \eqref{DSL}. To see the soliton with the spin 1 quantum number from the WZW term, we break the global symmetry to SO(5) and take the target space as in \eqref{so(5)}:     
\begin{eqnarray}
   S&&=\frac{4\pi i}{\text{Area}(S^4)}\int^1_0 du\int drd\phi dt \epsilon^{ijklm}n_{i}\partial_{r}n_j\partial_{\phi}n_k \partial_{t}n_{l}\partial_{u}n_{m} \nonumber\\
   &&=i\int dudt \epsilon^{ijk} N_i \partial_t N_j\partial_u N_k,
   \label{firstterm}
\end{eqnarray}
where we take the configuration of the five component $n_i$ as: $n_{i}=N_i(u,t) \cos\theta(r) \ (i=1,2,3)$ and $n_{i}=m_i(\phi) \sin\theta(r) \ (i=4,5)$,  when $r \to 0/\infty$, $\theta\to 0/\frac{\pi}{2}$. The three-component vector $\vec{N}$ is a unit vector: $N_1^2+N_2^2+N_3^2=1$, and the action \eqref{firstterm} describes the spin-1 quantum number, which is half of the action \eqref{quan_num_gauge}. And $\vec{m}$ is a unit vector with winding number 1: $ \int d\phi \epsilon^{ij} m_i\partial_{\phi}m_j=2\pi$.
\section{Macroscopic symmetry representation and quantum number of monopoles}
An important class of operators that is possible to destabilize the QED theory is the monopole operators\cite{PhysRevX.10.011033,song2019unifying,PhysRevB.77.224413}. Physically, they can serve as order parameters adjacent to the critical theory. The order represented by the monopole operators can be inferred by how the macroscopic symmetries (lattice, onsite spin rotation, etc) act on the monopoles and the quantum number of the monopole operators. Since there are eight zero modes for the nonrelativistic $\text{QED}$ theory \eqref{qbt_qed3} with $2\pi$ flux, the structure of monopoles is much richer than that of Dirac fermions\cite{PhysRevX.10.011033,song2019unifying,PhysRevB.77.224413}. Thus in this section, we discuss how the macroscopic symmetries act on the zero modes from the projective symmetry representation of the parton construction of our lattice models. Moreover, we can also clarify the quantum number of the bare monopole operator by group-theoretic considerations for the bipartite lattice models.
\subsection{Classification of monopole operators}
We have to fill half of the total zero modes in the background of 2$\pi$ flux to arrive at the gauge-invariant monopole operators \footnote{Strictly speaking, this statement can be made precise in the QED theory with Dirac fermions by the state-operator correspondence \cite{borokhov2003topological}. And the state-operator correspondence is not necessarily to hold in nonrelativistic $\text{QED}$ theory here. However, we can deform each fermion with quadratic band touching into two Dirac fermions. For example, if the $t_{\perp}$ is zero in the AB stacked bilayer honeycomb lattice model or the AA stacked bilayer triangular lattice model above, each QBT splits into two degenerate Dirac cones. Moreover, these two theories have the same kinematic properties, since their dual topological NLSM descriptions have the same WZW terms on the Grassmannian manifold. As a result, it is quite reasonable to conjecture that the gauge-invariant monopole operators of the nonrelativistic $\text{QED}$ here are the same as the $N_f=8$ relativistic $\text{QED}$ theory. And the gauge-invariant monopole operators of the latter theory correspond to the states which occupy half of the total zero modes in the presence of unit magnetic flux.}. As a result, since there are two zero modes for each valley and spin after introducing 2$\pi$ flux, there are in total $C^4_8=70$ gauge-invariant monopole operators. We label the first and second zero mode (per spin and valley DOF) as $(f^{\dagger})^1, (f^{\dagger})^2$ respectively and classify all the gauge-invariant monopole operators into three classes:

1.$M$ type:  $M_{ij}=\phi^1_{i}\phi^2_{j}M^+_{\textbf{bare}}, i,j=1,2...6$, where $(\phi_i)^{1,2}$ is the bilinear zero mode creation operator, similar to that in the $N_f=4$ relativistic $\text{QED}$ theory \cite{PhysRevX.10.011033}. The six components $(\phi_i)^{1,2}$ forms an $\text{SO}(6)$ vector representation of the $\text{SU}(4)$ symmetry in the valley and spin DOF. It can be written explicitly as follows:
\begin{equation}
    \begin{array}{l}\phi_{1 / 2 / 3}=f_{\alpha, s}^{\dagger}\left(\epsilon \tau^{1 / 2 / 3}\right)^{\alpha \beta} \epsilon^{s s^{\prime}} f_{\beta, s^{\prime}}^{\dagger},  \\ \phi_{4 / 5 / 6}=i f_{\alpha, s}^{\dagger}(\epsilon)^{\alpha \beta}\left(\epsilon s^{1 / 2 / 3}\right)^{s s^{\prime}} f_{\beta, s^{\prime}}^{\dagger}, \end{array}
\end{equation}
where we have neglected the zero mode type index for convenience. The $f^{\dagger}_{\alpha,s}$ operator creates a fermion in the zero mode with the valley and spin index $\alpha,s$ respectively. And $\mathcal{M}^{\dagger}_{\mathrm{bare}}$ creates a $2\pi$ flux, which is equivalent to insert a magnetic monopole in the path integral. There are $6\times6=36$ monopole operators of this kind. Each operator has two zero modes belonging to the first and second types respectively. 

2.$N$ type: $(f^{\dagger})^{k_1}_1(f^{\dagger})^{k_2}_2(f^{\dagger})^{k_3}_3(f^{\dagger})^{k_4}_4 M^+_{\textbf{bare}}$. Here either three of the four upper indexes $(k_1, k_2 ,k_3, k_4)$ are the same or all of them all the same. Thus there are $2+4\times2=10$ monopole operators. The lower index labels valley($1,2$) and spin($\uparrow,\downarrow$) as $1=(1,\uparrow), 2=(2,\uparrow),3=(1,\downarrow), 4=(2,\downarrow)$.

3.$K$ type: $(f^{\dagger})^{1}_{i}(f^{\dagger})^{2}_{i}(f^{\dagger})^{n}_{j}(f^{\dagger})^{n}_{k}M^+_{\textbf{bare}}$ where $i,j,k$ represent different spin and valley indices and $n=1,2$. There are $4\times3\times2=24$ monopole operators of this kind.

\subsection{AB stacked bilayer honeycomb lattice}
Let's discuss the Bernal stacked bilayer honeycomb lattice in this section. Under an appropriate basis, the low energy Lagrangian density is (Euclidean signature) :
\begin{equation}
    \mathcal{L}={\psi}^{\dagger}D_t\psi+{\psi}^{\dagger}[(D_x^2-D_y^2)\mu_1\otimes\tau_0\otimes\sigma_0+2D_xD_y\mu_2\otimes\tau_0\otimes\sigma_0]\psi,
\end{equation}
where $\tau, \sigma,\mu$ Pauli matrices act on the valley, spin and QBT DOF respectively. And the covariant derivative is: $D_{\mu}=\partial_{\mu}+ia_{\mu}$. The macroscopic symmetries are translation $T_{1/2}$ along basis vectors of the honeycomb lattice,  $2\pi/3$ rotation $C_3$ with the rotation axis on $A$ sublattice, and reflections along the bond of first(second) layer $R_{\theta}$($R'_{\theta}$). $\theta$ is the angle between the reflection bond and the horizontal line which can be $0$, $\frac{2\pi}{3}$ and $\frac{4\pi}{3}$.

First of all, let's consider how these macroscopic symmetries act on bare monopoles.  The reflection $R_0$ ($\theta=0$) and charge conjugation symmetry change the 2$\pi$ flux monopole to the -2$\pi$ flux monopole, as the gauge field $a$  is changed into: 
\begin{equation}
    \begin{aligned}
    &R_0: (a_x(x,y),a_y(x,y))\rightarrow (a_x(x,-y),-a_y(x,-y)),\\
    &C:(a_x(x,y),a_y(x,y))\rightarrow (-a_x(x,y),-a_y(x,y)).
    \end{aligned}
\end{equation}
 We remark that it is more convenient to
view this bare -2$\pi$ flux monopole operator as the “anti-particle”, or hermitian
conjugate, of the bare monopole operators defined above.
Therefore, we have

 \begin{equation}
    \begin{aligned}
   R(C) M_{\text{bare}}^{\dagger}=e^{i\theta_R(\theta_C)}M_{\text{bare}}R(C).
    \end{aligned}
\end{equation}

And the other symmetries map the bare monopole to itself with addition phases. These nontrivial phases (including $\theta_{R_0}$ and $\theta_C$)  are known as the Berry phases of bare monopoles on the lattice\cite{PhysRevX.10.011033},
which comes from the embedding of lattice symmetries to the topological U(1) symmetry. Thus we have that $\theta_C=0$ and the phases obtained by the action of lattice symmetries on the bare monopole operator forms a one dimensional representation. These phases must satisfy the algebraic relations between lattice symmetries:
\begin{equation}
\begin{aligned}
& C^3_3=1, T_1T_2C_3T_1=C_3,C_3T_2=T_1C_3,\\
&
  C_3R_{0}=R_{-\frac{\pi}{3}}, T_1 T_2=R'_{0}R_0, T_2=R'_{\frac{\pi}{3}}R_{\frac{\pi}{3}}. \end{aligned}
\end{equation}
  According to the first to third equations, we obtain that $\theta_{T_1}=\theta_{T_2}=\frac{2\pi}{3}\mathbb{Z}$ and $\theta_{C_3}=\frac{2\pi}{3}\mathbb{Z}$. Since the reflection is a $\mathbb{Z}_2$ symmetry, its phase is $0$ or $\pi$. Then the last three equations imply that $\theta_{T_1}=\theta_{T_2}=\theta_{C_3}=0$. Moreover, we can also obtain that reflections with respect to different bonds have the same phase. 
    
    To fix the Berry phases $\theta_R$ of all the reflection symmetries, we use the $CR$ operator trick similar to that in the relativistic $\text{QED}$ theory\cite{PhysRevX.10.011033}, which is the combination of charge conjugation and reflection symmetry. First, since the $\text{U}(1)_{\text{top}}$ phase can only be nontrivial for lattice symmetries, $\theta_R=\theta_{CR}$. Next, we add a quantum spin Hall mass $m\psi^{\dagger}\left(\mu_3\otimes\sigma_3\right)\psi$. Now fermions are fully gapped and only a pure Maxwell U(1) gauge theory is left in the IR theory. Thus the degeneracy among the seventy monopole operators is completely lifted, and there is only one gauge-invariant zero energy monopole operator left. To identify the surviving zero energy monopole, we introduce the $\text{SO}(2)$ gauge field of spin-$z$ rotation $A^{\text{SO}(2)}$, then the mutual spin-charge response theory is given by the second topological WZW term in \eqref{DSL} \footnote{The second topological WZW term can be understood as U(1) gauge field coupled with two spin-1 excitations. If we add a background SO(2) gauge field and integrate the $n$ field, the Berry phase of spin-1 excitation will give the flux of SO(2) gauge field.}:
    \begin{eqnarray}\label{spin-charge}
 \mathcal{L}=\frac{2}{2\pi}\textbf{sgn}(m) A^{\text{SO}(2)}da.
 \end{eqnarray}
 Thus if $m>0$, the monopole operator is charge-2 under $S_z$ rotation, which is just the operator $(\phi^1_{4}+i\phi^1_{5})(\phi^2_{4}+i\phi^2_{5})M^+_{\textbf{bare}}$. Moreover, Eq.\eqref{spin-charge} also implies that this QSH insulator is in a trivial SPT phase protected by $CR$ symmetry due to the absence of anomaly of its boundary theory. As a result, $CR$ acts on this monopole operator as the identity operator. Thus to see the phase $\theta_{R}$, we only need to study how $CR$ acts on fermion biquadratic terms.
 
Then let's look at how the lattice symmetry acts on the fermions in low energy:
    \begin{eqnarray}
 && T_{1/2}:\psi \to  \exp(-\frac{2\pi i}{3}\tau^3)\psi,\\
  &&R_{0}:\psi(k_x,k_y) \to  \mu^1\tau^2\psi(k_x,-k_y), \label{reflection}\\
 &&C_3:\psi(\vec{k}) \to   \exp(\frac{2\pi i}{3}\mu^3)\psi(R_{\frac{2\pi}{3}}\vec{k}),\\
 &&C:\psi(\vec{k})\to i\tau^1 \sigma^2\mu^2\psi^{*}(-\vec{k}).
 \end{eqnarray}
 From the representation of $C$ and $R_0$, we can obtain that the biquadratic fermion part  $(\phi^1_{4}+i\phi^1_{5})(\phi^2_{4}+i\phi^2_{5})$ is invariant under $CR_0$ transformation \footnote{Here we can think of $CR_0$ as obtained from Wick-rotating time-reversal symmetry $T$ in the Lorentz signature by CRT theorem and monopole operators is invariant under $T$ symmetry}, which means $\theta_{R}=0$.

Therefore,  we can conclude that the macroscopic symmetries only act on the biquadratic fermion part of monopole operators. These results are listed the Table~\ref{table1}.
	\begin{table*}[t]
\centering
\begin{tabular}{|c|c|c|c|c|c|c|c|c|}
\hline
 &$\phi_{+}$ &$\phi_{-}$ &$\phi_3 $&$\phi_{4/5/6}$&$N$&$\quad \mathcal{K}^{1/2;3/4}_A \quad$&$\quad \mathcal{K}^{1/2;3/4}_B \quad$&$\quad \mathcal{K}^{5/6;7/8}_B \quad $ \\ \hline
$\quad T_{1/2}\quad $ &$e^{\frac{2\pi i}{3}} \phi_{+}$&$e^{-\frac{2\pi i}{3}} \phi_{-}$ &$\quad \phi_3\quad $&$\phi_{4/5/6}$&$N$&$\mathbb{I}_{2\times 2}$&$e^{-\frac{2\pi i}{3}\sigma^3}$&$e^{\frac{-2\pi i}{3}\sigma^3}$\\ \hline
$R$ &$\phi_{-}$&$\phi_{+}$&$\phi_3$&$-\phi_{4/5/6}$&$\quad N \quad $&$\quad -\sigma^1 \quad$&$\sigma^1$&$\sigma^1$\\ \hline
$C_3$ &$\phi_{+}$&$\phi_{-}$&$\phi_3$&$\phi_{4/5/6}$&$N$&$\mathbb{I}_{2\times 2}$&$\mathbb{I}_{2\times 2}$&$\mathbb{I}_{2\times 2}$\\ \hline
\end{tabular}
	\caption{The transformation of the zero mode part of the monopoles on the honeycomb lattice under the macroscopic symmetries. For $M$ type, we only list transformation of  fermion bilinear part  and $\phi_{\pm}=\phi_1\pm i \phi_{2}$ . }\label{table1}
	\end{table*}
	
Firstly, for the $M$ type monopoles,  the effect acted by the macroscopic symmetries on the biquadratic fermion part can be arrived by the tensor product of the results of fermion quadratic terms $\phi_i$ in the Table~\ref{table1}.

Secondly, for the $N$ type monopoles, since the  macroscopic symmetries do not flip the first and second zero mode in each spin and valley, we can neglect the upper index of zero modes. Thus the $N$ type monopole operators  can be rewritten as:
\begin{equation}
    N=[(f^{\dagger})^T_{\uparrow, \alpha}(i\tau^2)^{\alpha\beta}f^{\dagger}_{\uparrow,\beta}] [(f^{\dagger})^T_{\downarrow,\gamma}(i\tau^2)^{\gamma,\delta}f^{\dagger}_{\downarrow,\delta} ]M_{\text{bare}}^{\dagger}.
\end{equation}
It can be directly verified that they are invariant under lattice symmetries. 

Finally, for the $K$ type monopoles,  we can divide these operators into two classes:

1) The first class includes four two dimensional invariant subspaces under lattice symmetries:
\begin{equation}
     \begin{aligned}
 \mathcal{K}^{1/2}_A:\   &(f^{\dagger})^1_{1,\uparrow}(f^{\dagger})^2_{1,\uparrow}(f^{\dagger})^{1/2}_{2,\uparrow}(f^{\dagger})^{1/2}_{2,\downarrow},\\ &(f^{\dagger})^1_{2,\downarrow}(f^{\dagger})^2_{2,\downarrow}(f^{\dagger})^{1/2}_{1,\uparrow}(f^{\dagger})^{1/2}_{1,\downarrow},
     \end{aligned}
 \end{equation}
 \begin{equation}
 \begin{aligned}
 \mathcal{K}^{3/4}_A:\   &(f^{\dagger})^1_{2,\uparrow}(f^{\dagger})^2_{2,\uparrow}(f^{\dagger})^{1/2}_{1,\uparrow}(f^{\dagger})^{1/2}_{1,\downarrow}, \\ &(f^{\dagger})^1_{1,\downarrow}(f^{\dagger})^2_{1,\downarrow}(f^{\dagger})^{1/2}_{2,\uparrow}(f^{\dagger})^{1/2}_{2,\downarrow}.
     \end{aligned}
 \end{equation}
In each subspace, translation acts as identity while $R_0$ acts as $-\sigma^1$.

2) The second class includes eight invariant subspaces which are spanned by:
\begin{equation}
     \begin{aligned}
 \mathcal{K}^{1/2}_B:\  &(f^{\dagger})^1_{1,\uparrow}(f^{\dagger})^2_{1,\uparrow}(f^{\dagger})^{1/2}_{1,\downarrow}(f^{\dagger})^{1/2}_{2,\uparrow},\\ &(f^{\dagger})^1_{2,\downarrow}(f^{\dagger})^2_{2,\downarrow}(f^{\dagger})^{1/2}_{1,\downarrow}(f^{\dagger})^{1/2}_{2,\uparrow},
     \end{aligned}
 \end{equation}
\begin{equation}
     \begin{aligned}
 \mathcal{K}^{3/4}_B: \ &(f^{\dagger})^1_{1,\downarrow}(f^{\dagger})^2_{1,\downarrow}(f^{\dagger})^{1/2}_{1,\uparrow}(f^{\dagger})^{1/2}_{2,\downarrow},\\ &(f^{\dagger})^1_{2,\uparrow}(f^{\dagger})^2_{2,\uparrow}(f^{\dagger})^{1/2}_{1,\uparrow}(f^{\dagger})^{1/2}_{2,\downarrow},
     \end{aligned}
 \end{equation}
 \begin{equation}
     \begin{aligned}
 \mathcal{K}^{5/6}_B: \ &(f^{\dagger})^1_{1,\downarrow}(f^{\dagger})^2_{1,\downarrow}(f^{\dagger})^{1/2}_{1,\uparrow}(f^{\dagger})^{1/2}_{2,\uparrow},\\ &(f^{\dagger})^1_{2,\uparrow}(f^{\dagger})^2_{2,\uparrow}(f^{\dagger})^{1/2}_{1,\downarrow}(f^{\dagger})^{1/2}_{2,\downarrow},
     \end{aligned}
 \end{equation}
  \begin{equation}
     \begin{aligned}
 \mathcal{K}^{7/8}_B: \ &(f^{\dagger})^1_{1,\uparrow}(f^{\dagger})^2_{1,\uparrow}(f^{\dagger})^{1/2}_{1,\downarrow}(f^{\dagger})^{1/2}_{2,\downarrow},\\&(f^{\dagger})^1_{2,\downarrow}(f^{\dagger})^2_{2,\downarrow}(f^{\dagger})^{1/2}_{1,\uparrow}(f^{\dagger})^{1/2}_{2,\uparrow} .
  \end{aligned}
 \end{equation}
  In each two dimensional subspace, translation acts as $\exp(-\frac{2\pi i}{3}\sigma^3)$ while $R_0$ acts as $\sigma^1$. 

\subsection{AA stacked bilayer checkerboard lattice}
On the bilayer checkerboard lattice, the macroscopic symmetries are transition, $x,y$-axis reflection, $C_{4z}$ rotation and particle-hole symmetry. Their action on the low energy fermions are:
  \begin{eqnarray}
 && T_{1/2}:\psi \to -\psi,\nonumber\\
 &&R_{x,y}:\psi(\vec{k}) \to\tau^1\mu^1\psi(R_{x,y}\vec{k}), \nonumber\\
 &&C_{4z}:\psi(\vec{k}) \to   \mu^3\psi(R_{\frac{\pi}{2}}\vec{k}),\nonumber\\
 &&C:\psi(\vec{k}) \to   i\mu^2 \sigma^2\psi^{*}(-\vec{k}),
 \end{eqnarray}
 where $\mu,\tau,\sigma$ acts on the QBT, layer (valley) and spin DOF respectively. The reflection symmetry defined here exchanges the two layers besides the standard reflection symmetry, as the standard one acts trivially on all the monopole operators, which is shown below.
 
First of all, similar to the previous section, we have $R_{x/y}M_{\text{bare}}^{\dagger}=e^{i\theta_{R_{x/y}}}M_{\text{bare}}R_{x/y}$ and $CM_{\text{bare}}^{\dagger}=M_{\text{bare}}C$  as the gauge field $a$ is transformed as: 
\begin{equation}
    \begin{aligned}
    &R_{x}: (a_x(x,y),a_y(x,y))\rightarrow (a_x(x,-y),-a_y(x,-y)),\\
    &R_{y}: (a_x(x,y),a_y(x,y))\rightarrow (-a_x(-x,y),a_y(-x,y)),\\
    &C: (a_x(x,y),a_y(x,y))\rightarrow (-a_x(x,y),-a_y(x,y)).
    \end{aligned}
\end{equation}
Next we determine the Berry phases of these macroscopic symmetries through the algebraic relations of the lattice symmetries. To begin with,  we have $\theta_{C_{4z}}=\frac{\pi}{2}\mathbb{Z},\theta_{R_{x,y}}=\pi \mathbb{Z}$ from the direct relations $C_{4z}^4=1,R_{x,y}^2=1$. Secondly, the algebraic relations among the different symmetries are:
 \begin{equation}
 \begin{aligned}
     &C_{4z}R_{x,y} C_{4z}=R_{x,y},\quad R_x R_y=C_{4z}^2,\\
     & C_{4z} T_1=T_2C_{4z},\quad T_1T_2^{-1}C_{4z}T_1T_2=C_{4z},\\
     &R_y=T_1 (C_{4z})^2 R_x .
     \end{aligned}
 \end{equation}
 From the first equation, we immediately arrive at $\theta_{C_{4z}}=\pi \mathbb{Z}$. The second equation further gives $\theta_{R_x}=\theta_{R_y}$. The final three equations constraint the $\text{U}(1)_{\text{top}}$ phases of translation symmetries as: $\theta_{T_1}=\theta_{T_2}=0$. To determine the phases of reflection and rotation symmetries, we consider another reflection symmetry:
 $R_3=C_{4z}R_y$. 
  Similar to the argument of the bilayer honeycomb lattice, since $CR_{x/y/3}$ maps the monopole operator $(\phi^1_4 \pm i \phi^1_5)(\phi^2_4 \pm i \phi^2_5)$ to itself, we can arrive at the conclusion that $\theta_{R_x}=\theta_{R_y}=\theta_{R_3}=\theta_{C_{4z}}=0$.

Since all $\text{U}(1)_{\text{top}}$ phases of macroscopic symmetries are trivial,  the action of macroscopic symmetries on the whole monopole operators is determined by that of fermion zero mode part. We list the result in
Table~\ref{table2}.
	\begin{table*}[t]
\centering
\begin{tabular}{|c|c|c|c|c|c|c|c|c|}
\hline
 &$\phi_{1}$ &$\phi_{2}$ &$\phi_3 $&$\quad\phi_{4}\quad $&$\quad\phi_{5}\quad $&$\quad\phi_{6}\quad $&$N$&$\mathcal{K}$ \\ \hline
$\quad T_{1/2}\quad $ &$ \quad \phi_{1}\quad $&$\quad \phi_{2}\quad$  &$\quad \phi_3\quad $&$\phi_{4}$&$\phi_{5}$&$\phi_{6}$&$N$&$\mathbb{I}_{2\times 2}$\\ \hline
$R_{x/y}$ &$-\phi_{1}$&$\phi_{2}$&$-\phi_3$&$\phi_{4}$&$\phi_{5}$&$\phi_{6}$&$\quad N \quad $&$\quad \sigma^1 \quad$\\ \hline
$C_{4z}$ &$\phi_{1}$&$\phi_{2}$&$\phi_3$&$\phi_{4}$&$\phi_{5}$&$\phi_{6}$&$N$&$\mathbb{I}_{2\times 2}$\\ \hline
\end{tabular}
	\caption{The transformation of monopoles on bilayer checkerboard lattice. For $M$ type, we only list transformation of  fermion bilinear terms.}\label{table2}
	\end{table*}
	
For the $M$ type monopole operators,  the action of  macroscopic symmetries on the fermion biquadratic terms are given by the tensor product of that of  fermion bilinear terms $\phi_i$:

(1)$T_{1,2}$:
The translation symmetries act on the fermion bilinear trivially.

(2)$R_{x,y}$:\quad $\vec{\phi}\rightarrow(-\phi_1,\phi_2,\phi_3,-\phi_4,-\phi_5,-\phi_6)$.

 (3)$C_{4z}$:
 This symmetry acts trivially on the fermion bilinear.
 
Secondly, all the symmetries act trivially on the fermion part of $N$ type monopoles.

Finally, for the last $K$ type monopoles, there are twelve two-dimensional invariant subspaces:
 \begin{equation}
     \begin{aligned}
 K^{1/2}: \quad &(f^{\dagger})^1_{1,\uparrow}(f^{\dagger})^2_{1,\uparrow}(f^{\dagger})^{1/2}_{2,\uparrow}(f^{\dagger})^{1/2}_{2,\downarrow},\\& (f^{\dagger})^1_{2,\uparrow}(f^{\dagger})^2_{2,\uparrow}(f^{\dagger})^{1/2}_{1,\uparrow}(f^{\dagger})^{1/2}_{1,\downarrow}.
     \end{aligned}
 \end{equation}
 
  \begin{equation}
     \begin{aligned}
 K^{3/4}: \quad &(f^{\dagger})^1_{1,\downarrow}(f^{\dagger})^2_{1,\downarrow}(f^{\dagger})^{1/2}_{2,\uparrow}(f^{\dagger})^{1/2}_{2,\downarrow},\\& (f^{\dagger})^1_{2,\downarrow}(f^{\dagger})^2_{2,\downarrow}(f^{\dagger})^{1/2}_{1,\uparrow}(f^{\dagger})^{1/2}_{1,\downarrow},
     \end{aligned}
 \end{equation}
 
 \begin{equation}
     \begin{aligned}
  K^{5/6}: \quad & (f^{\dagger})^1_{1,\uparrow}(f^{\dagger})^2_{1,\uparrow}(f^{\dagger})^{1/2}_{1,\downarrow}(f^{\dagger})^{1/2}_{2,\uparrow}, \\ &(f^{\dagger})^1_{2,\uparrow}(f^{\dagger})^2_{2,\uparrow}(f^{\dagger})^{1/2}_{2,\downarrow}(f^{\dagger})^{1/2}_{1,\uparrow},
     \end{aligned}
 \end{equation}
 
  \begin{equation}
     \begin{aligned}
  K^{7/8}: \quad &(f^{\dagger})^1_{1,\downarrow}(f^{\dagger})^2_{1,\downarrow}(f^{\dagger})^{1/2}_{1,\uparrow}(f^{\dagger})^{1/2}_{2,\downarrow},\\ &(f^{\dagger})^1_{2,\downarrow}(f^{\dagger})^2_{2,\downarrow}(f^{\dagger})^{1/2}_{2,\uparrow}(f^{\dagger})^{1/2}_{1,\downarrow},
     \end{aligned}
 \end{equation}
 \begin{equation}
     \begin{aligned}
     K^{9/10}:\quad &(f^{\dagger})^1_{1,\uparrow}(f^{\dagger})^2_{1,\uparrow}(f^{\dagger})^{1/2}_{1,\downarrow}(f^{\dagger})^{1/2}_{2,\downarrow},\\&(f^{\dagger})^1_{2,\uparrow}(f^{\dagger})^2_{2,\uparrow}(f^{\dagger})^{1/2}_{2,\downarrow}(f^{\dagger})^{1/2}_{1,\downarrow},
     \end{aligned}
 \end{equation}
 
 \begin{equation}
     \begin{aligned}
    K^{11/12}:\quad &(f^{\dagger})^1_{1,\downarrow}(f^{\dagger})^2_{1,\downarrow}(f^{\dagger})^{1/2}_{1,\uparrow}(f^{\dagger})^{1/2}_{2,\uparrow},\\& (f^{\dagger})^1_{2,\downarrow}(f^{\dagger})^2_{2,\downarrow}(f^{\dagger})^{1/2}_{2,\uparrow}(f^{\dagger})^{1/2}_{1,\uparrow},
     \end{aligned}
 \end{equation}
 where the reflection symmetry acts as  $\sigma^1$  and all other symmetries act trivially on each subspace. 
 
\subsection{AA stacked bilayer triangular lattice}
For the bilayer triangular lattice model Eq.(\ref{bitri}), the macroscopic symmetry includes the translation $T_1,2T_2$ ($T_1,2T_2$ are the translation in the directions of unit vectors of a triangular lattice), two-fold rotation $C_2$ and the reflection $R_x$. Under an appropriate basis, the low energy parton Hamiltonian takes the form of two QBT valleys Eq.(\ref{qbt}). The projective symmetry representation in the QBT basis is:
\begin{equation}
\begin{aligned}
&T_1:\psi(k_x,k_y)\rightarrow -i\tau_3\psi(k_x,k_y),\\
&2T_2:
\psi(k_x,k_y)\rightarrow -\mu_0\otimes\tau_0\psi(k_x,k_y),\\
& C_2^A:
   \psi(k_x,k_y)\rightarrow i\mu_2\otimes\tau_3 s_2\psi^{*}(k_x,k_y),\\
   &R_x:
  \psi(k_x,k_y)\rightarrow \mu_3 s_2\psi^{*}(-k_x,k_y),
 \end{aligned}
 \end{equation}
 where $C_2^A$ is the two fold rotation centered at $A$ sublattice. Similar to the previous section, the $\mu$ matrices act on the QBT DOF, $\tau$ act on the valley freedom and $\sigma$ act on the spin freedom. The details of the derivation are listed in the Appendix \ref{psg}. 

Now let's consider how the macroscopic symmetries act on the monopole operators. Since the translation $2T_2$ acts trivially on all kinds of monopole operators, we neglect this symmetry. Moreover, we should note that the $\text{U}(1)$ Berry phase is generally a theoretical challenge on the nonbipartite lattice. Thus we temporarily focus on how the symmetries act on the fermionic zero modes and leave the investigation of the corresponding Berry phases for future work.

For the $M$ type monopole operators,  the action of  macroscopic symmetries on the zero mode part is given by the tensor product of the fermion bilinear terms $\phi_i$ listed in Table\eqref{table3}.
\begin{table*}[t]
\centering
\begin{tabular}{|c|c|c|c|c|c|c|c|c|c|c|c|c|}
\hline
 &$\phi_{1}$ &$\phi_{2}$ &$\phi_3 $&$\quad\phi_{4}\quad $&$\quad\phi_{5}\quad $&$\quad\phi_{6}\quad $&$N$&$\mathcal{K}^{1}_A$&$\quad \mathcal{K}^{2}_A \quad$&$\quad \mathcal{K}^{1/2/3/4}_B \quad $&$\quad \mathcal{K}^{1/2/3/4}_C \quad $ \\ \hline
$\quad T_{1}\quad $ &$ \quad -\phi_{1}\quad $&$\quad -\phi_{2}\quad$  &$\quad \phi_3\quad $&$\phi_{4}$&$\phi_{5}$&$\phi_{6}$&$N$&$\mathbb{I}_{2\times 2}$&$\mathbb{I}_{2\times 2}$&$-\mathbb{I}_{2\times 2}$&$-\mathbb{I}_{2\times 2}$\\ \hline
$R^A_{x}$ &$-\phi_{1}$&$-\phi_{2}$&$-\phi_3$&$\phi_{4}$&$-\phi_{5}$&$\phi_{6}$&$\quad N \quad $&$\quad \mathbb{I}_{2\times 2} \quad$&$-\mathbb{I}_{2\times 2}$&$\sigma^1$&$-\sigma^1$\\ \hline
$C^A_{2}$ &$\phi_{1}$&$\phi_{2}$&$-\phi_3$&$\phi_{4}$&$-\phi_{5}$&$\phi_{6}$&$N$&$\mathbb{I}_{2\times 2}$&$\mathbb{I}_{2\times 2}$&$-\sigma^1$&$\sigma^1$\\ \hline

\end{tabular}
	\caption{The transformation of monopoles on the bilayer triangular lattice. For $M$ type, we only list transformation of  fermion bilinear terms. }\label{table3}
	\end{table*}
	



Next, for the $N$ type monopole operators,  all the symmetries act on the zero modes as identity. 

Finally, let's consider the $K$ type of monopole operators. They can be classified into three classes under the lattice symmetries:

1) The first class consists of eight monopole operators, which can be further divided into two four-dimensional invariant  subspaces :
\begin{equation}
\begin{aligned}
\mathcal{K}^{1}_A  :  &&(f^{\dagger}_{1})^T[\frac{\tau_0\pm\tau_3}{2}\otimes \sigma_0]f^{\dagger}_2(f^{\dagger}_{1/2})^T(\frac{\tau_0\mp\tau_3}{2}\otimes\frac{i\sigma_2+\sigma_1}{2})f_{1/2}^{\dagger},\\
\mathcal{K}^{2}_A  :     &&(f^{\dagger}_{1})^T[\frac{\tau_0\pm\tau_3}{2}\otimes \sigma_3]f^{\dagger}_2(f^{\dagger}_{1/2})^T(\frac{\tau_0\mp\tau_3}{2}\otimes\frac{i\sigma_2+\sigma_1}{2})f_{1/2}^{\dagger},
    \end{aligned}
\end{equation}

In the $\mathcal{K}^{1}_A$ subspace, all the symmetries act as identity operators. In the $\mathcal{K}^{1}_B$ subspace, $R_x$ assigns an extra minus sign to each element and other symmetries still act as identity.
 
 2) There are eight monopole operators in the second class, which take the form as:

 \begin{equation}
 \begin{aligned}
     &(f_1^{\dagger})^TM_{ab}f_2^{\dagger}(f^{\dagger}_{1/2})^TN_{b}f^{\dagger}_{1/2},
  \end{aligned}
 \end{equation}     
where     
  \begin{equation}
 \begin{aligned}    
     &M_{ab}=\frac{\tau_0+(-1)^a\tau_3}{2}\otimes\frac{\sigma_0+(-1)^b \sigma_3}{2},\\
     &N_{c}=\frac{i
     \tau_2+\tau_1}{2}\otimes\frac{\sigma_0+(-1)^{c+1}\sigma_3}{2},
     \end{aligned}
 \end{equation}
and $a,b,c$\ $\in$$\{0,1\}$. 

Moreover, they can be classified into four two-dimensional invariant subspaces:
 \begin{equation}\label{basis-K}
\begin{aligned}
\mathcal{K}^{1/2}_B  :  &&(f_1^{\dagger})^TM_{00}f_2^{\dagger}(f^{\dagger}_{1/2})^TN_{0}f^{\dagger}_{1/2},\\ &&(f_1^{\dagger})^TM_{01}f_2^{\dagger}(f^{\dagger}_{1/2})^TN_{1}f^{\dagger}_{1/2},\\
\mathcal{K}^{3/4}_B  :     &&(f_1^{\dagger})^TM_{10}f_2^{\dagger}(f^{\dagger}_{1/2})^TN_{1}f^{\dagger}_{1/2},\\&& (f_1^{\dagger})^TM_{11}f_2^{\dagger}(f^{\dagger}_{1/2})^TN_{1}f^{\dagger}_{1/2}.
    \end{aligned}
\end{equation}

In each subspace, $T_1$  assigns a minus sign to each element while $C^A_2$ and $R_x$ maps $M_{ab}N_{b}$ to $\mp M_{a,1-b}N_{1-b}$ respectively which act as the $\mp \sigma^x$ in the basis \eqref{basis-K}.

(3) The final eight monopole operators take the form as:

\begin{equation}
 \begin{aligned}
     &(f_1^{\dagger})^TM_{ab}f_2^{\dagger}(f^{\dagger}_{1/2})^TN'_{ab}f^{\dagger}_{1/2},
  \end{aligned}
 \end{equation}   
where  \begin{equation}
 \begin{aligned}    
     &M_{ab}=\frac{\tau_0+(-1)^a\tau_3}{2}\otimes\frac{\sigma_0+(-1)^b \sigma_3}{2},\\
     &N'_{ab}=\frac{i
     \tau_2+\tau_1}{2}\otimes\frac{\sigma_1+(-1)^{a+b+1}i\sigma_2}{2},
     \end{aligned}
 \end{equation}
and $a,b$\ $\in$$\{0,1\}$.

They can also be further divided into four two-dimensional invariant subspaces:
 \begin{equation}\label{basis1}
\begin{aligned}
\mathcal{K}^{1/2}_C  :  &&(f_1^{\dagger})^TM_{00}f_2^{\dagger}(f^{\dagger}_{1,2})^TN'_{00}f_{1,2}^{\dagger},\\ &&(f_1^{\dagger})^TM_{01}f_2^{\dagger}(f^{\dagger}_{1/2})^TN'_{01}f^{\dagger}_{1/2},\\
\mathcal{K}^{3/4}_C  :     &&(f_1^{\dagger})^TM_{10}f_2^{\dagger}(f^{\dagger}_{1/2})^TN'_{10}f^{\dagger}_{1/2},\\&& (f_1^{\dagger})^TM_{11}f_2^{\dagger}(f^{\dagger}_{1/2})^TN'_{11}f^{\dagger}_{1/2}.
    \end{aligned}
\end{equation}

In each subspace, $T_1$  assigns a minus sign to each element while $C^A_2$ and $R_x$ maps $M_{ab}N_{ab}$ to $\pm M_{a,1-b}N'_{a,1-b}$ respectively which act as the $\pm \sigma^x$ in the basis \eqref{basis1}.

\section{Conclusions and discussions}
In this work, we have proposed a general dimensional reduction scheme to derive the duality between NLSM with the WZW term and QED theory with a general target space. The fermions of the QED theory are not limited to the Dirac fermions and can come from any lattice regularization, such as parton construction of lattice spin models, which means they can have any dispersion or even a Fermi surface in the low energy. Therefore our scheme is beyond the famous AW mechanism, whose applicability is limited to fermions with linear dispersion. As concrete applications of our general method, we have also constructed lattice models on the bilayer honeycomb, checkerboard, and triangular lattice, whose low energy descriptions are all $N_f=4$ $\text{QED}_3$ theory with quadratic dispersive fermions. We have confirmed that this QED theory is dual to the topological NLSM with level-2 WZW terms on the Grassmannian manifold. Further, we have conducted some nontrivial theoretical checks to verify the duality between the QED theory and topological NLSM. The 't Hooft anomalies of both theories are matched with each other. Moreover, both theories give the same physical results under the insertion of defects, such as interface, gauge monopole, vortex of valley rotation(valley vortex), etc. Finally, we have also discussed the behaviour of monopoles, which are gauge invariant physical observable, under the action of macroscopic symmetries. These results can guide the future investigations of ordered phases around this critical theory. 

There are a lot of future directions, and we discuss some of them here. First, the lattice models we have constructed are all parton mean field models with a fluctuating U(1) gauge field. The construction of lattice spin models, whose critical point or phases are described by our $N_f=4$ $\text{QED}_3$ theory with quadratic dispersive  fermions or dual topological NLSM with level-2 WZW terms on the Grassmannian manifold, is left for future work. It is of great interest to find out whether twisted bilayer \mr systems can host a natural platform for our QED theory\cite{doi:10.1073/pnas.2014691117,https://doi.org/10.48550/arxiv.2203.05480,https://doi.org/10.48550/arxiv.2207.11468,PhysRevB.101.035122,PhysRevB.99.205150}. Moreover, magnetic monopoles can drive confinement. Thus the result of  macroscopic symmetries on  monopole operators would help to arrive at a unifying description of competing orders in these quantum magnets.  Secondly, 
it's quite energizing how to apply our dimensional reduction scheme to systems with a Fermi surface  \cite{PhysRevX.11.021005,PhysRevB.84.165126}. As the exotic low energy symmetry is loop group LU(1), the 't Hooft anomaly matching, which  enables the system to satisfy the filling constraints, is quite an interesting question.  Finally, our formalism cannot determine the dynamics of the topological NLSM, which in principle can have tree-level dynamical exponent $z=1$ or $z=2$. The investigation of the dynamical properties of this dual model is also an important future direction.  
\section{Acknowledgements}
We sincerely thank Xin Chen, Hong Yao, Yunqin Zheng, and especially Liujun Zou for helpful discussions. L. L. is supported by Global Science Graduate Course (GSGC) program at the
University of Tokyo.

\appendix
\section{Proof of the Dimensinoal Reduction}\label{General Theory}
\subsection{The QED theory with fermions of general dispersion}

To begin with, we consider general  $n-1$ dimensional lattice QED theory as eq.\eqref{QED}:
\begin{equation}
\begin{aligned}
S_0&=S_e+S_B,\\
S_e&=\int dt  \sum_{\vec{j}}\sum_{l=1}^{D} \psi^{\dagger}_{l}(\vec{j})(\partial_{t}-iB_0) \psi_{l}(\vec{j})\\
&+\sum_{\vec{i},\vec{j}}\sum_{l,l^{\prime}}\psi^{\dagger}_l(\vec{i})e^{-i\int_{\vec{i}}^{\vec{j}} \vec{B}\cdot d\mathbf{l} }h_{l,l^{\prime}}(\vec{i},\vec{j})\psi_{l^{\prime}}(\vec{j}),
\end{aligned}
\end{equation}
where $f$ label the onsite freedom of the lattice fermion, such as sublattice or spin index, etc, and $\vec{j}$ is the lattice site. And $B_{\mu}$ is the dynamical $\text{U}(1)$ gauge field.

For the convention convenience, we take $x$ as the coarse grain coordinate of the lattice site. $h_{l,l^{\prime}}(x)$ is the $n$-1 dimensional real space lattice Bloch Hamiltonian, and can have nodal Fermi points or Fermi surface in the low energy.

For a general spacetime dimension $n$, we couple the fermion field with the dynamical bosonic field $M(x,t)$, physically the order parameter field, with a general target space as the manifold $\mathbf{M}$:
\begin{equation}
   S_c= \int dt d^{n-1}x \psi^{\dagger}(x,t) \gamma_0 M(x,t)\psi(x,t),
\end{equation}
where $M(x,t)$ is a matrix representation of $\mathbf{M}$ and satisfies $M^{\dagger}M=1$ \cite{PhysRevB.102.245113}. 

Theoretically, topological phases (or WZW terms) composed of the bosonic field $M(t,x)$ and the dynamical gauge field $B_{\mu}$ may arise if we integrate out the fermions. And there will be a series of WZW terms labeled by an integer $k$: $k=[\frac{n}{2}],[\frac{n}{2}]+1,...n$, as $M(t,x)$ can couple with the gauge field \cite{PhysRevB.102.245113}. To determine whether we can obtain the $k$-th WZW term from integrating the lattice fermions of the Hamiltonian $H[\vec{k} ; M(t, \vec{x})]$, where $ \vec{x}=(x_1,x_2,...x_{n-1})$, we take the following steps for a given $k$ ($[\frac{n}{2}]\le k \le n$). And for simplicity, we first discuss $k=n$, where there is no gauge field, and then turn to the general $k$.

First, we create a $2n$ dimensional Bloch hamiltonian $H_{2n}$.  We replace $M(x,t)$ with $M(p_1,p_2,...p_{n+1})$ , which still belongs to the target manifold $\mathbf{M}$, and  $(p_1,p_2,...p_{n+1})$ lives on $n+1$ torus $T^{n+1}$.  Besides, we require the Pontryagin index $P$ of the map $T^{n+1}\rightarrow M(p_1,p_2,...p_{n+1})$ is nonzero:
\begin{equation}
   P_{n+1}=\theta_{n+1}\int_{T^{n+1}} d^{n+1}p \mathbf{Tr}[M^{-1}(dM)^{n+1}],
\end{equation}
where $2\pi\theta_{n+1}=\frac{2^{\frac{n-1}{2}}}{4^{n}\pi^{\frac{n-1}{2}}\Gamma(\frac{n+1}{2})(n+1)}$. This kind of map exists if $\pi_{n+1}(\mathbf{M})=\mathbb{Z}$.  As we can compose another map $g$ from $T^{n+1}$ to $S^{n+1}$, which covers $S^{n+1}$ only once \cite{RevModPhys.83.1057}. An example is the map $g$ from $T^{n+1}$ to $S^{n+1}$ constructed in \cite{PhysRevB.82.245117}. Thus we obtain:
\begin{equation}
\begin{aligned}
   P_{n+1}&=\theta_{n+1}\int_{T^{n+1}}  \mathbf{Tr}[M^{-1}(dM)^{n+1}]\\
   &=\theta_{n+1}\int_{S^{n+1}}  \mathbf{Tr}[M^{-1}(dM)^{n+1}].
\end{aligned}
\end{equation}

We can consider the following simplest example: $\mathbb{M}=S^{n+1}$. It can be represented as: $M=n_{i}\Gamma^{i}$ and $n^2=1$. We can prove the above Pontryagin index is just the number of covering over the area of $S^{n+1}$: $\textbf{Area}(S^{n+1})=\frac{2\pi^{\frac{n+2}{2}}}{\Gamma(\frac{n+2}{2})}$. Actually, we have the following simple relation between $\Gamma$ functions: $\Gamma\left(\frac{n+1}{2}\right) \Gamma\left(\frac{n+2}{2}\right)=\frac{\sqrt{\pi}}{2^{n}} n !$. So the Pontryagin index becomes:
\begin{equation}
\begin{aligned}
P_{n+1}&=\theta_{n+1} 2^{\frac{n+1}{2}}(n+1)!\\&\int_{T^{n+1}} \epsilon^{a_{1}\cdots a_{n+2}}n^{a_{1}}\partial_{p_{1}}n_{a_{2}}\cdots \partial_{p_{n+1}}n_{a_{n+2}}\\
&=\frac{1}{2\pi}\frac{2^{n}n!}{4^{n}\pi^{\frac{n-1}{2}}}\frac{\Gamma\left(\frac{n+2}{2}\right)2^{n}}{n !\sqrt{\pi}}\\
&\int_{T^{n+1}} \epsilon^{a_{1}\cdots a_{n+2}}n^{a_{1}}\partial_{p_{1}}n_{a_{2}}\cdots \partial_{p_{n+1}}n_{a_{n+2}}\\
&=\frac{\Gamma(\frac{n+2}{2})}{2\pi^{\frac{n+2}{2}}}\int_{T^{n+1}} \epsilon^{a_{1}\cdots a_{n+2}}n^{a_{1}}\partial_{p_{1}}n_{a_{2}}\cdots \partial_{p_{n+1}}n_{a_{n+2}}\\
&=\frac{1}{\textbf{Area}(S^{n+1})}\int_{T^{n+1}} \epsilon^{a_{1}\cdots a_{n+2}}n^{a_{1}}\cdots \partial_{p_{n+1}}n_{a_{n+2}},
\end{aligned}
\end{equation}
where the integrand is the number of covering over $\textbf{Area}(S^{n+1})$.

Thus the $2n$ dimensional Bloch Hamiltonian is $H_{2n}(k_1,..k_{n-1},p_1,...p_{n+1})=h_0({k_1,...k_{n-1}})+\lambda M(p_1,p_2,...p_{n+1})$ and we further demand $H_{2n}$ is fully gapped in the  Brillouin zone at least in a finite range of $\lambda$. We can calculate the $n$th Chern number $C_n$ of this Hamiltonian and then we can go to the second step and prove a level $\frac{C_n}{P_{n+1}}$ WZW term will be generated upon integrating out the fermions. We first couple $H_{2n}(k_1,..k_{n-1},p_1,...p_{n+1})$ with a U(1) gauge field and obtain the CS theory after integration on fermions:
\begin{equation}
  S_{\text{CS}}=\int dt d^{2k}x\frac{i C_{k}}{(k+1)!(2\pi)^k} \epsilon^{\mu_{1}\cdots\mu_{2k+1}}A_{\mu_1}\partial_{\mu_2}A_{\mu_3}\cdots .
\end{equation}
where $A$ is the total U(1) gauge field.

If we choose the total gauge field as:
\begin{equation}
\begin{aligned}
   &A_{\mu}\left(t, x_{1}, \ldots, x_{2 n}\right)=B_{\mu}\left(t, x_{1}, \ldots, x_{n-1}\right), \quad \\
    &\text { for } \mu=0, \ldots, n-1, \\ 
    &A_{i+n-1}\left(t, x_{1}, \ldots, x_{2 n}\right)=\theta_{i}\left(t, x_{1}, \ldots, x_{n-1}\right), \quad \\
    &\text { for } i=1, \ldots, n+1,
\end{aligned}
\end{equation}  
where $\theta$ is a background U(1) gauge field. Both of them only depend on the coordinate $t,x_1,\cdots,x_{n-1}$.

We introduce one extra dimension $u$ and the Chern-Simons term becomes: 
\begin{equation}
\begin{aligned}
   S_{\mathsf{CS}}=&\frac{i C_{n}}{(n+1)!(2\pi)^n}\prod_{i=n}^{2n}L_{i}\\
  &\int dt d^{k-1}x\int du \epsilon^{a_1\cdots a_{n+1}}\epsilon^{b_1\cdots b_{n+1}}\partial_{a_1}\theta_{b_1}\cdots \partial_{a_{n+1}}\theta_{b_{n+1}} .
  \end{aligned}
\end{equation}

Thus, the action can be rewritten as:
\begin{equation}
\begin{aligned}
  S_{\mathsf{CS}}&=\frac{i C_{n}\theta_{n+1}}{P_{n+1}(n+1)!(2\pi)^n}\prod_{i=n}^{2n}L_{i}\int dt d^{n-1}x\int du \epsilon^{a_1\cdots a_{n+1}}\\
  &\epsilon^{b_1\cdots b_{n+1}}\partial_{a_1}\theta_{b_1}\cdots \partial_{a_{n+1}}\theta_{b_{n+1}}\int_{T^{n+1}}  \mathbf{Tr}[M^{-1}(dM)^{n+1}]\\
&=\frac{i C_{n}\theta_{n+1}}{P(2\pi)^n (n+1)!}\prod_{i=n}^{2n}L_{i}\int dt d^{n-1}x\int du \epsilon^{a_1\cdots a_{n+1}}\\
&\partial_{a_1}\theta_{b_1}\cdots \partial_{a_{n+1}}\theta_{b_{n+1}}\epsilon^{b_1\cdots b_{n+1}}\epsilon^{j_1\cdots j_{n+1}}\\
&\int_{T^{n+1}}   \mathbf{Tr}[M^{-1}\partial_{\theta_{j_1}}M\cdots \partial_{\theta_{j_{n+1}}}M],
\end{aligned}
\end{equation}
where the $M$ in the integration is $M(\vec{p}+\vec{\theta}(t,x_1,x_2...x_{n-1}))$, as the Pontryagin index does not depends on the $\vec{\theta}(t,x_1,x_2...x_{n-1})$. What's more, the mass term $\gamma_0 M(\vec{p})$ in $H_{2n}(\vec{k},\vec{p})$ can also depends on the momentum $\vec{k}$, as long as the Pontryagin index is the same for any $\vec{k}$. Then the Pontryagin index $P$ we put the denominator of the Chern-Simons action can take any $M(\vec{k}_0,\vec{p})$, which is only the function of momentum $\vec{p}$, for a fixed momentum $\vec{k}_0$.

Further, the tensor $\epsilon^{a_1\cdots a_{n+1}}\partial_{a_1}\theta_{b_1}\cdots \partial_{a_{n+1}}\theta_{b_{n+1}}$ is an anti-symmetric tensor:
\begin{equation}
\begin{aligned}
   & \epsilon^{a_1\cdots a_{n+1}}\partial_{a_1}\theta_{b_1}\cdots \partial_{a_{n+1}}\theta_{b_{n+1}}\\
    &=\frac{1}{(n+1)!}\epsilon^{b_1\cdots b_{n+1}}[\epsilon^{a_1\cdots a_{n+1}}\partial_{a_1}\theta_{c_1}\cdots \partial_{a_{n+1}}\theta_{c_{n+1}}\epsilon^{c_1\cdots c_{n+1}}].
    \end{aligned}
\end{equation}
So we have:
\begin{eqnarray}
    &&\epsilon^{a_1\cdots a_{n+1}}\partial_{a_1}\theta_{b_1}\cdots \partial_{a_{n+1}}\theta_{b_{n+1}}\epsilon^{b_1\cdots b_{n+1}}\epsilon^{j_1\cdots j_{n+1}}\nonumber\\
    &&=\epsilon^{b_1\cdots b_{n+1}}\epsilon_{b_1\cdots b_{n+1}}\frac{\epsilon^{j_1\cdots j_{n+1}}}{(n+1)!}\epsilon^{a_1\cdots a_{n+1}}\epsilon^{c_1\cdots c_{n+1}}\nonumber\\
    &&\partial_{a_1}\theta_{c_1}\cdots \partial_{a_{n+1}}\theta_{c_{n+1}}\nonumber.\\
    &&=(n+1)!\epsilon^{a_1\cdots a_{n+1}}\partial_{a_1}\theta_{j_1}\cdots \partial_{a_{n+1}}\theta_{j_{n+1}}.
\end{eqnarray}
As a result, the Chern-Simons action becomes:
\begin{equation}
\begin{aligned}
S_{\mathsf{CS}}& =\frac{i C_{n}\theta_{n+1}}{P_{n+1}(2\pi)^n(n+1)!}\prod_{i=n}^{2n}L_{i}\int dt d^{n-1}x\int du \\
&\epsilon^{a_1\cdots a_{n+1}}\partial_{a_1}\theta_{j_1}\cdots \partial_{a_{n+1}}\theta_{j_{n+1}}(n+1)!\\
&\int \mathbf{Tr}[M^{-1}\partial_{\theta_{j_1}}M\cdots \partial_{\theta_{j_{n+1}}}M]\\
&=\int d^{n+1}p\frac{i C_{n}\theta_{n+1}}{(2\pi)^n P_{n+1}}\prod_{i=n}^{2n}L_{i}\int dt d^{n-1}x\int du\\
&=\sum_{\vec{p}}\frac{2\pi i C_{n}\theta_{n+1}}{ P_{n+1}}\int dt d^{n-1}x\int du
\mathbf{Tr}[M^{-1}(dM)^{n+1}].
\end{aligned}
\end{equation}

Since the total gauge field doesn't depend on the coordinate $(x_{n},\cdots,x_{2n+1})$,  the $2n$ dimensional system (coupled with gauge field) still preserves translation symmetry on these directions and the  momenta $p$ is still good quantum number. Therefore, the $2n$ dimensional system decouples into ($n$-1) dimensional subsystems labeled by different $\vec{p}$ and the Hamiltonian of each subsystem is  $H(\vec{k}; \lambda M(\vec{p} + \theta(t, \vec{x})))$. Moreover, for different
$\vec{p}$, the Hamiltonians $H(\vec{k}; \lambda M(\vec{p} + \theta(t, \vec{x})))$ can be adiabatically deformed into each other without closing the energy
gap, thus topological actions of subsystems $S_{(n-1)D}(\vec{p})$ corresponding different $\vec{p}$  should have the same level. Thus, we can rewrite the CS action as
\begin{equation}
\begin{aligned}
S_{\mathsf{CS}}=\sum_{\vec{p}} S_{(n-1)D}(\vec{p}=0)&=\sum_{\vec{p}} S^{k=n}_{\text{WZW}}(M(t,\vec{x})),\\&  \text{for}\quad M(t,x)= M(\vec{\theta}(t,\vec{x})).
\end{aligned}
\end{equation}

Then the $n+1$ dimensional action is given by :
\begin{equation}
   S^{k=n}_{\text{WZW}}=\frac{2\pi i C_{n}\theta_{n+1}}{P_{n+1}}\int\mathbf{Tr}[M^{-1}(dM)^{n+1}],
\end{equation}
where the $M$ here is $M(t,x_1,x_2...x_{n-1},u)$, which is replaced from the $M(\vec{\theta}(t,x_1,x_2...x_{n-1},u))$. Mathematically, the topological term classifies the ($n$+1)th homotopy group of the manifold $\mathbf{M}$.

Now we consider the general $k$ which classifies the lower homotopy group of the manifold $\mathbf{M}$. We follow the same routine of the above dimension reduction approach and define $M$ on $2k-n+1$ dimensional torus as $M(p_1, p_2,\cdots,p_{2k-n+1})$ where $k$ is from $[\frac{n}{2}]$ to $n$.  The total gauge field is chosen as:
\begin{equation}
    \begin{aligned}
    &A_{\mu}\left(t, x_{1}, \ldots, x_{2 n}\right)=B_{\mu}\left(t, x_{1}, \ldots, x_{n-1}\right), \quad \\
    &\text { for } \mu=0, \ldots, n-1, \\ 
    &A_{i+n-1}\left(t, x_{1}, \ldots, x_{2 n}\right)=\theta_{i}\left(t, x_{1}, \ldots, x_{n-1}\right), \quad \\
    &\text { for } i=1, \ldots, 2k-n+1,
    \end{aligned}
\end{equation}  
Thus the CS theory is given by
\begin{equation}
    \begin{aligned}
   S_{\text{CS}}&=\frac{i C_{k}}{(k+1)!(2\pi)^k}\prod_{i=n}^{2k}L_{i}\int dt d^{n-1}x\int du \\
   &\epsilon^{a_1\cdots a_{2k+2}}\partial_{a_1}A_{a_2}\cdots \partial_{a_{2k+1}}A_{a_{2k+2}}\\
   &=\frac{i C_{k}}{(2k-n+1)!(n-k)!(2\pi)^k}\prod_{i=n}^{2k}L_{i}\int dt d^{n-1}x\int du\\ 
   &\epsilon^{a_1 \cdots a_{2k-n+1}a_{2k-n+2}\cdots a_{n+1}}\epsilon^{b_1\cdots b_{2k-n+1}}\\
   &\partial_{a_1}\theta_{b_1}\cdots \partial_{a_{2k-n+1}}\theta_{b_{2k-n+1}} \partial_{a_{2k-n+2}}B_{a_{2k-n+3}}\cdots\partial_{a_{n}}B_{a_{n+1}},
   \end{aligned}
\end{equation}
where the number $\frac{1}{(2k-n+1)!(n-k)!}$ comes from $\frac{1}{(k+1)!}C^{2k-n+1}_{k+1}$. 
This is due to that we should choose $2k-n+1$ number from $k+1$ number $\{a_2 ,a_4, \cdots a_{2k+2}\}$ to put the index $b_a, a=1,2,3,...2k-n+1$.
We can still use the trick of putting the Pontryagin index in the denominator, and the action becomes: 
\begin{equation}
\begin{aligned}
   S_{\mathsf{CS}}&=\frac{i C_{k}\theta_{2k-n+1}}{P_{2k-n+1}(n-k)!(2\pi)^k}\int_{T^{2k-n+1}}\prod_{i=n}^{2k}L_{i}\\
   &\int dt d^{n-1}x\int du \mathbf{Tr}[M^{-1}(dM)^{2k-n+1} ](F)^{n-k}.
   \end{aligned}
\end{equation}
Then we obtain the $n+1$ dimensional action:
\begin{equation}
\begin{aligned}
   S^{k}_{\text{WZW}}&=\frac{2\pi i C_{k}\theta_{2k-n+1}}{P_{k}(n-k)!}\int dt d^{n-1}x\int du\\&\mathbf{Tr}[M^{-1}(dM)^{2k-n+1} ](\frac{F}{2\pi})^{n-k}.
   \end{aligned}
\end{equation}

Thus the total action is summed over different $k$:
\begin{equation}
\begin{aligned}
   S_{\text{WZW}}&=(\sum_k\frac{2\pi i C_{k}\theta^{2k-n+1}_0}{P_{2k-n+1}}\int dt d^{n-1}x\int du\\& \mathbf{Tr}[M^{-1}(dM)^{2k-n+1} ])(\frac{F}{2\pi})^{n-k}.
     \end{aligned}
\end{equation}

When $n$ is even, the order parameter should be: $M=M_{1}+i\gamma^{0\cdots n-1}M_{2}$. $M_1$ and $M_2$ only act on the flavor indices and commute with each other. We can also assume $M^{+}M=1$. Now the Pontryagin index $P$ of the map $T^{n+1}\rightarrow M(p_1,p_2,...p_{n+1})$ is given by:
\begin{equation}
\begin{aligned}
   P_{n+1}&=\frac{\theta_{n+1}}{2^{\frac{n}{2}}}\int_{T^{n+1}} d^{n+1}p \mathbf{Tr}[M^{-1}dM(d\!\!\!/M)^{n}]\\
   &=\frac{(\frac{n}{2})!}{(2\pi)^{n/2}}\sum_{j}\int_{T^{n+1}} d^{n+1}p \\
   &\frac{1}{(2j+1)!(n-2j)!}\mathbf{Tr}[M^+_{1}(dM_{1})^{n-2j}(dM_2)^{2j+1}]\\
   & -\frac{1}{(2j)!(n-2j+1)!}\mathbf{Tr}[M^+_{2}(dM_{1})^{n-2j+1}(dM_2)^{2j}].
   \end{aligned}
\end{equation}

The number $\frac{1}{(2j+1)!(n-2j)!}$ and $\frac{1}{(2j)!(n-2j+1)!}$ comes from $C^{2j+1}_{n+1}\frac{1}{(n+1)!}$ and $C^{2j}_{n+1}\frac{1}{(n+1)!}$ . The reason is similar to that in the discussion of odd $n$ case and here we choose $2j+1$ or $2j$ number to put $M_2$ in the $n+1$ digits (the number of $M$).

For example, when $\mathbb{M}=S^{n+1}$, we can take $M_1=n_i\Gamma^i$ and $M_2=n_{n+2}$, the Pontryagin index is
\begin{equation}
\begin{aligned}
   P_{n+1}&=\frac{(\frac{n}{2})!}{(2\pi)^{n/2}}\int_{T^{n+1}} d^{n+1}p \frac{1}{n!} \mathbf{Tr}[M^+_{1}(dM_{1})^{n}(dM_2)]\\
   &-\frac{1}{(n+1)!}\mathbf{Tr}[M^+_{2}(dM_{1})^{n+1}]\\
   &=\frac{2\pi }{\textbf{Area}(S^{n+1})} \int \epsilon^{a_{1}\cdots a_{n+2}}n^{a_{1}}\partial_{p_{1}}n_{a_{2}}\cdots \partial_{p_{n+1}}n_{a_{n+2}}.
   \end{aligned}
\end{equation}

For a given $k$, we can do a similar calculation and the final result is:
\begin{equation}
\begin{aligned}
   S^{k}_{\text{WZW}}&=2\pi i\frac{ C_{k}\theta_{2k-n+1}}{ P_{2k-n+1}(n-k)!}\int dt d^{n-1}x\int du \\
   &\sum_{j} \frac{1}{(2j+1)!(2k-n-2j)!}\\& \mathbf{Tr}[M^+_{1}(dM_{1})^{2k-n-2j}(dM_2)^{2j+1}]\\& -\frac{1}{(2j)!(2k-n-2j+1)!}\\
   &\mathbf{Tr}[M^+_{2}(dM_{1})^{2k-n-2j+1}(dM_2)^{2j}](\frac{F}{2\pi})^{n-k}.
\end{aligned}
\end{equation}
For example, we can take $M_1=\cos\alpha$ and $M_2=\sin\alpha$, thus we can obtain:
\begin{eqnarray}
\begin{aligned}
   S^{n/2}_{\text{WZW}}&=\frac{2\pi i C_{1}}{P_{1}(\frac{n}{2})!}\int dt d^{n-1}x\int du   \ d\alpha(\frac{F}{2\pi})^{n/2}\\&=\frac{2\pi i C_{1}}{P_{1}(\frac{n}{2})!}\int dt d^{n-1}x \ \alpha(\frac{F}{2\pi})^{n/2}.
   \end{aligned}
\end{eqnarray}
This equation is just the chiral anomaly for QED theory with fermion of general dispersion.

\subsection{Generalization to the Nambu Space}
The generalization to the Nambu space is straightforward. We just need to calculate the Chern number $C$ of the $2n$ dimensional BdG Hamiltonian. The gauge field of the CS term is coupled to the Bogliubov fermion, namely, the eigenbases of the BdG Hamiltonian. 

The target space with superconducting order can be the general manifold discussed above, but here we consider a simpler 1+1d model $S^3$, to illustrate our method. The effective tight binding Hamiltonian, after projecting out two gapped bands, is the same as (\ref{1+1}):
\begin{equation}
    H(k)=2\left(\begin{array}{cc}0 & \left(\cos \frac{k}{2}+i \delta t \sin \frac{k}{2}\right)^{2} \\ \left(\cos \frac{k}{2}-i \delta t \sin \frac{k}{2}\right)^{2} & 0\end{array}\right).
\end{equation}
In order to introduce the superconducting order, we take the Nambu space now. The annihilation operator is: $c=(c_{1,\sigma}(k),c_{2,\sigma}(k),c^{\dagger}_{1,\sigma}(-k),c^{\dagger}_{2,\sigma}(-k))^T$. The indices $1,2$ is the sublattice(or orbital) index. If we condense the domain wall of the dimerized phase, we can have charge $\pm2$ and spin $0$ quantum number, which corresponds to the superconducting order and charge density wave order. We label the superconducting order and charge density order as $m_1,m_2,m_3$. Then we introduce three extra dimensions $k_1,k_2,k_3$ and the 4d Hamiltonian is:
\begin{equation}
    \begin{aligned} 
H\left(k_{1}, k_{2}, k_{3}, k_{4}\right)&=-\left[\left(1-m_{4}^{2}\right)+\left(1+m_{4}^{2}\right) \cos \left(k_{4}\right)\right] \Gamma_{5}\\
&-m_{1} \Gamma_{1} -m_{2} \Gamma_{2}-m_{3} \Gamma_{3}-2 m_{4} \sin \left(k_{4}\right) \Gamma_{4}, \nonumber
\end{aligned}
\end{equation}
\begin{equation}
\begin{aligned}
&m_{1}=\sin k_{1} \\ 
&m_{2}=\sin k_{2} \\ 
&m_{3}=\sin k_{3} \\ 
&m_{4}=m+\cos k_{1}+\cos k_{2}+\cos k_{3}\\
&\Gamma_{1}=I\otimes \sigma_{2}\otimes\mu_x, \Gamma_{2}=I\otimes \sigma_{2}\otimes\mu_y,\Gamma_{3}=\tau_1\otimes I\otimes\mu_z,\\
&\Gamma_{4}=\tau_{2} \otimes I\otimes\mu_z, \quad \Gamma_{5}=\tau_{3} \otimes I\otimes\mu_z,
\end{aligned}
\end{equation}
where $\tau,\sigma,\mu$ act on the sublattice(orbital), spin and Nambu indices.

If we take $m=-2.5$, then the Pontryagin index form $T^3$ to $S^3$: $(k_1,k_2,k_3) \to (m_1,m_2,m_3,m_4)/\sqrt{\sum_{i=1}^4m_i^2}$ is one. We can calculate the Chern number of the ground state to arrive at the level of the WZW term directly. The Chern number is $-2$, so the level of the corresponding WZW term is 2.

\section{Zero modes of U(1) monopole}

In this section, we will calculate the zero modes of gauge monopole in the two dimensional plane. Now the monopole can be understood as the configuration of a gauge field with nonzero flux. Let's first consider the Dirac fermion theory in a disk: $r<R$.

The Hamiltonian of Dirac fermion with a gauge field on a plane  is given by
\begin{equation}
    \begin{aligned}
    H_{\mathsf{Dirac}}&=
    \left(\begin{array}{cc}0 &  \partial_x-i\partial_y-iA_x-A_y \\ \partial_x+i\partial_y-iA_x+A_y& 0 \end{array}\right) 
   \\ &=2\left(\begin{array}{cc}0 &  \partial_z-iA_z \\ \partial_{\bar{z}}-iA_{\bar{z}}& 0  \end{array}\right),
    \end{aligned}
\end{equation}
where $A_{z}=\frac{1}{2}(A_x-iA_y)=-i\partial_z \beta(z,\bar{z})$, $A_{\bar{z}}=\frac{1}{2}(A_x+iA_y)=i\partial_{\bar{z}} \beta(z,\bar{z})$ and $B=2i(\partial_{\bar{z}}A_{z}-\partial_z A_{\bar{z}})=4\partial_z\partial_{\bar{z}} \beta(z,\bar{z})=\nabla^2 \beta$

Thus we can obtain solutions of the zero modes are
\begin{equation}
 \begin{array}{l}
   f_n (z,\bar{z})=\left(\begin{array}{cc}  u_n (z,\bar{z}) \\  \nu_n(z,\bar{z}) \end{array}\right),
   \end{array}
\end{equation}
where 
\begin{equation}
  u_n=c_n e^{-\beta}r^n e^{in\theta},\quad\quad
    \nu_n=d_n e^{\beta}r^n e^{-in\theta}
    \end{equation}
Here $n$ must be a nonnegative integer since the wavefunction must be normalized near the $r=0$.

Next, we assume $\beta$ only depends on $r$. Then the flux is given by
\begin{equation}
\begin{aligned}
    \Phi(r)=\frac{1}{2\pi}\int_{D(r'\le r)} B=\frac{1}{2\pi}\int_{D(r'\le r)} \nabla^2 \beta=r\beta'(r).
    \end{aligned}
\end{equation}
If the total flux is $\Phi_0$ which is a positive integer, i.e., $\Phi(R)=\Phi_0$,  we can obtain the asymptotic behavior of $\beta$ near the boundary
\begin{equation}
\begin{aligned}
    \beta(r\to R)=\Phi_0\ln r,
    \end{aligned}
\end{equation}
and the above zero mode solution near the boundary behaves as
\begin{eqnarray}
    &&u_n (\Phi_0)|_{r\to R}=c_n r^{n-\Phi_0}e^{in\theta},\nonumber\\
    &&\nu_n (\Phi_0)|_{r\to R}=d_n r^{n+\Phi_0} e^{-in\theta}.
\end{eqnarray}
Moreover, in order to make the Hamiltonian hermitian:
\begin{equation}
\langle \psi, H\psi\rangle=\langle H \psi, \psi\rangle,
\end{equation}
We choose the following local boundary condition: $\nu|_R=0$.

Therefore we will have infinite $u$ solutions. To obtain the finite number, we must do the regularization by misusing the number of solutions when there is no gauge field:
\begin{eqnarray}
&&\quad \mathsf{lim}_{\epsilon\to 0}(\sum_{n=0}^{\infty}\int^{R-\epsilon}_0 |u_n (\Phi_0)|^2d^2 x-\sum_{n=0}^{\infty}\int^{R-\epsilon}_0 |u_n (0)|^2d^2 x)\nonumber\\&&= \mathsf{lim}_{\epsilon\to 0}(\sum_{n=N}^{\infty}\int^{R-\epsilon}_0 |u_n (\Phi_0)|^2d^2 x-\sum_{n=N}^{\infty}\int^{R-\epsilon}_0 |u_n (0)|^2d^2 x)\nonumber\\&&= \mathsf{lim}_{\epsilon\to 0}(\sum_{n=N}^{\infty}(1-\frac{\epsilon}{R})^{2n-2\Phi_0+2}-\sum_{n=N}^{\infty}(1-\frac{\epsilon}{R})^{2n+2})=\Phi_0 .\nonumber\\
~
\end{eqnarray}
Here we choose $N$ to make sure that when $n>N$, the boundary behaviour
dominates the wavefunctions.

For the QBT mode, the Hamiltonian with a gauge field on a plane  is given by
\begin{equation}
    \begin{aligned}
    H_{\mathsf{QBT}} =4\left(\begin{array}{cc}0 &  (\partial_z-iA_z)^2 \\ (\partial_{\bar{z}}-iA_{\bar{z}})^2& 0  \end{array}\right). 
    \end{aligned}
\end{equation}

There will be two kinds of solutions for zero modes  with local boundary condition $\nu|_R=0$. :
\begin{eqnarray}
 \begin{array}{l}
  f^{1/2}_n (z,\bar{z})=\left(\begin{array}{cc}  u^{1/2}_n (z,\bar{z}) \\ 0 \end{array}\right),
   \end{array}
\end{eqnarray}
where
\begin{eqnarray}
    u^1_n=c^1_n e^{-\beta}r^n e^{in\theta},\quad\quad
    u^2_n=c^2_n e^{-\beta}r^{n+1} e^{i(n-1)\theta}\nonumber\\~
\end{eqnarray}
 and $n$ is also a nonnegative integer.
 
If the total flux is $\Phi_0$, the behaviour of these two solutions near the boundary is similar to that of Dirac fermion:
\begin{equation}
    u^1_n (\Phi_0)=c^1_n r^{n-\Phi_0}e^{in\theta},\quad\quad
   u^2_n (\Phi_0)=c^2_n r^{n+1-\Phi_0}e^{i(n-1)\theta}.\quad\quad
\end{equation}
Since the first solution is the same as that of Dirac fermion, we only need to do regularization for the second solution:
\begin{eqnarray}
  && \mathsf{lim}_{\epsilon\to 0}(\sum_{n=0}^{\infty}\int^{R-\epsilon}_0 |u^2_n (\Phi_0)|^2d^2 x-\sum_{n=0}^{\infty}\int^{R-\epsilon}_0 |u^2_n (0)|^2d^2 x)\nonumber\\=&&\mathsf{lim}_{\epsilon\to 0}(\sum_{n=N}^{\infty}(1-\frac{\epsilon}{R})^{2n-2\Phi_0+4}-\sum_{n=N}^{\infty}(1-\frac{\epsilon}{R})^{2n+4})=\Phi_0 .\nonumber\\
  ~
\end{eqnarray}
Thus we conclude that there will be $2\Phi_0$ zero modes. When $\Phi_0=1$, i.e., adding a $2\pi$ flux, there will be two zero modes.
\section{Derivation of the Projective Symmetry Representation}
\label{psg}
In this appendix, we derive the projective symmetry representation of our bilayer triangular lattice model Eq.(\ref{bitri}). The bilayer triangular lattice model couples two triangular lattice models with staggered $\pi$-flux parton ansatz, so we begin with the staggered  $\pi$-flux parton mean field theory on the triangular lattice. The $\pi$-flux lattice Hamiltonian is \cite{PhysRevB.103.165138}:
\begin{equation}
   H_{\pi-\text{flux}}= \sum_{i,j}f^{\dagger}_iu_{i,j}f_j,
\end{equation}
where $u_{i,j}=u^*_{j,i}$ and the $f_i$ operators are the spinon operators on the lattice. The nonzero hopping amplitudes are:
\begin{equation}
    u_{i,i+\hat{x}}=1,\quad u_{i,i+\hat{y}}=(-1)^{i_x},\quad u_{i,i+\hat{y}-\hat{x}}=(-1)^{i_x},
\end{equation}
where $\hat{x},\hat{y}$ are the unit vectors of the triangular lattice, which are half the unit vectors of $\pi$-flux state: $\hat{x}=\frac{1}{2}r_1,\hat{y}=\frac{1}{2}r_2$. And we assume $i_{Ax}\equiv i_{Cx}\equiv 0\pmod2, \quad i_{Bx}\equiv i_{Dx}\equiv 1\pmod2$. 

We first consider the reflection symmetry $R_x$. The lattice coordinate $x\hat{x}+y\hat{y}$ transforms as:
\begin{equation}
\begin{aligned}
    R_x(x\hat{x}+y\hat{y})&= xR_x\hat{x}+yR_x\hat{y}\\
    &=x\hat{x}+y(-\hat{y}+\hat{x})=(x+y)\hat{x}-y\hat{y}.
    \end{aligned}
\end{equation}
So the operators transform as:
\begin{equation}
    f_{(x,y)}=(-1)^{x}(-1)^{Y}f^{\dagger}_{(x+y,-y)},
\end{equation}
where $Y$ is the $r_2$ coordinate in the $r_1,r_2$ basis; or in other words the function $(-1)^{Y}$ is period 4 in $\hat{x},\hat{y}$ basis: $1,1,-1,-1$. Written in the unit cell basis, the PSG of the low energy operators are:
\begin{equation}
R_x:\quad
      \left(\begin{array}{c}
f_A(k_x,k_y)  \\  
f_B(k_x,k_y) \\
f_C(k_x,k_y)  \\
 f_D(k_x,k_y) 
 \end{array}\right)
 \rightarrow \sigma_2
 \left(\begin{array}{c}
f_A^{\dagger} (-k_x,k_y)  \\  
-f_B^{\dagger}(-k_x,k_y) \\
-f_D^{\dagger}(-k_x,k_y)  \\
 -f_C^{\dagger}(-k_x,k_y) 
 \end{array}\right).
\end{equation}

Next we consider the rotation symmetry $C_6^A$. The coordinate transforms as:
\begin{equation}
    C_6(x\hat{x}+y\hat{y})=x\hat{y}+y(\hat{y}-\hat{x}).
\end{equation}
So the operators transform as:
\begin{equation}
\begin{aligned}
    &\text{A-sublattice}: f_{(x,y)}=(-1)^Y f^{\dagger}_{(-y,x+y)},\\
    &\text{B,C,D-sublattice}: f_{(x,y)}=-(-1)^Y f^{\dagger}_{(-y,x+y)}.
    \end{aligned}
\end{equation}
Written in the unit cell basis, the PSG of the low energy operators are:
\begin{equation}
C_6^A:\quad
      \left(\begin{array}{c}
f_A  \\  
f_B\\
f_C \\
 f_D
 \end{array}\right)
 \rightarrow\sigma_2
 \left(\begin{array}{c}
f_A^{\dagger} \\  
-f_C^{\dagger}\\
f_D^{\dagger} \\
 -f_B^{\dagger}
 \end{array}\right).
\end{equation}

Similarly, we have the projective symmetry representation of the two translation symmetries as:
\begin{equation}
\begin{aligned}
    &T_1:\\
&\left(\begin{array}{c}
f_A  \\  
f_B\\
f_C \\
 f_D
 \end{array}\right)
 \rightarrow
 \left(\begin{array}{c}
f_B  \\  
-f_A\\
-f_D \\
 f_C
 \end{array}\right),
 \end{aligned}\quad
 \begin{aligned}
  &T_2:\\
  &\left(\begin{array}{c}
f_A  \\  
f_B\\
f_C \\
 f_D
 \end{array}\right)
 \rightarrow
 \left(\begin{array}{c}
f_C \\  
f_D\\
-f_A \\
 -f_B
 \end{array}\right).
 \end{aligned}
\end{equation}

Now we move to the bilayer triangular lattice model Eq.(\ref{bitri}). The spinon on the A,B sublattices are gapped, and we will focus on the C,D sublattices hereafter. The lattice symmetries are translation $T_1,2T_2$, where the translation in $r_2$ is broken down into twice the lattice constants. The rotation symmetry is broken down into $C_2$ and there is another reflection symmetry $R_x$. We should do a basis transformation U from the low energy sublattice basis $(f_{C,a},f_{D,a})^T$ above, where $a=1,2$ is the layer index, to the sublattice basis of the low energy Hamiltonian Eq.(\ref{qbt}). The basis transformation reads: $U=\exp(\frac{\pi i}{4}\mu_z) \otimes I_{2\times 2}$, where the $\mu_z$ here acts on the C,D subspace and the latter identity matrix acts on the layer index. The projective symmetry representation should save the staggered $\pi$-flux pattern and the opposite hopping amplitude between the two layers on A,B sublattices. The projective symmetry representations are:
\begin{equation}
\begin{aligned}
&T_1:\\
&\left(\begin{array}{c}
f_{C,a} \\
 f_{D,a}
 \end{array}\right)
 \rightarrow i(-1)^a
 \left(\begin{array}{c}
f_{D,a} \\
 f_{C,a}
 \end{array}\right),\quad \chi\rightarrow -i\tau_3\chi,\\
&T_2:\\
&\left(\begin{array}{c}
f_C \\
 f_D
 \end{array}\right)
 \rightarrow
 \left(\begin{array}{c}
-f_C\\
 -f_D
 \end{array}\right),\quad \chi\rightarrow -\mu_0\otimes\tau_0\chi,\\
& C_2^A:\\
  & \left(\begin{array}{c}
f_{C,a} \\
 f_{D,a}
 \end{array}\right)
 \rightarrow
i(-1)^a \sigma_2\left(\begin{array}{c}
f_{C,a}^{\dagger} \\
 -f_{D,a}^{\dagger}
 \end{array}\right), \chi\rightarrow i\mu_2\otimes\tau_3 \sigma_2\chi^{*},\\
 & R_x^A:\\
  & \left(\begin{array}{c}
f_{C,a} \\
 f_{D,a}
 \end{array}\right)
 \rightarrow
(-1)^a \sigma_2\left(\begin{array}{c}
f_{D,a}^{\dagger} \\
 f_{C,a}^{\dagger} \end{array}\right),\chi\rightarrow \mu_3\sigma_2\chi^{*},
 \end{aligned}
 \end{equation}
 where the quadratic band touching basis $\chi$ is obtained from the sublattice basis $(f_{C,a},f_{D,a})^T$ through another basis transformation: $V=\exp(-i\pi\mu_y/4)\otimes \mathcal{I}_{2\times 2}$. The $\mu_y$ acts on the C,D sublattices and $V$ is the identity in the layer index.
\bibliography{bib}

\begin{thebibliography}{103}%
\makeatletter
\providecommand \@ifxundefined [1]{%
 \@ifx{#1\undefined}
}%
\providecommand \@ifnum [1]{%
 \ifnum #1\expandafter \@firstoftwo
 \else \expandafter \@secondoftwo
 \fi
}%
\providecommand \@ifx [1]{%
 \ifx #1\expandafter \@firstoftwo
 \else \expandafter \@secondoftwo
 \fi
}%
\providecommand \natexlab [1]{#1}%
\providecommand \enquote  [1]{``#1''}%
\providecommand \bibnamefont  [1]{#1}%
\providecommand \bibfnamefont [1]{#1}%
\providecommand \citenamefont [1]{#1}%
\providecommand \href@noop [0]{\@secondoftwo}%
\providecommand \href [0]{\begingroup \@sanitize@url \@href}%
\providecommand \@href[1]{\@@startlink{#1}\@@href}%
\providecommand \@@href[1]{\endgroup#1\@@endlink}%
\providecommand \@sanitize@url [0]{\catcode `\\12\catcode `\$12\catcode
  `\&12\catcode `\#12\catcode `\^12\catcode `\_12\catcode `\%12\relax}%
\providecommand \@@startlink[1]{}%
\providecommand \@@endlink[0]{}%
\providecommand \url  [0]{\begingroup\@sanitize@url \@url }%
\providecommand \@url [1]{\endgroup\@href {#1}{\urlprefix }}%
\providecommand \urlprefix  [0]{URL }%
\providecommand \Eprint [0]{\href }%
\providecommand \doibase [0]{https://doi.org/}%
\providecommand \selectlanguage [0]{\@gobble}%
\providecommand \bibinfo  [0]{\@secondoftwo}%
\providecommand \bibfield  [0]{\@secondoftwo}%
\providecommand \translation [1]{[#1]}%
\providecommand \BibitemOpen [0]{}%
\providecommand \bibitemStop [0]{}%
\providecommand \bibitemNoStop [0]{.\EOS\space}%
\providecommand \EOS [0]{\spacefactor3000\relax}%
\providecommand \BibitemShut  [1]{\csname bibitem#1\endcsname}%
\let\auto@bib@innerbib\@empty
\bibitem [{\citenamefont {Kitaev}(2003)}]{KITAEV20032}%
  \BibitemOpen
  \bibfield  {author} {\bibinfo {author} {\bibfnamefont {A.}~\bibnamefont
  {Kitaev}},\ }\bibfield  {title} {\bibinfo {title} {Fault-tolerant quantum
  computation by anyons},\ }\href
  {https://doi.org/https://doi.org/10.1016/S0003-4916(02)00018-0} {\bibfield
  {journal} {\bibinfo  {journal} {Annals of Physics}\ }\textbf {\bibinfo
  {volume} {303}},\ \bibinfo {pages} {2} (\bibinfo {year} {2003})}\BibitemShut
  {NoStop}%
\bibitem [{\citenamefont {Kitaev}(2006)}]{KITAEV20062}%
  \BibitemOpen
  \bibfield  {author} {\bibinfo {author} {\bibfnamefont {A.}~\bibnamefont
  {Kitaev}},\ }\bibfield  {title} {\bibinfo {title} {Anyons in an exactly
  solved model and beyond},\ }\href
  {https://doi.org/https://doi.org/10.1016/j.aop.2005.10.005} {\bibfield
  {journal} {\bibinfo  {journal} {Annals of Physics}\ }\textbf {\bibinfo
  {volume} {321}},\ \bibinfo {pages} {2} (\bibinfo {year} {2006})},\ \bibinfo
  {note} {january Special Issue}\BibitemShut {NoStop}%
\bibitem [{\citenamefont {Wen}(1995)}]{doi:10.1080/00018739500101566}%
  \BibitemOpen
  \bibfield  {author} {\bibinfo {author} {\bibfnamefont {X.-G.}\ \bibnamefont
  {Wen}},\ }\bibfield  {title} {\bibinfo {title} {Topological orders and edge
  excitations in fractional quantum hall states},\ }\href
  {https://doi.org/10.1080/00018739500101566} {\bibfield  {journal} {\bibinfo
  {journal} {Advances in Physics}\ }\textbf {\bibinfo {volume} {44}},\ \bibinfo
  {pages} {405} (\bibinfo {year} {1995})},\ \Eprint
  {https://arxiv.org/abs/https://doi.org/10.1080/00018739500101566}
  {https://doi.org/10.1080/00018739500101566} \BibitemShut {NoStop}%
\bibitem [{\citenamefont {WEN}(1990)}]{doi:10.1142/S0217979290000139}%
  \BibitemOpen
  \bibfield  {author} {\bibinfo {author} {\bibfnamefont {X.~G.}\ \bibnamefont
  {WEN}},\ }\bibfield  {title} {\bibinfo {title} {Topological orders in rigid
  states},\ }\href {https://doi.org/10.1142/S0217979290000139} {\bibfield
  {journal} {\bibinfo  {journal} {International Journal of Modern Physics B}\
  }\textbf {\bibinfo {volume} {04}},\ \bibinfo {pages} {239} (\bibinfo {year}
  {1990})},\ \Eprint
  {https://arxiv.org/abs/https://doi.org/10.1142/S0217979290000139}
  {https://doi.org/10.1142/S0217979290000139} \BibitemShut {NoStop}%
\bibitem [{\citenamefont {Wen}(2002)}]{PhysRevB.65.165113}%
  \BibitemOpen
  \bibfield  {author} {\bibinfo {author} {\bibfnamefont {X.-G.}\ \bibnamefont
  {Wen}},\ }\bibfield  {title} {\bibinfo {title} {Quantum orders and symmetric
  spin liquids},\ }\href {https://doi.org/10.1103/PhysRevB.65.165113}
  {\bibfield  {journal} {\bibinfo  {journal} {Phys. Rev. B}\ }\textbf {\bibinfo
  {volume} {65}},\ \bibinfo {pages} {165113} (\bibinfo {year}
  {2002})}\BibitemShut {NoStop}%
\bibitem [{\citenamefont {Wen}\ \emph {et~al.}(1989)\citenamefont {Wen},
  \citenamefont {Wilczek},\ and\ \citenamefont {Zee}}]{PhysRevB.39.11413}%
  \BibitemOpen
  \bibfield  {author} {\bibinfo {author} {\bibfnamefont {X.~G.}\ \bibnamefont
  {Wen}}, \bibinfo {author} {\bibfnamefont {F.}~\bibnamefont {Wilczek}},\ and\
  \bibinfo {author} {\bibfnamefont {A.}~\bibnamefont {Zee}},\ }\bibfield
  {title} {\bibinfo {title} {Chiral spin states and superconductivity},\ }\href
  {https://doi.org/10.1103/PhysRevB.39.11413} {\bibfield  {journal} {\bibinfo
  {journal} {Phys. Rev. B}\ }\textbf {\bibinfo {volume} {39}},\ \bibinfo
  {pages} {11413} (\bibinfo {year} {1989})}\BibitemShut {NoStop}%
\bibitem [{\citenamefont {Levin}\ and\ \citenamefont
  {Wen}(2006)}]{PhysRevLett.96.110405}%
  \BibitemOpen
  \bibfield  {author} {\bibinfo {author} {\bibfnamefont {M.}~\bibnamefont
  {Levin}}\ and\ \bibinfo {author} {\bibfnamefont {X.-G.}\ \bibnamefont
  {Wen}},\ }\bibfield  {title} {\bibinfo {title} {Detecting topological order
  in a ground state wave function},\ }\href
  {https://doi.org/10.1103/PhysRevLett.96.110405} {\bibfield  {journal}
  {\bibinfo  {journal} {Phys. Rev. Lett.}\ }\textbf {\bibinfo {volume} {96}},\
  \bibinfo {pages} {110405} (\bibinfo {year} {2006})}\BibitemShut {NoStop}%
\bibitem [{\citenamefont {Wen}(2013)}]{Wen2013}%
  \BibitemOpen
  \bibfield  {author} {\bibinfo {author} {\bibfnamefont {X.-G.}\ \bibnamefont
  {Wen}},\ }\bibfield  {title} {\bibinfo {title} {Topological order: From
  long-range entangled quantum matter to a unified origin of light and
  electrons},\ }\href {https://doi.org/10.1155/2013/198710} {\bibfield
  {journal} {\bibinfo  {journal} {ISRN Condensed Matter Physics}\ }\textbf
  {\bibinfo {volume} {2013}},\ \bibinfo {pages} {198710} (\bibinfo {year}
  {2013})}\BibitemShut {NoStop}%
\bibitem [{\citenamefont {Wen}(2017)}]{RevModPhys.89.041004}%
  \BibitemOpen
  \bibfield  {author} {\bibinfo {author} {\bibfnamefont {X.-G.}\ \bibnamefont
  {Wen}},\ }\bibfield  {title} {\bibinfo {title} {Colloquium: Zoo of
  quantum-topological phases of matter},\ }\href
  {https://doi.org/10.1103/RevModPhys.89.041004} {\bibfield  {journal}
  {\bibinfo  {journal} {Rev. Mod. Phys.}\ }\textbf {\bibinfo {volume} {89}},\
  \bibinfo {pages} {041004} (\bibinfo {year} {2017})}\BibitemShut {NoStop}%
\bibitem [{\citenamefont {Gang}(2007)}]{Xiao:803748}%
  \BibitemOpen
  \bibfield  {author} {\bibinfo {author} {\bibfnamefont {W.~X.}\ \bibnamefont
  {Gang}},\ }\href {https://doi.org/10.1093/acprof:oso/9780199227259.001.0001}
  {\emph {\bibinfo {title} {{Quantum field theory of many-body systems: from
  the origin of sound to an origin of light and electrons}}}}\ (\bibinfo
  {publisher} {Oxford University Press},\ \bibinfo {address} {Oxford},\
  \bibinfo {year} {2007})\BibitemShut {NoStop}%
\bibitem [{\citenamefont {Vijay}\ \emph {et~al.}(2016)\citenamefont {Vijay},
  \citenamefont {Haah},\ and\ \citenamefont {Fu}}]{vijay2016fracton}%
  \BibitemOpen
  \bibfield  {author} {\bibinfo {author} {\bibfnamefont {S.}~\bibnamefont
  {Vijay}}, \bibinfo {author} {\bibfnamefont {J.}~\bibnamefont {Haah}},\ and\
  \bibinfo {author} {\bibfnamefont {L.}~\bibnamefont {Fu}},\ }\bibfield
  {title} {\bibinfo {title} {Fracton topological order, generalized lattice
  gauge theory, and duality},\ }\href
  {https://doi.org/10.1103/PhysRevB.94.235157} {\bibfield  {journal} {\bibinfo
  {journal} {Phys. Rev. B}\ }\textbf {\bibinfo {volume} {94}},\ \bibinfo
  {pages} {235157} (\bibinfo {year} {2016})}\BibitemShut {NoStop}%
\bibitem [{\citenamefont {Haah}(2011)}]{Haah2011Local}%
  \BibitemOpen
  \bibfield  {author} {\bibinfo {author} {\bibfnamefont {J.}~\bibnamefont
  {Haah}},\ }\bibfield  {title} {\bibinfo {title} {Local stabilizer codes in
  three dimensions without string logical operators},\ }\href
  {https://doi.org/10.1103/PhysRevA.83.042330} {\bibfield  {journal} {\bibinfo
  {journal} {Phys. Rev. A}\ }\textbf {\bibinfo {volume} {83}},\ \bibinfo
  {pages} {042330} (\bibinfo {year} {2011})}\BibitemShut {NoStop}%
\bibitem [{\citenamefont {Pretko}\ \emph {et~al.}(2020)\citenamefont {Pretko},
  \citenamefont {Chen},\ and\ \citenamefont {You}}]{pretko2020fracton}%
  \BibitemOpen
  \bibfield  {author} {\bibinfo {author} {\bibfnamefont {M.}~\bibnamefont
  {Pretko}}, \bibinfo {author} {\bibfnamefont {X.}~\bibnamefont {Chen}},\ and\
  \bibinfo {author} {\bibfnamefont {Y.}~\bibnamefont {You}},\ }\bibfield
  {title} {\bibinfo {title} {Fracton phases of matter},\ }\href
  {https://doi.org/10.1142/S0217751X20300033} {\bibfield  {journal} {\bibinfo
  {journal} {International Journal of Modern Physics A}\ }\textbf {\bibinfo
  {volume} {35}},\ \bibinfo {pages} {2030003} (\bibinfo {year}
  {2020})}\BibitemShut {NoStop}%
\bibitem [{\citenamefont {Shen}\ \emph {et~al.}(2022)\citenamefont {Shen},
  \citenamefont {Wu}, \citenamefont {Li}, \citenamefont {Qin},\ and\
  \citenamefont {Yao}}]{PhysRevResearch.4.L032008}%
  \BibitemOpen
  \bibfield  {author} {\bibinfo {author} {\bibfnamefont {X.}~\bibnamefont
  {Shen}}, \bibinfo {author} {\bibfnamefont {Z.}~\bibnamefont {Wu}}, \bibinfo
  {author} {\bibfnamefont {L.}~\bibnamefont {Li}}, \bibinfo {author}
  {\bibfnamefont {Z.}~\bibnamefont {Qin}},\ and\ \bibinfo {author}
  {\bibfnamefont {H.}~\bibnamefont {Yao}},\ }\bibfield  {title} {\bibinfo
  {title} {Fracton topological order at finite temperature},\ }\href
  {https://doi.org/10.1103/PhysRevResearch.4.L032008} {\bibfield  {journal}
  {\bibinfo  {journal} {Phys. Rev. Research}\ }\textbf {\bibinfo {volume}
  {4}},\ \bibinfo {pages} {L032008} (\bibinfo {year} {2022})}\BibitemShut
  {NoStop}%
\bibitem [{\citenamefont
  {McGreevy}(2022)}]{https://doi.org/10.48550/arxiv.2204.03045}%
  \BibitemOpen
  \bibfield  {author} {\bibinfo {author} {\bibfnamefont {J.}~\bibnamefont
  {McGreevy}},\ }\href {https://doi.org/10.48550/ARXIV.2204.03045} {\bibinfo
  {title} {Generalized symmetries in condensed matter}} (\bibinfo {year}
  {2022})\BibitemShut {NoStop}%
\bibitem [{\citenamefont {Cordova}\ \emph {et~al.}(2022)\citenamefont
  {Cordova}, \citenamefont {Dumitrescu}, \citenamefont {Intriligator},\ and\
  \citenamefont {Shao}}]{https://doi.org/10.48550/arxiv.2205.09545}%
  \BibitemOpen
  \bibfield  {author} {\bibinfo {author} {\bibfnamefont {C.}~\bibnamefont
  {Cordova}}, \bibinfo {author} {\bibfnamefont {T.~T.}\ \bibnamefont
  {Dumitrescu}}, \bibinfo {author} {\bibfnamefont {K.}~\bibnamefont
  {Intriligator}},\ and\ \bibinfo {author} {\bibfnamefont {S.-H.}\ \bibnamefont
  {Shao}},\ }\href {https://doi.org/10.48550/ARXIV.2205.09545} {\bibinfo
  {title} {Snowmass white paper: Generalized symmetries in quantum field theory
  and beyond}} (\bibinfo {year} {2022})\BibitemShut {NoStop}%
\bibitem [{\citenamefont {Pollmann}\ \emph {et~al.}(2010)\citenamefont
  {Pollmann}, \citenamefont {Turner}, \citenamefont {Berg},\ and\ \citenamefont
  {Oshikawa}}]{PhysRevB.81.064439}%
  \BibitemOpen
  \bibfield  {author} {\bibinfo {author} {\bibfnamefont {F.}~\bibnamefont
  {Pollmann}}, \bibinfo {author} {\bibfnamefont {A.~M.}\ \bibnamefont
  {Turner}}, \bibinfo {author} {\bibfnamefont {E.}~\bibnamefont {Berg}},\ and\
  \bibinfo {author} {\bibfnamefont {M.}~\bibnamefont {Oshikawa}},\ }\bibfield
  {title} {\bibinfo {title} {Entanglement spectrum of a topological phase in
  one dimension},\ }\href {https://doi.org/10.1103/PhysRevB.81.064439}
  {\bibfield  {journal} {\bibinfo  {journal} {Phys. Rev. B}\ }\textbf {\bibinfo
  {volume} {81}},\ \bibinfo {pages} {064439} (\bibinfo {year}
  {2010})}\BibitemShut {NoStop}%
\bibitem [{\citenamefont {Chen}\ \emph {et~al.}(2011)\citenamefont {Chen},
  \citenamefont {Gu},\ and\ \citenamefont {Wen}}]{chen2011complete}%
  \BibitemOpen
  \bibfield  {author} {\bibinfo {author} {\bibfnamefont {X.}~\bibnamefont
  {Chen}}, \bibinfo {author} {\bibfnamefont {Z.-C.}\ \bibnamefont {Gu}},\ and\
  \bibinfo {author} {\bibfnamefont {X.-G.}\ \bibnamefont {Wen}},\ }\bibfield
  {title} {\bibinfo {title} {Complete classification of one-dimensional gapped
  quantum phases in interacting spin systems},\ }\href
  {https://doi.org/10.1103/PhysRevB.84.235128} {\bibfield  {journal} {\bibinfo
  {journal} {Phys. Rev. B}\ }\textbf {\bibinfo {volume} {84}},\ \bibinfo
  {pages} {235128} (\bibinfo {year} {2011})}\BibitemShut {NoStop}%
\bibitem [{\citenamefont {Hasan}\ and\ \citenamefont
  {Kane}(2010)}]{hasan2010colloquium}%
  \BibitemOpen
  \bibfield  {author} {\bibinfo {author} {\bibfnamefont {M.~Z.}\ \bibnamefont
  {Hasan}}\ and\ \bibinfo {author} {\bibfnamefont {C.~L.}\ \bibnamefont
  {Kane}},\ }\bibfield  {title} {\bibinfo {title} {Colloquium: Topological
  insulators},\ }\href {https://doi.org/10.1103/RevModPhys.82.3045} {\bibfield
  {journal} {\bibinfo  {journal} {Rev. Mod. Phys.}\ }\textbf {\bibinfo {volume}
  {82}},\ \bibinfo {pages} {3045} (\bibinfo {year} {2010})}\BibitemShut
  {NoStop}%
\bibitem [{\citenamefont {Qi}\ and\ \citenamefont
  {Zhang}(2011)}]{RevModPhys.83.1057}%
  \BibitemOpen
  \bibfield  {author} {\bibinfo {author} {\bibfnamefont {X.-L.}\ \bibnamefont
  {Qi}}\ and\ \bibinfo {author} {\bibfnamefont {S.-C.}\ \bibnamefont {Zhang}},\
  }\bibfield  {title} {\bibinfo {title} {Topological insulators and
  superconductors},\ }\href {https://doi.org/10.1103/RevModPhys.83.1057}
  {\bibfield  {journal} {\bibinfo  {journal} {Rev. Mod. Phys.}\ }\textbf
  {\bibinfo {volume} {83}},\ \bibinfo {pages} {1057} (\bibinfo {year}
  {2011})}\BibitemShut {NoStop}%
\bibitem [{\citenamefont {Chen}\ \emph {et~al.}(2012)\citenamefont {Chen},
  \citenamefont {Gu}, \citenamefont {Liu},\ and\ \citenamefont
  {Wen}}]{doi:10.1126/science.1227224}%
  \BibitemOpen
  \bibfield  {author} {\bibinfo {author} {\bibfnamefont {X.}~\bibnamefont
  {Chen}}, \bibinfo {author} {\bibfnamefont {Z.-C.}\ \bibnamefont {Gu}},
  \bibinfo {author} {\bibfnamefont {Z.-X.}\ \bibnamefont {Liu}},\ and\ \bibinfo
  {author} {\bibfnamefont {X.-G.}\ \bibnamefont {Wen}},\ }\bibfield  {title}
  {\bibinfo {title} {Symmetry-protected topological orders in interacting
  bosonic systems},\ }\href {https://doi.org/10.1126/science.1227224}
  {\bibfield  {journal} {\bibinfo  {journal} {Science}\ }\textbf {\bibinfo
  {volume} {338}},\ \bibinfo {pages} {1604} (\bibinfo {year} {2012})},\ \Eprint
  {https://arxiv.org/abs/https://www.science.org/doi/pdf/10.1126/science.1227224}
  {https://www.science.org/doi/pdf/10.1126/science.1227224} \BibitemShut
  {NoStop}%
\bibitem [{\citenamefont {Senthil}\ \emph
  {et~al.}(2004{\natexlab{a}})\citenamefont {Senthil}, \citenamefont
  {Vishwanath}, \citenamefont {Balents}, \citenamefont {Sachdev},\ and\
  \citenamefont {Fisher}}]{doi:10.1126/science.1091806}%
  \BibitemOpen
  \bibfield  {author} {\bibinfo {author} {\bibfnamefont {T.}~\bibnamefont
  {Senthil}}, \bibinfo {author} {\bibfnamefont {A.}~\bibnamefont {Vishwanath}},
  \bibinfo {author} {\bibfnamefont {L.}~\bibnamefont {Balents}}, \bibinfo
  {author} {\bibfnamefont {S.}~\bibnamefont {Sachdev}},\ and\ \bibinfo {author}
  {\bibfnamefont {M.~P.~A.}\ \bibnamefont {Fisher}},\ }\bibfield  {title}
  {\bibinfo {title} {Deconfined quantum critical points},\ }\href
  {https://doi.org/10.1126/science.1091806} {\bibfield  {journal} {\bibinfo
  {journal} {Science}\ }\textbf {\bibinfo {volume} {303}},\ \bibinfo {pages}
  {1490} (\bibinfo {year} {2004}{\natexlab{a}})},\ \Eprint
  {https://arxiv.org/abs/https://www.science.org/doi/pdf/10.1126/science.1091806}
  {https://www.science.org/doi/pdf/10.1126/science.1091806} \BibitemShut
  {NoStop}%
\bibitem [{\citenamefont {Senthil}\ \emph
  {et~al.}(2004{\natexlab{b}})\citenamefont {Senthil}, \citenamefont {Balents},
  \citenamefont {Sachdev}, \citenamefont {Vishwanath},\ and\ \citenamefont
  {Fisher}}]{PhysRevB.70.144407}%
  \BibitemOpen
  \bibfield  {author} {\bibinfo {author} {\bibfnamefont {T.}~\bibnamefont
  {Senthil}}, \bibinfo {author} {\bibfnamefont {L.}~\bibnamefont {Balents}},
  \bibinfo {author} {\bibfnamefont {S.}~\bibnamefont {Sachdev}}, \bibinfo
  {author} {\bibfnamefont {A.}~\bibnamefont {Vishwanath}},\ and\ \bibinfo
  {author} {\bibfnamefont {M.~P.~A.}\ \bibnamefont {Fisher}},\ }\bibfield
  {title} {\bibinfo {title} {Quantum criticality beyond the
  landau-ginzburg-wilson paradigm},\ }\href
  {https://doi.org/10.1103/PhysRevB.70.144407} {\bibfield  {journal} {\bibinfo
  {journal} {Phys. Rev. B}\ }\textbf {\bibinfo {volume} {70}},\ \bibinfo
  {pages} {144407} (\bibinfo {year} {2004}{\natexlab{b}})}\BibitemShut
  {NoStop}%
\bibitem [{\citenamefont {Sandvik}\ \emph {et~al.}(2002)\citenamefont
  {Sandvik}, \citenamefont {Daul}, \citenamefont {Singh},\ and\ \citenamefont
  {Scalapino}}]{PhysRevLett.89.247201}%
  \BibitemOpen
  \bibfield  {author} {\bibinfo {author} {\bibfnamefont {A.~W.}\ \bibnamefont
  {Sandvik}}, \bibinfo {author} {\bibfnamefont {S.}~\bibnamefont {Daul}},
  \bibinfo {author} {\bibfnamefont {R.~R.~P.}\ \bibnamefont {Singh}},\ and\
  \bibinfo {author} {\bibfnamefont {D.~J.}\ \bibnamefont {Scalapino}},\
  }\bibfield  {title} {\bibinfo {title} {Striped phase in a quantum $xy$ model
  with ring exchange},\ }\href {https://doi.org/10.1103/PhysRevLett.89.247201}
  {\bibfield  {journal} {\bibinfo  {journal} {Phys. Rev. Lett.}\ }\textbf
  {\bibinfo {volume} {89}},\ \bibinfo {pages} {247201} (\bibinfo {year}
  {2002})}\BibitemShut {NoStop}%
\bibitem [{\citenamefont {Lou}\ \emph {et~al.}(2009)\citenamefont {Lou},
  \citenamefont {Sandvik},\ and\ \citenamefont
  {Kawashima}}]{PhysRevB.80.180414}%
  \BibitemOpen
  \bibfield  {author} {\bibinfo {author} {\bibfnamefont {J.}~\bibnamefont
  {Lou}}, \bibinfo {author} {\bibfnamefont {A.~W.}\ \bibnamefont {Sandvik}},\
  and\ \bibinfo {author} {\bibfnamefont {N.}~\bibnamefont {Kawashima}},\
  }\bibfield  {title} {\bibinfo {title} {Antiferromagnetic to
  valence-bond-solid transitions in two-dimensional $\text{SU}(n)$ heisenberg
  models with multispin interactions},\ }\href
  {https://doi.org/10.1103/PhysRevB.80.180414} {\bibfield  {journal} {\bibinfo
  {journal} {Phys. Rev. B}\ }\textbf {\bibinfo {volume} {80}},\ \bibinfo
  {pages} {180414} (\bibinfo {year} {2009})}\BibitemShut {NoStop}%
\bibitem [{\citenamefont {Sen}\ and\ \citenamefont
  {Sandvik}(2010)}]{PhysRevB.82.174428}%
  \BibitemOpen
  \bibfield  {author} {\bibinfo {author} {\bibfnamefont {A.}~\bibnamefont
  {Sen}}\ and\ \bibinfo {author} {\bibfnamefont {A.~W.}\ \bibnamefont
  {Sandvik}},\ }\bibfield  {title} {\bibinfo {title} {Example of a first-order
  n\'eel to valence-bond-solid transition in two dimensions},\ }\href
  {https://doi.org/10.1103/PhysRevB.82.174428} {\bibfield  {journal} {\bibinfo
  {journal} {Phys. Rev. B}\ }\textbf {\bibinfo {volume} {82}},\ \bibinfo
  {pages} {174428} (\bibinfo {year} {2010})}\BibitemShut {NoStop}%
\bibitem [{\citenamefont {Nahum}\ \emph {et~al.}(2011)\citenamefont {Nahum},
  \citenamefont {Chalker}, \citenamefont {Serna}, \citenamefont {Ortu\~no},\
  and\ \citenamefont {Somoza}}]{PhysRevLett.107.110601}%
  \BibitemOpen
  \bibfield  {author} {\bibinfo {author} {\bibfnamefont {A.}~\bibnamefont
  {Nahum}}, \bibinfo {author} {\bibfnamefont {J.~T.}\ \bibnamefont {Chalker}},
  \bibinfo {author} {\bibfnamefont {P.}~\bibnamefont {Serna}}, \bibinfo
  {author} {\bibfnamefont {M.}~\bibnamefont {Ortu\~no}},\ and\ \bibinfo
  {author} {\bibfnamefont {A.~M.}\ \bibnamefont {Somoza}},\ }\bibfield  {title}
  {\bibinfo {title} {3d loop models and the ${\mathrm{cp}}^{n\ensuremath{-}1}$
  sigma model},\ }\href {https://doi.org/10.1103/PhysRevLett.107.110601}
  {\bibfield  {journal} {\bibinfo  {journal} {Phys. Rev. Lett.}\ }\textbf
  {\bibinfo {volume} {107}},\ \bibinfo {pages} {110601} (\bibinfo {year}
  {2011})}\BibitemShut {NoStop}%
\bibitem [{\citenamefont {Pujari}\ \emph {et~al.}(2013)\citenamefont {Pujari},
  \citenamefont {Damle},\ and\ \citenamefont {Alet}}]{PhysRevLett.111.087203}%
  \BibitemOpen
  \bibfield  {author} {\bibinfo {author} {\bibfnamefont {S.}~\bibnamefont
  {Pujari}}, \bibinfo {author} {\bibfnamefont {K.}~\bibnamefont {Damle}},\ and\
  \bibinfo {author} {\bibfnamefont {F.}~\bibnamefont {Alet}},\ }\bibfield
  {title} {\bibinfo {title} {N\'eel-state to valence-bond-solid transition on
  the honeycomb lattice: Evidence for deconfined criticality},\ }\href
  {https://doi.org/10.1103/PhysRevLett.111.087203} {\bibfield  {journal}
  {\bibinfo  {journal} {Phys. Rev. Lett.}\ }\textbf {\bibinfo {volume} {111}},\
  \bibinfo {pages} {087203} (\bibinfo {year} {2013})}\BibitemShut {NoStop}%
\bibitem [{\citenamefont {Nahum}\ \emph {et~al.}(2015)\citenamefont {Nahum},
  \citenamefont {Chalker}, \citenamefont {Serna}, \citenamefont {Ortu\~no},\
  and\ \citenamefont {Somoza}}]{PhysRevX.5.041048}%
  \BibitemOpen
  \bibfield  {author} {\bibinfo {author} {\bibfnamefont {A.}~\bibnamefont
  {Nahum}}, \bibinfo {author} {\bibfnamefont {J.~T.}\ \bibnamefont {Chalker}},
  \bibinfo {author} {\bibfnamefont {P.}~\bibnamefont {Serna}}, \bibinfo
  {author} {\bibfnamefont {M.}~\bibnamefont {Ortu\~no}},\ and\ \bibinfo
  {author} {\bibfnamefont {A.~M.}\ \bibnamefont {Somoza}},\ }\bibfield  {title}
  {\bibinfo {title} {Deconfined quantum criticality, scaling violations, and
  classical loop models},\ }\href {https://doi.org/10.1103/PhysRevX.5.041048}
  {\bibfield  {journal} {\bibinfo  {journal} {Phys. Rev. X}\ }\textbf {\bibinfo
  {volume} {5}},\ \bibinfo {pages} {041048} (\bibinfo {year}
  {2015})}\BibitemShut {NoStop}%
\bibitem [{\citenamefont {Shao}\ \emph {et~al.}(2016)\citenamefont {Shao},
  \citenamefont {Guo},\ and\ \citenamefont
  {Sandvik}}]{doi:10.1126/science.aad5007}%
  \BibitemOpen
  \bibfield  {author} {\bibinfo {author} {\bibfnamefont {H.}~\bibnamefont
  {Shao}}, \bibinfo {author} {\bibfnamefont {W.}~\bibnamefont {Guo}},\ and\
  \bibinfo {author} {\bibfnamefont {A.~W.}\ \bibnamefont {Sandvik}},\
  }\bibfield  {title} {\bibinfo {title} {Quantum criticality with two length
  scales},\ }\href {https://doi.org/10.1126/science.aad5007} {\bibfield
  {journal} {\bibinfo  {journal} {Science}\ }\textbf {\bibinfo {volume}
  {352}},\ \bibinfo {pages} {213} (\bibinfo {year} {2016})},\ \Eprint
  {https://arxiv.org/abs/https://www.science.org/doi/pdf/10.1126/science.aad5007}
  {https://www.science.org/doi/pdf/10.1126/science.aad5007} \BibitemShut
  {NoStop}%
\bibitem [{\citenamefont {Wang}\ \emph {et~al.}(2017)\citenamefont {Wang},
  \citenamefont {Nahum}, \citenamefont {Metlitski}, \citenamefont {Xu},\ and\
  \citenamefont {Senthil}}]{PhysRevX.7.031051}%
  \BibitemOpen
  \bibfield  {author} {\bibinfo {author} {\bibfnamefont {C.}~\bibnamefont
  {Wang}}, \bibinfo {author} {\bibfnamefont {A.}~\bibnamefont {Nahum}},
  \bibinfo {author} {\bibfnamefont {M.~A.}\ \bibnamefont {Metlitski}}, \bibinfo
  {author} {\bibfnamefont {C.}~\bibnamefont {Xu}},\ and\ \bibinfo {author}
  {\bibfnamefont {T.}~\bibnamefont {Senthil}},\ }\bibfield  {title} {\bibinfo
  {title} {Deconfined quantum critical points: Symmetries and dualities},\
  }\href {https://doi.org/10.1103/PhysRevX.7.031051} {\bibfield  {journal}
  {\bibinfo  {journal} {Phys. Rev. X}\ }\textbf {\bibinfo {volume} {7}},\
  \bibinfo {pages} {031051} (\bibinfo {year} {2017})}\BibitemShut {NoStop}%
\bibitem [{\citenamefont {Huang}\ and\ \citenamefont
  {Lee}(2022{\natexlab{a}})}]{huang2022competing}%
  \BibitemOpen
  \bibfield  {author} {\bibinfo {author} {\bibfnamefont {Y.-T.}\ \bibnamefont
  {Huang}}\ and\ \bibinfo {author} {\bibfnamefont {D.-H.}\ \bibnamefont
  {Lee}},\ }\href {https://doi.org/10.48550/ARXIV.2204.12485} {\bibinfo {title}
  {Competing orders, the wess-zumino-witten term, and spin liquids}} (\bibinfo
  {year} {2022}{\natexlab{a}})\BibitemShut {NoStop}%
\bibitem [{\citenamefont {Huang}\ and\ \citenamefont
  {Lee}(2022{\natexlab{b}})}]{huang2022non}%
  \BibitemOpen
  \bibfield  {author} {\bibinfo {author} {\bibfnamefont {Y.-T.}\ \bibnamefont
  {Huang}}\ and\ \bibinfo {author} {\bibfnamefont {D.-H.}\ \bibnamefont
  {Lee}},\ }\href {https://doi.org/10.48550/ARXIV.2206.09393} {\bibinfo {title}
  {Non-abelian bosonization of fermion symmetry-protected topological states}}
  (\bibinfo {year} {2022}{\natexlab{b}})\BibitemShut {NoStop}%
\bibitem [{\citenamefont {Lee}\ \emph {et~al.}(2006)\citenamefont {Lee},
  \citenamefont {Nagaosa},\ and\ \citenamefont {Wen}}]{RevModPhys.78.17}%
  \BibitemOpen
  \bibfield  {author} {\bibinfo {author} {\bibfnamefont {P.~A.}\ \bibnamefont
  {Lee}}, \bibinfo {author} {\bibfnamefont {N.}~\bibnamefont {Nagaosa}},\ and\
  \bibinfo {author} {\bibfnamefont {X.-G.}\ \bibnamefont {Wen}},\ }\bibfield
  {title} {\bibinfo {title} {Doping a mott insulator: Physics of
  high-temperature superconductivity},\ }\href
  {https://doi.org/10.1103/RevModPhys.78.17} {\bibfield  {journal} {\bibinfo
  {journal} {Rev. Mod. Phys.}\ }\textbf {\bibinfo {volume} {78}},\ \bibinfo
  {pages} {17} (\bibinfo {year} {2006})}\BibitemShut {NoStop}%
\bibitem [{\citenamefont {Hastings}(2000)}]{PhysRevB.63.014413}%
  \BibitemOpen
  \bibfield  {author} {\bibinfo {author} {\bibfnamefont {M.~B.}\ \bibnamefont
  {Hastings}},\ }\bibfield  {title} {\bibinfo {title} {Dirac structure, rvb,
  and goldstone modes in the kagom\'e antiferromagnet},\ }\href
  {https://doi.org/10.1103/PhysRevB.63.014413} {\bibfield  {journal} {\bibinfo
  {journal} {Phys. Rev. B}\ }\textbf {\bibinfo {volume} {63}},\ \bibinfo
  {pages} {014413} (\bibinfo {year} {2000})}\BibitemShut {NoStop}%
\bibitem [{\citenamefont {Hermele}\ \emph {et~al.}(2005)\citenamefont
  {Hermele}, \citenamefont {Senthil},\ and\ \citenamefont
  {Fisher}}]{PhysRevB.72.104404}%
  \BibitemOpen
  \bibfield  {author} {\bibinfo {author} {\bibfnamefont {M.}~\bibnamefont
  {Hermele}}, \bibinfo {author} {\bibfnamefont {T.}~\bibnamefont {Senthil}},\
  and\ \bibinfo {author} {\bibfnamefont {M.~P.~A.}\ \bibnamefont {Fisher}},\
  }\bibfield  {title} {\bibinfo {title} {Algebraic spin liquid as the mother of
  many competing orders},\ }\href {https://doi.org/10.1103/PhysRevB.72.104404}
  {\bibfield  {journal} {\bibinfo  {journal} {Phys. Rev. B}\ }\textbf {\bibinfo
  {volume} {72}},\ \bibinfo {pages} {104404} (\bibinfo {year}
  {2005})}\BibitemShut {NoStop}%
\bibitem [{\citenamefont {Hermele}\ \emph {et~al.}(2008)\citenamefont
  {Hermele}, \citenamefont {Ran}, \citenamefont {Lee},\ and\ \citenamefont
  {Wen}}]{PhysRevB.77.224413}%
  \BibitemOpen
  \bibfield  {author} {\bibinfo {author} {\bibfnamefont {M.}~\bibnamefont
  {Hermele}}, \bibinfo {author} {\bibfnamefont {Y.}~\bibnamefont {Ran}},
  \bibinfo {author} {\bibfnamefont {P.~A.}\ \bibnamefont {Lee}},\ and\ \bibinfo
  {author} {\bibfnamefont {X.-G.}\ \bibnamefont {Wen}},\ }\bibfield  {title}
  {\bibinfo {title} {Properties of an algebraic spin liquid on the kagome
  lattice},\ }\href {https://doi.org/10.1103/PhysRevB.77.224413} {\bibfield
  {journal} {\bibinfo  {journal} {Phys. Rev. B}\ }\textbf {\bibinfo {volume}
  {77}},\ \bibinfo {pages} {224413} (\bibinfo {year} {2008})}\BibitemShut
  {NoStop}%
\bibitem [{\citenamefont {Ran}\ \emph {et~al.}(2007)\citenamefont {Ran},
  \citenamefont {Hermele}, \citenamefont {Lee},\ and\ \citenamefont
  {Wen}}]{PhysRevLett.98.117205}%
  \BibitemOpen
  \bibfield  {author} {\bibinfo {author} {\bibfnamefont {Y.}~\bibnamefont
  {Ran}}, \bibinfo {author} {\bibfnamefont {M.}~\bibnamefont {Hermele}},
  \bibinfo {author} {\bibfnamefont {P.~A.}\ \bibnamefont {Lee}},\ and\ \bibinfo
  {author} {\bibfnamefont {X.-G.}\ \bibnamefont {Wen}},\ }\bibfield  {title}
  {\bibinfo {title} {Projected-wave-function study of the spin-$1/2$ heisenberg
  model on the kagom\'e lattice},\ }\href
  {https://doi.org/10.1103/PhysRevLett.98.117205} {\bibfield  {journal}
  {\bibinfo  {journal} {Phys. Rev. Lett.}\ }\textbf {\bibinfo {volume} {98}},\
  \bibinfo {pages} {117205} (\bibinfo {year} {2007})}\BibitemShut {NoStop}%
\bibitem [{\citenamefont {Iqbal}\ \emph {et~al.}(2016)\citenamefont {Iqbal},
  \citenamefont {Hu}, \citenamefont {Thomale}, \citenamefont {Poilblanc},\ and\
  \citenamefont {Becca}}]{PhysRevB.93.144411}%
  \BibitemOpen
  \bibfield  {author} {\bibinfo {author} {\bibfnamefont {Y.}~\bibnamefont
  {Iqbal}}, \bibinfo {author} {\bibfnamefont {W.-J.}\ \bibnamefont {Hu}},
  \bibinfo {author} {\bibfnamefont {R.}~\bibnamefont {Thomale}}, \bibinfo
  {author} {\bibfnamefont {D.}~\bibnamefont {Poilblanc}},\ and\ \bibinfo
  {author} {\bibfnamefont {F.}~\bibnamefont {Becca}},\ }\bibfield  {title}
  {\bibinfo {title} {Spin liquid nature in the heisenberg
  ${J}_{1}\ensuremath{-}{J}_{2}$ triangular antiferromagnet},\ }\href
  {https://doi.org/10.1103/PhysRevB.93.144411} {\bibfield  {journal} {\bibinfo
  {journal} {Phys. Rev. B}\ }\textbf {\bibinfo {volume} {93}},\ \bibinfo
  {pages} {144411} (\bibinfo {year} {2016})}\BibitemShut {NoStop}%
\bibitem [{\citenamefont {Zhu}\ \emph {et~al.}(2018)\citenamefont {Zhu},
  \citenamefont {Chen}, \citenamefont {He},\ and\ \citenamefont
  {Witczak-Krempa}}]{doi:10.1126/sciadv.aat5535}%
  \BibitemOpen
  \bibfield  {author} {\bibinfo {author} {\bibfnamefont {W.}~\bibnamefont
  {Zhu}}, \bibinfo {author} {\bibfnamefont {X.}~\bibnamefont {Chen}}, \bibinfo
  {author} {\bibfnamefont {Y.-C.}\ \bibnamefont {He}},\ and\ \bibinfo {author}
  {\bibfnamefont {W.}~\bibnamefont {Witczak-Krempa}},\ }\bibfield  {title}
  {\bibinfo {title} {Entanglement signatures of emergent dirac fermions: Kagome
  spin liquid and quantum criticality},\ }\href
  {https://doi.org/10.1126/sciadv.aat5535} {\bibfield  {journal} {\bibinfo
  {journal} {Science Advances}\ }\textbf {\bibinfo {volume} {4}},\ \bibinfo
  {pages} {eaat5535} (\bibinfo {year} {2018})},\ \Eprint
  {https://arxiv.org/abs/https://www.science.org/doi/pdf/10.1126/sciadv.aat5535}
  {https://www.science.org/doi/pdf/10.1126/sciadv.aat5535} \BibitemShut
  {NoStop}%
\bibitem [{\citenamefont {Song}\ \emph
  {et~al.}(2020{\natexlab{a}})\citenamefont {Song}, \citenamefont {He},
  \citenamefont {Vishwanath},\ and\ \citenamefont {Wang}}]{PhysRevX.10.011033}%
  \BibitemOpen
  \bibfield  {author} {\bibinfo {author} {\bibfnamefont {X.-Y.}\ \bibnamefont
  {Song}}, \bibinfo {author} {\bibfnamefont {Y.-C.}\ \bibnamefont {He}},
  \bibinfo {author} {\bibfnamefont {A.}~\bibnamefont {Vishwanath}},\ and\
  \bibinfo {author} {\bibfnamefont {C.}~\bibnamefont {Wang}},\ }\bibfield
  {title} {\bibinfo {title} {From spinon band topology to the symmetry quantum
  numbers of monopoles in dirac spin liquids},\ }\href
  {https://doi.org/10.1103/PhysRevX.10.011033} {\bibfield  {journal} {\bibinfo
  {journal} {Phys. Rev. X}\ }\textbf {\bibinfo {volume} {10}},\ \bibinfo
  {pages} {011033} (\bibinfo {year} {2020}{\natexlab{a}})}\BibitemShut
  {NoStop}%
\bibitem [{\citenamefont {Song}\ \emph {et~al.}(2019)\citenamefont {Song},
  \citenamefont {Wang}, \citenamefont {Vishwanath},\ and\ \citenamefont
  {He}}]{song2019unifying}%
  \BibitemOpen
  \bibfield  {author} {\bibinfo {author} {\bibfnamefont {X.-Y.}\ \bibnamefont
  {Song}}, \bibinfo {author} {\bibfnamefont {C.}~\bibnamefont {Wang}}, \bibinfo
  {author} {\bibfnamefont {A.}~\bibnamefont {Vishwanath}},\ and\ \bibinfo
  {author} {\bibfnamefont {Y.-C.}\ \bibnamefont {He}},\ }\bibfield  {title}
  {\bibinfo {title} {Unifying description of competing orders in
  two-dimensional quantum magnets},\ }\href
  {https://doi.org/10.1038/s41467-019-11727-3} {\bibfield  {journal} {\bibinfo
  {journal} {Nature Communications}\ }\textbf {\bibinfo {volume} {10}},\
  \bibinfo {pages} {4254} (\bibinfo {year} {2019})}\BibitemShut {NoStop}%
\bibitem [{\citenamefont {Calvera}\ and\ \citenamefont
  {Wang}(2021)}]{https://doi.org/10.48550/arxiv.2103.13405}%
  \BibitemOpen
  \bibfield  {author} {\bibinfo {author} {\bibfnamefont {V.}~\bibnamefont
  {Calvera}}\ and\ \bibinfo {author} {\bibfnamefont {C.}~\bibnamefont {Wang}},\
  }\href {https://doi.org/10.48550/ARXIV.2103.13405} {\bibinfo {title} {Theory
  of dirac spin-orbital liquids: monopoles, anomalies, and applications to
  $su(4)$ honeycomb models}} (\bibinfo {year} {2021})\BibitemShut {NoStop}%
\bibitem [{\citenamefont {He}\ \emph {et~al.}(2017)\citenamefont {He},
  \citenamefont {Zaletel}, \citenamefont {Oshikawa},\ and\ \citenamefont
  {Pollmann}}]{PhysRevX.7.031020}%
  \BibitemOpen
  \bibfield  {author} {\bibinfo {author} {\bibfnamefont {Y.-C.}\ \bibnamefont
  {He}}, \bibinfo {author} {\bibfnamefont {M.~P.}\ \bibnamefont {Zaletel}},
  \bibinfo {author} {\bibfnamefont {M.}~\bibnamefont {Oshikawa}},\ and\
  \bibinfo {author} {\bibfnamefont {F.}~\bibnamefont {Pollmann}},\ }\bibfield
  {title} {\bibinfo {title} {Signatures of dirac cones in a dmrg study of the
  kagome heisenberg model},\ }\href {https://doi.org/10.1103/PhysRevX.7.031020}
  {\bibfield  {journal} {\bibinfo  {journal} {Phys. Rev. X}\ }\textbf {\bibinfo
  {volume} {7}},\ \bibinfo {pages} {031020} (\bibinfo {year}
  {2017})}\BibitemShut {NoStop}%
\bibitem [{\citenamefont {Ma}\ and\ \citenamefont
  {Wang}(2020)}]{PhysRevB.102.020407}%
  \BibitemOpen
  \bibfield  {author} {\bibinfo {author} {\bibfnamefont {R.}~\bibnamefont
  {Ma}}\ and\ \bibinfo {author} {\bibfnamefont {C.}~\bibnamefont {Wang}},\
  }\bibfield  {title} {\bibinfo {title} {Theory of deconfined
  pseudocriticality},\ }\href {https://doi.org/10.1103/PhysRevB.102.020407}
  {\bibfield  {journal} {\bibinfo  {journal} {Phys. Rev. B}\ }\textbf {\bibinfo
  {volume} {102}},\ \bibinfo {pages} {020407} (\bibinfo {year}
  {2020})}\BibitemShut {NoStop}%
\bibitem [{\citenamefont {Karthik}\ and\ \citenamefont
  {Narayanan}(2016{\natexlab{a}})}]{PhysRevD.94.065026}%
  \BibitemOpen
  \bibfield  {author} {\bibinfo {author} {\bibfnamefont {N.}~\bibnamefont
  {Karthik}}\ and\ \bibinfo {author} {\bibfnamefont {R.}~\bibnamefont
  {Narayanan}},\ }\bibfield  {title} {\bibinfo {title} {Scale invariance of
  parity-invariant three-dimensional qed},\ }\href
  {https://doi.org/10.1103/PhysRevD.94.065026} {\bibfield  {journal} {\bibinfo
  {journal} {Phys. Rev. D}\ }\textbf {\bibinfo {volume} {94}},\ \bibinfo
  {pages} {065026} (\bibinfo {year} {2016}{\natexlab{a}})}\BibitemShut
  {NoStop}%
\bibitem [{\citenamefont {Karthik}\ and\ \citenamefont
  {Narayanan}(2016{\natexlab{b}})}]{PhysRevD.93.045020}%
  \BibitemOpen
  \bibfield  {author} {\bibinfo {author} {\bibfnamefont {N.}~\bibnamefont
  {Karthik}}\ and\ \bibinfo {author} {\bibfnamefont {R.}~\bibnamefont
  {Narayanan}},\ }\bibfield  {title} {\bibinfo {title} {No evidence for
  bilinear condensate in parity-invariant three-dimensional qed with massless
  fermions},\ }\href {https://doi.org/10.1103/PhysRevD.93.045020} {\bibfield
  {journal} {\bibinfo  {journal} {Phys. Rev. D}\ }\textbf {\bibinfo {volume}
  {93}},\ \bibinfo {pages} {045020} (\bibinfo {year}
  {2016}{\natexlab{b}})}\BibitemShut {NoStop}%
\bibitem [{\citenamefont {Karthik}\ and\ \citenamefont
  {Narayanan}(2019)}]{PhysRevD.100.054514}%
  \BibitemOpen
  \bibfield  {author} {\bibinfo {author} {\bibfnamefont {N.}~\bibnamefont
  {Karthik}}\ and\ \bibinfo {author} {\bibfnamefont {R.}~\bibnamefont
  {Narayanan}},\ }\bibfield  {title} {\bibinfo {title} {Numerical determination
  of monopole scaling dimension in parity-invariant three-dimensional
  noncompact qed},\ }\href {https://doi.org/10.1103/PhysRevD.100.054514}
  {\bibfield  {journal} {\bibinfo  {journal} {Phys. Rev. D}\ }\textbf {\bibinfo
  {volume} {100}},\ \bibinfo {pages} {054514} (\bibinfo {year}
  {2019})}\BibitemShut {NoStop}%
\bibitem [{\citenamefont {Senthil}\ and\ \citenamefont
  {Fisher}(2006)}]{PhysRevB.74.064405}%
  \BibitemOpen
  \bibfield  {author} {\bibinfo {author} {\bibfnamefont {T.}~\bibnamefont
  {Senthil}}\ and\ \bibinfo {author} {\bibfnamefont {M.~P.~A.}\ \bibnamefont
  {Fisher}},\ }\bibfield  {title} {\bibinfo {title} {Competing orders,
  nonlinear sigma models, and topological terms in quantum magnets},\ }\href
  {https://doi.org/10.1103/PhysRevB.74.064405} {\bibfield  {journal} {\bibinfo
  {journal} {Phys. Rev. B}\ }\textbf {\bibinfo {volume} {74}},\ \bibinfo
  {pages} {064405} (\bibinfo {year} {2006})}\BibitemShut {NoStop}%
\bibitem [{\citenamefont {Zou}\ \emph {et~al.}(2021)\citenamefont {Zou},
  \citenamefont {He},\ and\ \citenamefont {Wang}}]{PhysRevX.11.031043}%
  \BibitemOpen
  \bibfield  {author} {\bibinfo {author} {\bibfnamefont {L.}~\bibnamefont
  {Zou}}, \bibinfo {author} {\bibfnamefont {Y.-C.}\ \bibnamefont {He}},\ and\
  \bibinfo {author} {\bibfnamefont {C.}~\bibnamefont {Wang}},\ }\bibfield
  {title} {\bibinfo {title} {Stiefel liquids: Possible non-lagrangian quantum
  criticality from intertwined orders},\ }\href
  {https://doi.org/10.1103/PhysRevX.11.031043} {\bibfield  {journal} {\bibinfo
  {journal} {Phys. Rev. X}\ }\textbf {\bibinfo {volume} {11}},\ \bibinfo
  {pages} {031043} (\bibinfo {year} {2021})}\BibitemShut {NoStop}%
\bibitem [{\citenamefont {Abanov}\ and\ \citenamefont
  {Wiegmann}(2000)}]{ABANOV2000685}%
  \BibitemOpen
  \bibfield  {author} {\bibinfo {author} {\bibfnamefont {A.}~\bibnamefont
  {Abanov}}\ and\ \bibinfo {author} {\bibfnamefont {P.}~\bibnamefont
  {Wiegmann}},\ }\bibfield  {title} {\bibinfo {title} {Theta-terms in nonlinear
  sigma-models},\ }\href
  {https://doi.org/https://doi.org/10.1016/S0550-3213(99)00820-2} {\bibfield
  {journal} {\bibinfo  {journal} {Nuclear Physics B}\ }\textbf {\bibinfo
  {volume} {570}},\ \bibinfo {pages} {685} (\bibinfo {year}
  {2000})}\BibitemShut {NoStop}%
\bibitem [{\citenamefont {Jian}\ \emph {et~al.}(2018)\citenamefont {Jian},
  \citenamefont {Thomson}, \citenamefont {Rasmussen}, \citenamefont {Bi},\ and\
  \citenamefont {Xu}}]{PhysRevB.97.195115}%
  \BibitemOpen
  \bibfield  {author} {\bibinfo {author} {\bibfnamefont {C.-M.}\ \bibnamefont
  {Jian}}, \bibinfo {author} {\bibfnamefont {A.}~\bibnamefont {Thomson}},
  \bibinfo {author} {\bibfnamefont {A.}~\bibnamefont {Rasmussen}}, \bibinfo
  {author} {\bibfnamefont {Z.}~\bibnamefont {Bi}},\ and\ \bibinfo {author}
  {\bibfnamefont {C.}~\bibnamefont {Xu}},\ }\bibfield  {title} {\bibinfo
  {title} {Deconfined quantum critical point on the triangular lattice},\
  }\href {https://doi.org/10.1103/PhysRevB.97.195115} {\bibfield  {journal}
  {\bibinfo  {journal} {Phys. Rev. B}\ }\textbf {\bibinfo {volume} {97}},\
  \bibinfo {pages} {195115} (\bibinfo {year} {2018})}\BibitemShut {NoStop}%
\bibitem [{\citenamefont {Fisher}\ and\ \citenamefont
  {Lee}(1989)}]{PhysRevB.39.2756}%
  \BibitemOpen
  \bibfield  {author} {\bibinfo {author} {\bibfnamefont {M.~P.~A.}\
  \bibnamefont {Fisher}}\ and\ \bibinfo {author} {\bibfnamefont {D.~H.}\
  \bibnamefont {Lee}},\ }\bibfield  {title} {\bibinfo {title} {Correspondence
  between two-dimensional bosons and a bulk superconductor in a magnetic
  field},\ }\href {https://doi.org/10.1103/PhysRevB.39.2756} {\bibfield
  {journal} {\bibinfo  {journal} {Phys. Rev. B}\ }\textbf {\bibinfo {volume}
  {39}},\ \bibinfo {pages} {2756} (\bibinfo {year} {1989})}\BibitemShut
  {NoStop}%
\bibitem [{\citenamefont {Dasgupta}\ and\ \citenamefont
  {Halperin}(1981)}]{PhysRevLett.47.1556}%
  \BibitemOpen
  \bibfield  {author} {\bibinfo {author} {\bibfnamefont {C.}~\bibnamefont
  {Dasgupta}}\ and\ \bibinfo {author} {\bibfnamefont {B.~I.}\ \bibnamefont
  {Halperin}},\ }\bibfield  {title} {\bibinfo {title} {Phase transition in a
  lattice model of superconductivity},\ }\href
  {https://doi.org/10.1103/PhysRevLett.47.1556} {\bibfield  {journal} {\bibinfo
   {journal} {Phys. Rev. Lett.}\ }\textbf {\bibinfo {volume} {47}},\ \bibinfo
  {pages} {1556} (\bibinfo {year} {1981})}\BibitemShut {NoStop}%
\bibitem [{\citenamefont {Peskin}(1978)}]{PESKIN1978122}%
  \BibitemOpen
  \bibfield  {author} {\bibinfo {author} {\bibfnamefont {M.~E.}\ \bibnamefont
  {Peskin}},\ }\bibfield  {title} {\bibinfo {title} {Mandelstam-'t hooft
  duality in abelian lattice models},\ }\href
  {https://doi.org/https://doi.org/10.1016/0003-4916(78)90252-X} {\bibfield
  {journal} {\bibinfo  {journal} {Annals of Physics}\ }\textbf {\bibinfo
  {volume} {113}},\ \bibinfo {pages} {122} (\bibinfo {year}
  {1978})}\BibitemShut {NoStop}%
\bibitem [{\citenamefont {Karch}\ and\ \citenamefont
  {Tong}(2016)}]{PhysRevX.6.031043}%
  \BibitemOpen
  \bibfield  {author} {\bibinfo {author} {\bibfnamefont {A.}~\bibnamefont
  {Karch}}\ and\ \bibinfo {author} {\bibfnamefont {D.}~\bibnamefont {Tong}},\
  }\bibfield  {title} {\bibinfo {title} {Particle-vortex duality from 3d
  bosonization},\ }\href {https://doi.org/10.1103/PhysRevX.6.031043} {\bibfield
   {journal} {\bibinfo  {journal} {Phys. Rev. X}\ }\textbf {\bibinfo {volume}
  {6}},\ \bibinfo {pages} {031043} (\bibinfo {year} {2016})}\BibitemShut
  {NoStop}%
\bibitem [{\citenamefont {Seiberg}\ \emph {et~al.}(2016)\citenamefont
  {Seiberg}, \citenamefont {Senthil}, \citenamefont {Wang},\ and\ \citenamefont
  {Witten}}]{SEIBERG2016395}%
  \BibitemOpen
  \bibfield  {author} {\bibinfo {author} {\bibfnamefont {N.}~\bibnamefont
  {Seiberg}}, \bibinfo {author} {\bibfnamefont {T.}~\bibnamefont {Senthil}},
  \bibinfo {author} {\bibfnamefont {C.}~\bibnamefont {Wang}},\ and\ \bibinfo
  {author} {\bibfnamefont {E.}~\bibnamefont {Witten}},\ }\bibfield  {title}
  {\bibinfo {title} {A duality web in 2+1 dimensions and condensed matter
  physics},\ }\href {https://doi.org/https://doi.org/10.1016/j.aop.2016.08.007}
  {\bibfield  {journal} {\bibinfo  {journal} {Annals of Physics}\ }\textbf
  {\bibinfo {volume} {374}},\ \bibinfo {pages} {395} (\bibinfo {year}
  {2016})}\BibitemShut {NoStop}%
\bibitem [{\citenamefont {Son}(2015)}]{PhysRevX.5.031027}%
  \BibitemOpen
  \bibfield  {author} {\bibinfo {author} {\bibfnamefont {D.~T.}\ \bibnamefont
  {Son}},\ }\bibfield  {title} {\bibinfo {title} {Is the composite fermion a
  dirac particle?},\ }\href {https://doi.org/10.1103/PhysRevX.5.031027}
  {\bibfield  {journal} {\bibinfo  {journal} {Phys. Rev. X}\ }\textbf {\bibinfo
  {volume} {5}},\ \bibinfo {pages} {031027} (\bibinfo {year}
  {2015})}\BibitemShut {NoStop}%
\bibitem [{\citenamefont {Metlitski}\ and\ \citenamefont
  {Vishwanath}(2016)}]{PhysRevB.93.245151}%
  \BibitemOpen
  \bibfield  {author} {\bibinfo {author} {\bibfnamefont {M.~A.}\ \bibnamefont
  {Metlitski}}\ and\ \bibinfo {author} {\bibfnamefont {A.}~\bibnamefont
  {Vishwanath}},\ }\bibfield  {title} {\bibinfo {title} {Particle-vortex
  duality of two-dimensional dirac fermion from electric-magnetic duality of
  three-dimensional topological insulators},\ }\href
  {https://doi.org/10.1103/PhysRevB.93.245151} {\bibfield  {journal} {\bibinfo
  {journal} {Phys. Rev. B}\ }\textbf {\bibinfo {volume} {93}},\ \bibinfo
  {pages} {245151} (\bibinfo {year} {2016})}\BibitemShut {NoStop}%
\bibitem [{\citenamefont {Wang}\ and\ \citenamefont
  {Senthil}(2015)}]{PhysRevX.5.041031}%
  \BibitemOpen
  \bibfield  {author} {\bibinfo {author} {\bibfnamefont {C.}~\bibnamefont
  {Wang}}\ and\ \bibinfo {author} {\bibfnamefont {T.}~\bibnamefont {Senthil}},\
  }\bibfield  {title} {\bibinfo {title} {Dual dirac liquid on the surface of
  the electron topological insulator},\ }\href
  {https://doi.org/10.1103/PhysRevX.5.041031} {\bibfield  {journal} {\bibinfo
  {journal} {Phys. Rev. X}\ }\textbf {\bibinfo {volume} {5}},\ \bibinfo {pages}
  {041031} (\bibinfo {year} {2015})}\BibitemShut {NoStop}%
\bibitem [{\citenamefont {POLYAKOV}(1988)}]{polyakov1988fermi}%
  \BibitemOpen
  \bibfield  {author} {\bibinfo {author} {\bibfnamefont {A.}~\bibnamefont
  {POLYAKOV}},\ }\bibfield  {title} {\bibinfo {title} {Fermi-bose
  transmutations induced by gauge fields},\ }\href
  {https://doi.org/10.1142/S0217732388000398} {\bibfield  {journal} {\bibinfo
  {journal} {Modern Physics Letters A}\ }\textbf {\bibinfo {volume} {03}},\
  \bibinfo {pages} {325} (\bibinfo {year} {1988})},\ \Eprint
  {https://arxiv.org/abs/https://doi.org/10.1142/S0217732388000398}
  {https://doi.org/10.1142/S0217732388000398} \BibitemShut {NoStop}%
\bibitem [{\citenamefont {SHAJI}\ \emph {et~al.}(1990)\citenamefont {SHAJI},
  \citenamefont {SHANKAR},\ and\ \citenamefont {SIVAKUMAR}}]{shaji1990bose}%
  \BibitemOpen
  \bibfield  {author} {\bibinfo {author} {\bibfnamefont {N.}~\bibnamefont
  {SHAJI}}, \bibinfo {author} {\bibfnamefont {R.}~\bibnamefont {SHANKAR}},\
  and\ \bibinfo {author} {\bibfnamefont {M.}~\bibnamefont {SIVAKUMAR}},\
  }\bibfield  {title} {\bibinfo {title} {On bose-fermi equivalence in a u(1)
  gauge theory with chern-simons action},\ }\href
  {https://doi.org/10.1142/S0217732390000664} {\bibfield  {journal} {\bibinfo
  {journal} {Modern Physics Letters A}\ }\textbf {\bibinfo {volume} {05}},\
  \bibinfo {pages} {593} (\bibinfo {year} {1990})},\ \Eprint
  {https://arxiv.org/abs/https://doi.org/10.1142/S0217732390000664}
  {https://doi.org/10.1142/S0217732390000664} \BibitemShut {NoStop}%
\bibitem [{\citenamefont {Chen}\ \emph {et~al.}(2018)\citenamefont {Chen},
  \citenamefont {Son}, \citenamefont {Wang},\ and\ \citenamefont
  {Raghu}}]{PhysRevLett.120.016602}%
  \BibitemOpen
  \bibfield  {author} {\bibinfo {author} {\bibfnamefont {J.-Y.}\ \bibnamefont
  {Chen}}, \bibinfo {author} {\bibfnamefont {J.~H.}\ \bibnamefont {Son}},
  \bibinfo {author} {\bibfnamefont {C.}~\bibnamefont {Wang}},\ and\ \bibinfo
  {author} {\bibfnamefont {S.}~\bibnamefont {Raghu}},\ }\bibfield  {title}
  {\bibinfo {title} {Exact boson-fermion duality on a 3d euclidean lattice},\
  }\href {https://doi.org/10.1103/PhysRevLett.120.016602} {\bibfield  {journal}
  {\bibinfo  {journal} {Phys. Rev. Lett.}\ }\textbf {\bibinfo {volume} {120}},\
  \bibinfo {pages} {016602} (\bibinfo {year} {2018})}\BibitemShut {NoStop}%
\bibitem [{\citenamefont {Nagaosa}\ and\ \citenamefont
  {Oshikawa}(1996)}]{nagaosa1996chiral}%
  \BibitemOpen
  \bibfield  {author} {\bibinfo {author} {\bibfnamefont {N.}~\bibnamefont
  {Nagaosa}}\ and\ \bibinfo {author} {\bibfnamefont {M.}~\bibnamefont
  {Oshikawa}},\ }\bibfield  {title} {\bibinfo {title} {Chiral anomaly and spin
  gap in one-dimensional interacting fermions},\ }\href
  {https://doi.org/10.1143/JPSJ.65.2241} {\bibfield  {journal} {\bibinfo
  {journal} {Journal of the Physical Society of Japan}\ }\textbf {\bibinfo
  {volume} {65}},\ \bibinfo {pages} {2241} (\bibinfo {year} {1996})},\ \Eprint
  {https://arxiv.org/abs/https://doi.org/10.1143/JPSJ.65.2241}
  {https://doi.org/10.1143/JPSJ.65.2241} \BibitemShut {NoStop}%
\bibitem [{\citenamefont {Yao}\ and\ \citenamefont
  {Lee}(2010)}]{PhysRevB.82.245117}%
  \BibitemOpen
  \bibfield  {author} {\bibinfo {author} {\bibfnamefont {H.}~\bibnamefont
  {Yao}}\ and\ \bibinfo {author} {\bibfnamefont {D.-H.}\ \bibnamefont {Lee}},\
  }\bibfield  {title} {\bibinfo {title} {Topological insulators and topological
  nonlinear $\ensuremath{\sigma}$ models},\ }\href
  {https://doi.org/10.1103/PhysRevB.82.245117} {\bibfield  {journal} {\bibinfo
  {journal} {Phys. Rev. B}\ }\textbf {\bibinfo {volume} {82}},\ \bibinfo
  {pages} {245117} (\bibinfo {year} {2010})}\BibitemShut {NoStop}%
\bibitem [{Note1()}]{Note1}%
  \BibitemOpen
  \bibinfo {note} {Strictly speaking, the duality here means the two theories
  have the same kinematic properties, i.e. the same local operators, same
  global symmetries and same 't Hooft anomalies, which is also known as the
  'weak duality' in the literature \cite
  {PhysRevX.7.031051,SEIBERG2016395,SENTHIL20191}. If the topological NLSM
  model and the \protect \text {QED} theory flow to the same IR fixed point,
  these two theories have the so-called 'strong duality'\cite
  {PhysRevX.7.031051,SEIBERG2016395,SENTHIL20191}. The determining answer of
  the IR dynamical behaviour of the topological NLSM is beyond the scope of
  this work. But we comment that the topological NLSM is theoretically possible
  to have the same IR behaviour with the $\protect \text {QED}$ theory since
  they share the same kinematic properties, if all the symmetry-allowed
  couplings in the topological NLSM are allowed to tune.}\BibitemShut {Stop}%
\bibitem [{\citenamefont {Qi}\ \emph {et~al.}(2008)\citenamefont {Qi},
  \citenamefont {Hughes},\ and\ \citenamefont {Zhang}}]{PhysRevB.78.195424}%
  \BibitemOpen
  \bibfield  {author} {\bibinfo {author} {\bibfnamefont {X.-L.}\ \bibnamefont
  {Qi}}, \bibinfo {author} {\bibfnamefont {T.~L.}\ \bibnamefont {Hughes}},\
  and\ \bibinfo {author} {\bibfnamefont {S.-C.}\ \bibnamefont {Zhang}},\
  }\bibfield  {title} {\bibinfo {title} {Topological field theory of
  time-reversal invariant insulators},\ }\href
  {https://doi.org/10.1103/PhysRevB.78.195424} {\bibfield  {journal} {\bibinfo
  {journal} {Phys. Rev. B}\ }\textbf {\bibinfo {volume} {78}},\ \bibinfo
  {pages} {195424} (\bibinfo {year} {2008})}\BibitemShut {NoStop}%
\bibitem [{\citenamefont {Ryu}\ \emph {et~al.}(2010)\citenamefont {Ryu},
  \citenamefont {Schnyder}, \citenamefont {Furusaki},\ and\ \citenamefont
  {Ludwig}}]{ryu2010topological}%
  \BibitemOpen
  \bibfield  {author} {\bibinfo {author} {\bibfnamefont {S.}~\bibnamefont
  {Ryu}}, \bibinfo {author} {\bibfnamefont {A.~P.}\ \bibnamefont {Schnyder}},
  \bibinfo {author} {\bibfnamefont {A.}~\bibnamefont {Furusaki}},\ and\
  \bibinfo {author} {\bibfnamefont {A.~W.~W.}\ \bibnamefont {Ludwig}},\
  }\bibfield  {title} {\bibinfo {title} {Topological insulators and
  superconductors: tenfold way and dimensional hierarchy},\ }\href
  {https://doi.org/10.1088/1367-2630/12/6/065010} {\bibfield  {journal}
  {\bibinfo  {journal} {New Journal of Physics}\ }\textbf {\bibinfo {volume}
  {12}},\ \bibinfo {pages} {065010} (\bibinfo {year} {2010})}\BibitemShut
  {NoStop}%
\bibitem [{\citenamefont {Song}\ \emph {et~al.}(2017)\citenamefont {Song},
  \citenamefont {Huang}, \citenamefont {Fu},\ and\ \citenamefont
  {Hermele}}]{PhysRevX.7.011020}%
  \BibitemOpen
  \bibfield  {author} {\bibinfo {author} {\bibfnamefont {H.}~\bibnamefont
  {Song}}, \bibinfo {author} {\bibfnamefont {S.-J.}\ \bibnamefont {Huang}},
  \bibinfo {author} {\bibfnamefont {L.}~\bibnamefont {Fu}},\ and\ \bibinfo
  {author} {\bibfnamefont {M.}~\bibnamefont {Hermele}},\ }\bibfield  {title}
  {\bibinfo {title} {Topological phases protected by point group symmetry},\
  }\href {https://doi.org/10.1103/PhysRevX.7.011020} {\bibfield  {journal}
  {\bibinfo  {journal} {Phys. Rev. X}\ }\textbf {\bibinfo {volume} {7}},\
  \bibinfo {pages} {011020} (\bibinfo {year} {2017})}\BibitemShut {NoStop}%
\bibitem [{\citenamefont {Tantivasadakarn}(2017)}]{PhysRevB.96.195101}%
  \BibitemOpen
  \bibfield  {author} {\bibinfo {author} {\bibfnamefont {N.}~\bibnamefont
  {Tantivasadakarn}},\ }\bibfield  {title} {\bibinfo {title} {Dimensional
  reduction and topological invariants of symmetry-protected topological
  phases},\ }\href {https://doi.org/10.1103/PhysRevB.96.195101} {\bibfield
  {journal} {\bibinfo  {journal} {Phys. Rev. B}\ }\textbf {\bibinfo {volume}
  {96}},\ \bibinfo {pages} {195101} (\bibinfo {year} {2017})}\BibitemShut
  {NoStop}%
\bibitem [{\citenamefont {Rasmussen}\ and\ \citenamefont
  {Lu}(2020)}]{PhysRevB.101.085137}%
  \BibitemOpen
  \bibfield  {author} {\bibinfo {author} {\bibfnamefont {A.}~\bibnamefont
  {Rasmussen}}\ and\ \bibinfo {author} {\bibfnamefont {Y.-M.}\ \bibnamefont
  {Lu}},\ }\bibfield  {title} {\bibinfo {title} {Classification and
  construction of higher-order symmetry-protected topological phases of
  interacting bosons},\ }\href {https://doi.org/10.1103/PhysRevB.101.085137}
  {\bibfield  {journal} {\bibinfo  {journal} {Phys. Rev. B}\ }\textbf {\bibinfo
  {volume} {101}},\ \bibinfo {pages} {085137} (\bibinfo {year}
  {2020})}\BibitemShut {NoStop}%
\bibitem [{\citenamefont {Ning}\ \emph {et~al.}(2021)\citenamefont {Ning},
  \citenamefont {Mao}, \citenamefont {Li},\ and\ \citenamefont
  {Wang}}]{PhysRevB.104.075111}%
  \BibitemOpen
  \bibfield  {author} {\bibinfo {author} {\bibfnamefont {S.-Q.}\ \bibnamefont
  {Ning}}, \bibinfo {author} {\bibfnamefont {B.-B.}\ \bibnamefont {Mao}},
  \bibinfo {author} {\bibfnamefont {Z.}~\bibnamefont {Li}},\ and\ \bibinfo
  {author} {\bibfnamefont {C.}~\bibnamefont {Wang}},\ }\bibfield  {title}
  {\bibinfo {title} {Anomaly indicators and bulk-boundary correspondences for
  three-dimensional interacting topological crystalline phases with mirror and
  continuous symmetries},\ }\href {https://doi.org/10.1103/PhysRevB.104.075111}
  {\bibfield  {journal} {\bibinfo  {journal} {Phys. Rev. B}\ }\textbf {\bibinfo
  {volume} {104}},\ \bibinfo {pages} {075111} (\bibinfo {year}
  {2021})}\BibitemShut {NoStop}%
\bibitem [{\citenamefont {Song}\ \emph
  {et~al.}(2020{\natexlab{b}})\citenamefont {Song}, \citenamefont {Xiong},\
  and\ \citenamefont {Huang}}]{PhysRevB.101.165129}%
  \BibitemOpen
  \bibfield  {author} {\bibinfo {author} {\bibfnamefont {H.}~\bibnamefont
  {Song}}, \bibinfo {author} {\bibfnamefont {C.~Z.}\ \bibnamefont {Xiong}},\
  and\ \bibinfo {author} {\bibfnamefont {S.-J.}\ \bibnamefont {Huang}},\
  }\bibfield  {title} {\bibinfo {title} {Bosonic crystalline symmetry protected
  topological phases beyond the group cohomology proposal},\ }\href
  {https://doi.org/10.1103/PhysRevB.101.165129} {\bibfield  {journal} {\bibinfo
   {journal} {Phys. Rev. B}\ }\textbf {\bibinfo {volume} {101}},\ \bibinfo
  {pages} {165129} (\bibinfo {year} {2020}{\natexlab{b}})}\BibitemShut
  {NoStop}%
\bibitem [{\citenamefont {Goldstone}\ and\ \citenamefont
  {Wilczek}(1981)}]{PhysRevLett.47.986}%
  \BibitemOpen
  \bibfield  {author} {\bibinfo {author} {\bibfnamefont {J.}~\bibnamefont
  {Goldstone}}\ and\ \bibinfo {author} {\bibfnamefont {F.}~\bibnamefont
  {Wilczek}},\ }\bibfield  {title} {\bibinfo {title} {Fractional quantum
  numbers on solitons},\ }\href {https://doi.org/10.1103/PhysRevLett.47.986}
  {\bibfield  {journal} {\bibinfo  {journal} {Phys. Rev. Lett.}\ }\textbf
  {\bibinfo {volume} {47}},\ \bibinfo {pages} {986} (\bibinfo {year}
  {1981})}\BibitemShut {NoStop}%
\bibitem [{\citenamefont {Fujikawa}(1979)}]{PhysRevLett.42.1195}%
  \BibitemOpen
  \bibfield  {author} {\bibinfo {author} {\bibfnamefont {K.}~\bibnamefont
  {Fujikawa}},\ }\bibfield  {title} {\bibinfo {title} {Path-integral measure
  for gauge-invariant fermion theories},\ }\href
  {https://doi.org/10.1103/PhysRevLett.42.1195} {\bibfield  {journal} {\bibinfo
   {journal} {Phys. Rev. Lett.}\ }\textbf {\bibinfo {volume} {42}},\ \bibinfo
  {pages} {1195} (\bibinfo {year} {1979})}\BibitemShut {NoStop}%
\bibitem [{Note2()}]{Note2}%
  \BibitemOpen
  \bibinfo {note} {The fermionic sigma models here are defined as fermions
  coupled with order parameters, as in \cite {PhysRevB.82.245117}.}\BibitemShut
  {Stop}%
\bibitem [{Note3()}]{Note3}%
  \BibitemOpen
  \bibinfo {note} {Here the broken symmetries of order parameters can be
  determined in the IR theory.}\BibitemShut {Stop}%
\bibitem [{Note4()}]{Note4}%
  \BibitemOpen
  \bibinfo {note} {Here for the convenience of notation, we denote the mass
  term as onsite coupling, but our following formalism can be directly applied
  to non-onsite mass terms as long as it preserves the lattice translation
  symmetries.}\BibitemShut {Stop}%
\bibitem [{Note5()}]{Note5}%
  \BibitemOpen
  \bibinfo {note} {For the Dirac fermions and quadratic dispersive fermions in
  IR, this kind of mass term is the only choice, if we impose the continuous
  spatial rotation symmetry.}\BibitemShut {Stop}%
\bibitem [{Note6()}]{Note6}%
  \BibitemOpen
  \bibinfo {note} {Actually, the mass in Eq.\protect \textup {\hbox
  {\mathsurround \z@ \protect \normalfont (\ignorespaces \ref {Bloch
  hamiltonian2}\unskip \@@italiccorr )}} can also depend on $\protect \vec
  {k}$, as long as the Pontryagin indexes are the same for each $\protect \vec
  {k}$. We give the details about this point in the Appendix \ref {General
  Theory}.}\BibitemShut {Stop}%
\bibitem [{\citenamefont {Hsin}\ \emph {et~al.}(2020)\citenamefont {Hsin},
  \citenamefont {Kapustin},\ and\ \citenamefont
  {Thorngren}}]{PhysRevB.102.245113}%
  \BibitemOpen
  \bibfield  {author} {\bibinfo {author} {\bibfnamefont {P.-S.}\ \bibnamefont
  {Hsin}}, \bibinfo {author} {\bibfnamefont {A.}~\bibnamefont {Kapustin}},\
  and\ \bibinfo {author} {\bibfnamefont {R.}~\bibnamefont {Thorngren}},\
  }\bibfield  {title} {\bibinfo {title} {Berry phase in quantum field theory:
  Diabolical points and boundary phenomena},\ }\href
  {https://doi.org/10.1103/PhysRevB.102.245113} {\bibfield  {journal} {\bibinfo
   {journal} {Phys. Rev. B}\ }\textbf {\bibinfo {volume} {102}},\ \bibinfo
  {pages} {245113} (\bibinfo {year} {2020})}\BibitemShut {NoStop}%
\bibitem [{\citenamefont {Huang}\ and\ \citenamefont
  {Lee}(2021)}]{huang2021non}%
  \BibitemOpen
  \bibfield  {author} {\bibinfo {author} {\bibfnamefont {Y.-T.}\ \bibnamefont
  {Huang}}\ and\ \bibinfo {author} {\bibfnamefont {D.-H.}\ \bibnamefont
  {Lee}},\ }\bibfield  {title} {\bibinfo {title} {Non-abelian bosonization in
  two and three spatial dimensions and applications},\ }\href
  {https://doi.org/https://doi.org/10.1016/j.nuclphysb.2021.115565} {\bibfield
  {journal} {\bibinfo  {journal} {Nuclear Physics B}\ }\textbf {\bibinfo
  {volume} {972}},\ \bibinfo {pages} {115565} (\bibinfo {year}
  {2021})}\BibitemShut {NoStop}%
\bibitem [{Note7()}]{Note7}%
  \BibitemOpen
  \bibinfo {note} {They can describe the usual Neel order and VBS order around
  the DQCP point. Besides, they can also describe the noncollinear order around
  Dirac spin liquids on the triangular lattice and kagome lattice.}\BibitemShut
  {Stop}%
\bibitem [{Note8()}]{Note8}%
  \BibitemOpen
  \bibinfo {note} {The mass term of the Hamiltonian \protect \textup {\hbox
  {\mathsurround \z@ \protect \normalfont (\ignorespaces \ref {dirac_6}\unskip
  \@@italiccorr )}} can also be written in the spinon basis $ f(\protect
  \mathbf {k})=(c_A(\protect \mathbf {k}+\protect \mathbf {K}),c_B(\protect
  \mathbf {k}+\protect \mathbf {K}),c_A(\protect \mathbf {k}-\protect \mathbf
  {K}),c_B(\protect \mathbf {k}-\protect \mathbf {K}))^T$, where $\pm \protect
  \mathbf {K}$ are the momentum of the two Dirac cones. The mass term under
  this new basis gives a clear physical meaning of the order parameter as
  density wave on the lattice. The Chern number is the same as the Hamiltonian
  \protect \textup {\hbox {\mathsurround \z@ \protect \normalfont
  (\ignorespaces \ref {dirac_6}\unskip \@@italiccorr )}}, since both mass terms
  gap out the Dirac cones and the gap is not closed when we deform one Chern
  insulator into the other. As a result, these two Chern insulators in turn
  give the same quantized levels of the WZW terms as $\protect \frac
  {C_3}{P_4}$. This also indicates the level of the WZW term is intrinsically
  determined by the low energy fermions in the QED theory.}\BibitemShut {Stop}%
\bibitem [{\citenamefont {Sun}\ \emph {et~al.}(2009)\citenamefont {Sun},
  \citenamefont {Yao}, \citenamefont {Fradkin},\ and\ \citenamefont
  {Kivelson}}]{PhysRevLett.103.046811}%
  \BibitemOpen
  \bibfield  {author} {\bibinfo {author} {\bibfnamefont {K.}~\bibnamefont
  {Sun}}, \bibinfo {author} {\bibfnamefont {H.}~\bibnamefont {Yao}}, \bibinfo
  {author} {\bibfnamefont {E.}~\bibnamefont {Fradkin}},\ and\ \bibinfo {author}
  {\bibfnamefont {S.~A.}\ \bibnamefont {Kivelson}},\ }\bibfield  {title}
  {\bibinfo {title} {Topological insulators and nematic phases from spontaneous
  symmetry breaking in 2d fermi systems with a quadratic band crossing},\
  }\href {https://doi.org/10.1103/PhysRevLett.103.046811} {\bibfield  {journal}
  {\bibinfo  {journal} {Phys. Rev. Lett.}\ }\textbf {\bibinfo {volume} {103}},\
  \bibinfo {pages} {046811} (\bibinfo {year} {2009})}\BibitemShut {NoStop}%
\bibitem [{\citenamefont {Li}\ \emph {et~al.}(2021)\citenamefont {Li},
  \citenamefont {He},\ and\ \citenamefont
  {Yao}}]{https://doi.org/10.48550/arxiv.2111.12107}%
  \BibitemOpen
  \bibfield  {author} {\bibinfo {author} {\bibfnamefont {M.-R.}\ \bibnamefont
  {Li}}, \bibinfo {author} {\bibfnamefont {A.-L.}\ \bibnamefont {He}},\ and\
  \bibinfo {author} {\bibfnamefont {H.}~\bibnamefont {Yao}},\ }\href
  {https://doi.org/10.48550/ARXIV.2111.12107} {\bibinfo {title} {Magic-angle
  twisted bilayer systems with quadratic-band-touching: Exactly flat bands with
  high-chern number}} (\bibinfo {year} {2021})\BibitemShut {NoStop}%
\bibitem [{Note9()}]{Note9}%
  \BibitemOpen
  \bibinfo {note} {Here take unit vectors as $a_1= (-\protect \frac {\protect
  \sqrt {3}}{2},\protect \frac {1}{2})a$ and $a_2=(0,1)a$. And $k_1=-\protect
  \frac {\protect \sqrt {3}}{2}k_x+\protect \frac {1}{2}k_y,k_2=k_y$
  here.}\BibitemShut {Stop}%
\bibitem [{\citenamefont {Komargodski}\ \emph {et~al.}(2019)\citenamefont
  {Komargodski}, \citenamefont {Sharon}, \citenamefont {Thorngren},\ and\
  \citenamefont {Zhou}}]{10.21468/SciPostPhys.6.1.003}%
  \BibitemOpen
  \bibfield  {author} {\bibinfo {author} {\bibfnamefont {Z.}~\bibnamefont
  {Komargodski}}, \bibinfo {author} {\bibfnamefont {A.}~\bibnamefont {Sharon}},
  \bibinfo {author} {\bibfnamefont {R.}~\bibnamefont {Thorngren}},\ and\
  \bibinfo {author} {\bibfnamefont {X.}~\bibnamefont {Zhou}},\ }\bibfield
  {title} {\bibinfo {title} {{Comments on Abelian Higgs Models and Persistent
  Order}},\ }\href {https://doi.org/10.21468/SciPostPhys.6.1.003} {\bibfield
  {journal} {\bibinfo  {journal} {SciPost Phys.}\ }\textbf {\bibinfo {volume}
  {6}},\ \bibinfo {pages} {3} (\bibinfo {year} {2019})}\BibitemShut {NoStop}%
\bibitem [{\citenamefont {Imura}\ \emph {et~al.}(2012)\citenamefont {Imura},
  \citenamefont {Yoshimura}, \citenamefont {Takane},\ and\ \citenamefont
  {Fukui}}]{imura2012spherical}%
  \BibitemOpen
  \bibfield  {author} {\bibinfo {author} {\bibfnamefont {K.-I.}\ \bibnamefont
  {Imura}}, \bibinfo {author} {\bibfnamefont {Y.}~\bibnamefont {Yoshimura}},
  \bibinfo {author} {\bibfnamefont {Y.}~\bibnamefont {Takane}},\ and\ \bibinfo
  {author} {\bibfnamefont {T.}~\bibnamefont {Fukui}},\ }\bibfield  {title}
  {\bibinfo {title} {Spherical topological insulator},\ }\href
  {https://doi.org/10.1103/PhysRevB.86.235119} {\bibfield  {journal} {\bibinfo
  {journal} {Phys. Rev. B}\ }\textbf {\bibinfo {volume} {86}},\ \bibinfo
  {pages} {235119} (\bibinfo {year} {2012})}\BibitemShut {NoStop}%
\bibitem [{\citenamefont {Lu}\ and\ \citenamefont {Herbut}(2012)}]{2012Zero}%
  \BibitemOpen
  \bibfield  {author} {\bibinfo {author} {\bibfnamefont {C.-K.}\ \bibnamefont
  {Lu}}\ and\ \bibinfo {author} {\bibfnamefont {I.~F.}\ \bibnamefont
  {Herbut}},\ }\bibfield  {title} {\bibinfo {title} {Zero modes and charged
  skyrmions in graphene bilayer},\ }\href
  {https://doi.org/10.1103/PhysRevLett.108.266402} {\bibfield  {journal}
  {\bibinfo  {journal} {Phys. Rev. Lett.}\ }\textbf {\bibinfo {volume} {108}},\
  \bibinfo {pages} {266402} (\bibinfo {year} {2012})}\BibitemShut {NoStop}%
\bibitem [{Note10()}]{Note10}%
  \BibitemOpen
  \bibinfo {note} {Strictly speaking, this statement can be made precise in the
  QED theory with Dirac fermions by the state-operator correspondence \cite
  {borokhov2003topological}. And the state-operator correspondence is not
  necessarily to hold in nonrelativistic $\protect \text {QED}$ theory here.
  However, we can deform each fermion with quadratic band touching into two
  Dirac fermions. For example, if the $t_{\perp }$ is zero in the AB stacked
  bilayer honeycomb lattice model or the AA stacked bilayer triangular lattice
  model above, each QBT splits into two degenerate Dirac cones. Moreover, these
  two theories have the same kinematic properties, since their dual topological
  NLSM descriptions have the same WZW terms on the Grassmannian manifold. As a
  result, it is quite reasonable to conjecture that the gauge-invariant
  monopole operators of the nonrelativistic $\protect \text {QED}$ here are the
  same as the $N_f=8$ relativistic $\protect \text {QED}$ theory. And the
  gauge-invariant monopole operators of the latter theory correspond to the
  states which occupy half of the total zero modes in the presence of unit
  magnetic flux.}\BibitemShut {Stop}%
\bibitem [{Note11()}]{Note11}%
  \BibitemOpen
  \bibinfo {note} {The second topological WZW term can be understood as U(1)
  gauge field coupled with two spin-1 excitations. If we add a background SO(2)
  gauge field and integrate the $n$ field, the Berry phase of spin-1 excitation
  will give the flux of SO(2) gauge field.}\BibitemShut {Stop}%
\bibitem [{Note12()}]{Note12}%
  \BibitemOpen
  \bibinfo {note} {Here we can think of $CR_0$ as obtained from Wick-rotating
  time-reversal symmetry $T$ in the Lorentz signature by CRT theorem and
  monopole operators is invariant under $T$ symmetry}\BibitemShut {NoStop}%
\bibitem [{\citenamefont {Christos}\ \emph {et~al.}(2020)\citenamefont
  {Christos}, \citenamefont {Sachdev},\ and\ \citenamefont
  {Scheurer}}]{doi:10.1073/pnas.2014691117}%
  \BibitemOpen
  \bibfield  {author} {\bibinfo {author} {\bibfnamefont {M.}~\bibnamefont
  {Christos}}, \bibinfo {author} {\bibfnamefont {S.}~\bibnamefont {Sachdev}},\
  and\ \bibinfo {author} {\bibfnamefont {M.~S.}\ \bibnamefont {Scheurer}},\
  }\bibfield  {title} {\bibinfo {title} {Superconductivity, correlated
  insulators, and wess-zumino-witten terms in twisted bilayer graphene},\
  }\href {https://doi.org/10.1073/pnas.2014691117} {\bibfield  {journal}
  {\bibinfo  {journal} {Proceedings of the National Academy of Sciences}\
  }\textbf {\bibinfo {volume} {117}},\ \bibinfo {pages} {29543} (\bibinfo
  {year} {2020})},\ \Eprint
  {https://arxiv.org/abs/https://www.pnas.org/doi/pdf/10.1073/pnas.2014691117}
  {https://www.pnas.org/doi/pdf/10.1073/pnas.2014691117} \BibitemShut {NoStop}%
\bibitem [{\citenamefont {Wu}\ \emph {et~al.}(2022{\natexlab{a}})\citenamefont
  {Wu}, \citenamefont {Wu},\ and\ \citenamefont
  {Yao}}]{https://doi.org/10.48550/arxiv.2203.05480}%
  \BibitemOpen
  \bibfield  {author} {\bibinfo {author} {\bibfnamefont {Y.-M.}\ \bibnamefont
  {Wu}}, \bibinfo {author} {\bibfnamefont {Z.}~\bibnamefont {Wu}},\ and\
  \bibinfo {author} {\bibfnamefont {H.}~\bibnamefont {Yao}},\ }\href
  {https://doi.org/10.48550/ARXIV.2203.05480} {\bibinfo {title}
  {Pair-density-wave and chiral superconductivity in twisted bilayer
  transition-metal-dichalcogenides}} (\bibinfo {year}
  {2022}{\natexlab{a}})\BibitemShut {NoStop}%
\bibitem [{\citenamefont {Wu}\ \emph {et~al.}(2022{\natexlab{b}})\citenamefont
  {Wu}, \citenamefont {Wu},\ and\ \citenamefont
  {Wu}}]{https://doi.org/10.48550/arxiv.2207.11468}%
  \BibitemOpen
  \bibfield  {author} {\bibinfo {author} {\bibfnamefont {Z.}~\bibnamefont
  {Wu}}, \bibinfo {author} {\bibfnamefont {Y.-m.}\ \bibnamefont {Wu}},\ and\
  \bibinfo {author} {\bibfnamefont {F.}~\bibnamefont {Wu}},\ }\href
  {https://doi.org/10.48550/ARXIV.2207.11468} {\bibinfo {title} {Pair density
  wave and loop current promoted by van hove singularities in moiré systems}}
  (\bibinfo {year} {2022}{\natexlab{b}})\BibitemShut {NoStop}%
\bibitem [{\citenamefont {Zhang}\ and\ \citenamefont
  {Mao}(2020)}]{PhysRevB.101.035122}%
  \BibitemOpen
  \bibfield  {author} {\bibinfo {author} {\bibfnamefont {Y.-H.}\ \bibnamefont
  {Zhang}}\ and\ \bibinfo {author} {\bibfnamefont {D.}~\bibnamefont {Mao}},\
  }\bibfield  {title} {\bibinfo {title} {Spin liquids and pseudogap metals in
  the su(4) hubbard model in a moir\'e superlattice},\ }\href
  {https://doi.org/10.1103/PhysRevB.101.035122} {\bibfield  {journal} {\bibinfo
   {journal} {Phys. Rev. B}\ }\textbf {\bibinfo {volume} {101}},\ \bibinfo
  {pages} {035122} (\bibinfo {year} {2020})}\BibitemShut {NoStop}%
\bibitem [{\citenamefont {Zhang}\ and\ \citenamefont
  {Senthil}(2019)}]{PhysRevB.99.205150}%
  \BibitemOpen
  \bibfield  {author} {\bibinfo {author} {\bibfnamefont {Y.-H.}\ \bibnamefont
  {Zhang}}\ and\ \bibinfo {author} {\bibfnamefont {T.}~\bibnamefont
  {Senthil}},\ }\bibfield  {title} {\bibinfo {title} {Bridging hubbard model
  physics and quantum hall physics in trilayer
  $\text{graphene}/h\ensuremath{-}\mathrm{BN}$ moir\'e superlattice},\ }\href
  {https://doi.org/10.1103/PhysRevB.99.205150} {\bibfield  {journal} {\bibinfo
  {journal} {Phys. Rev. B}\ }\textbf {\bibinfo {volume} {99}},\ \bibinfo
  {pages} {205150} (\bibinfo {year} {2019})}\BibitemShut {NoStop}%
\bibitem [{\citenamefont {Else}\ \emph {et~al.}(2021)\citenamefont {Else},
  \citenamefont {Thorngren},\ and\ \citenamefont
  {Senthil}}]{PhysRevX.11.021005}%
  \BibitemOpen
  \bibfield  {author} {\bibinfo {author} {\bibfnamefont {D.~V.}\ \bibnamefont
  {Else}}, \bibinfo {author} {\bibfnamefont {R.}~\bibnamefont {Thorngren}},\
  and\ \bibinfo {author} {\bibfnamefont {T.}~\bibnamefont {Senthil}},\
  }\bibfield  {title} {\bibinfo {title} {Non-fermi liquids as ersatz fermi
  liquids: General constraints on compressible metals},\ }\href
  {https://doi.org/10.1103/PhysRevX.11.021005} {\bibfield  {journal} {\bibinfo
  {journal} {Phys. Rev. X}\ }\textbf {\bibinfo {volume} {11}},\ \bibinfo
  {pages} {021005} (\bibinfo {year} {2021})}\BibitemShut {NoStop}%
\bibitem [{\citenamefont {Mross}\ and\ \citenamefont
  {Senthil}(2011)}]{PhysRevB.84.165126}%
  \BibitemOpen
  \bibfield  {author} {\bibinfo {author} {\bibfnamefont {D.~F.}\ \bibnamefont
  {Mross}}\ and\ \bibinfo {author} {\bibfnamefont {T.}~\bibnamefont
  {Senthil}},\ }\bibfield  {title} {\bibinfo {title} {Decohering the fermi
  liquid: A dual approach to the mott transition},\ }\href
  {https://doi.org/10.1103/PhysRevB.84.165126} {\bibfield  {journal} {\bibinfo
  {journal} {Phys. Rev. B}\ }\textbf {\bibinfo {volume} {84}},\ \bibinfo
  {pages} {165126} (\bibinfo {year} {2011})}\BibitemShut {NoStop}%
\bibitem [{\citenamefont {Song}\ \emph {et~al.}(2021)\citenamefont {Song},
  \citenamefont {Vishwanath},\ and\ \citenamefont
  {Zhang}}]{PhysRevB.103.165138}%
  \BibitemOpen
  \bibfield  {author} {\bibinfo {author} {\bibfnamefont {X.-Y.}\ \bibnamefont
  {Song}}, \bibinfo {author} {\bibfnamefont {A.}~\bibnamefont {Vishwanath}},\
  and\ \bibinfo {author} {\bibfnamefont {Y.-H.}\ \bibnamefont {Zhang}},\
  }\bibfield  {title} {\bibinfo {title} {Doping the chiral spin liquid:
  Topological superconductor or chiral metal},\ }\href
  {https://doi.org/10.1103/PhysRevB.103.165138} {\bibfield  {journal} {\bibinfo
   {journal} {Phys. Rev. B}\ }\textbf {\bibinfo {volume} {103}},\ \bibinfo
  {pages} {165138} (\bibinfo {year} {2021})}\BibitemShut {NoStop}%
\bibitem [{\citenamefont {Senthil}\ \emph {et~al.}(2019)\citenamefont
  {Senthil}, \citenamefont {Son}, \citenamefont {Wang},\ and\ \citenamefont
  {Xu}}]{SENTHIL20191}%
  \BibitemOpen
  \bibfield  {author} {\bibinfo {author} {\bibfnamefont {T.}~\bibnamefont
  {Senthil}}, \bibinfo {author} {\bibfnamefont {D.~T.}\ \bibnamefont {Son}},
  \bibinfo {author} {\bibfnamefont {C.}~\bibnamefont {Wang}},\ and\ \bibinfo
  {author} {\bibfnamefont {C.}~\bibnamefont {Xu}},\ }\bibfield  {title}
  {\bibinfo {title} {Duality between (2+1)d quantum critical points},\ }\href
  {https://doi.org/https://doi.org/10.1016/j.physrep.2019.09.001} {\bibfield
  {journal} {\bibinfo  {journal} {Physics Reports}\ }\textbf {\bibinfo {volume}
  {827}},\ \bibinfo {pages} {1} (\bibinfo {year} {2019})},\ \bibinfo {note}
  {duality between (2+1)d quantum critical points}\BibitemShut {NoStop}%
\bibitem [{\citenamefont {Borokhov}\ \emph {et~al.}(2002)\citenamefont
  {Borokhov}, \citenamefont {Kapustin},\ and\ \citenamefont
  {Wu}}]{borokhov2003topological}%
  \BibitemOpen
  \bibfield  {author} {\bibinfo {author} {\bibfnamefont {V.}~\bibnamefont
  {Borokhov}}, \bibinfo {author} {\bibfnamefont {A.}~\bibnamefont {Kapustin}},\
  and\ \bibinfo {author} {\bibfnamefont {X.}~\bibnamefont {Wu}},\ }\bibfield
  {title} {\bibinfo {title} {Topological disorder operators in
  three-dimensional conformal field theory},\ }\href
  {https://doi.org/10.1088/1126-6708/2002/11/049} {\bibfield  {journal}
  {\bibinfo  {journal} {Journal of High Energy Physics}\ }\textbf {\bibinfo
  {volume} {2002}},\ \bibinfo {pages} {049} (\bibinfo {year}
  {2002})}\BibitemShut {NoStop}%
\end{thebibliography}%

\end{document}